\documentclass{aa}  
\usepackage{natbib}
\bibpunct{(}{)}{;}{a}{}{,} 
\usepackage{graphicx}
\usepackage{txfonts}
\usepackage{xcolor}
\usepackage{multirow}
\usepackage{tablefootnote}
\usepackage{amsmath}
\usepackage{footnote}
\usepackage{pdflscape}
\usepackage{longtable}
\usepackage[flushleft]{threeparttable}
\usepackage{dirtytalk}
\usepackage[final]{hyperref}
\hypersetup{
	colorlinks=true, 
	linkcolor=blue,
	citecolor=blue,
	filecolor=magenta,
	urlcolor=blue         
}
\usepackage{pifont}
\newcommand{\cmark}{\ding{51}}%
\newcommand{\xmark}{\ding{55}}%
\begin{document} 
\authorrunning{Tong et al.}

   \title{Disk survey in the Serpens star-forming region: Environmental effects in nearby star-forming regions}
   \author{
          Simin Tong \inst{1,2}, Nienke van der Marel \inst{1}, Jonathan P. Williams \inst{3}, Alexa R. Anderson \inst{3}
          }
   \institute{Leiden Observatory, Leiden University, Niels Bohrweg 2, 2333 CA Leiden, The Netherlands
    \and 
    School of Physics \& Astronomy, University of Leicester, University Road, Leicester, LE1 7RH, UK \\
    \email{st547@leicester.ac.uk, astro.stong@gmail.com} 
    \and 
    Institute for Astronomy, University of Hawai`i at Mānoa, 2680 Woodlawn Drive, Honolulu, HI 96822, USA}
         
   \date{Received Month Date Year ; accepted Month Date Year}

  \abstract
   {The external environment where protoplanetary disks are embedded regulates disk evolution. External irradiation, which heats and evaporates gas, as well as frequent stellar encounters, which truncate disks, can reduce disk sizes and masses. Disk surveys in nearby star-forming regions with various environments can help us understand how external environments impact disk evolution. The Serpens star-forming region, which is thought to be dense but not highly irradiated by massive stars, is an ideal laboratory to study the dynamical effects on disk properties.}
   {We aim to study how dynamical interactions modify disk dust masses by comparing the Serpens star-forming region (distance $\sim400~$pc) with other nearby star-forming regions through their disk masses and 3D stellar densities.}
   {We survey 321 young stellar objects from Class I to III in Serpens using the Atacama Large Millimeter/submillimeter Array (ALMA) at a resolution of $\sim$ 0.\arcsec25. We measure disk dust masses from millimeter observations under the optically thin assumption and recompute the 3D stellar density using Gaia data. We apply the same method to other nearby star-forming regions for direct comparison with Serpens.}
   {The cumulative disk dust mass distribution of Serpens is similar to those of similarly aged nearby star-forming regions, such as Lupus and Taurus. Along with re-assessment of the 3D stellar density, it suggests that Serpens is not likely to have experienced strong tidal truncation capable of producing lower disk dust masses. We also find that the low disk dust mass tension in Ophiuchus and Corona Australis can be resolved when only disks that have been explicitly identified as members of young (1-2 Myr) sub-clusters are considered. Disks without Gaia identifications, especially in Ophiuchus, are spatially clustered in the 2D projected sky plane and less massive than nearby disks with Gaia identifications, possibly tracing the effects of tidal truncation in denser environments at earlier evolutionary stages.} %
   {}

   \maketitle
   
\section{Introduction}
Protoplanetary disks are widely acknowledged as the cradles of planets. Characterizing the reservoirs of gas and dust and their radial extent is important to understand the demographics of exoplanets that can form in them \citep[e.g.][and references therein]{2023ASPC..534..539M}. 

The homogeneous ALMA disk survey towards nearby star-forming regions is an efficient approach to measure disk masses and sizes at a population level. Recent disk surveys show that disk dust masses decrease over time. Within a given star-forming region, the early-stage Class 0/I disks are generally more massive than the more evolved Class II and III disks \citep[e.g.][]{2019ApJ...875L...9W, 2021ApJ...913..123G, 2021ApJ...913..149E, 2021MNRAS.500.4878L, 2022ApJ93855A}. Intra-region comparison further suggests that dust disks from younger star-forming regions are typically more massive than those from older regions \citep[e.g.][]{2016ApJ...827..142B, 2017AJ....153..240A, 2023ASPC..534..539M, 2025ApJ...989....1Z}. However, disk evolution is not governed by age alone. External environments, such as external radiation, dynamical interactions and infall, can also significantly modify disk masses and sizes \citep[e.g.][and references therein]{2021ApJ...923..221O, 2022EPJP..137.1132W, 2023EPJP..138..272K}, and in some cases, reshape the expected age-dependent trends and disk evolution.

Disks born in a harsher environment, where external irradiation or tidal truncation is strong, tend to have smaller disk sizes and apparently lower masses under the optically thin assumption \citep[e.g.][]{2012ApJ...751..115H, 2014ApJ...784...82M, 2019ApJ...872..158A}. Disks exposed to external radiation fields have been extensively studied both theoretically \citep[e.g.][]{1998ApJ...499..758J, 2018MNRAS.478.2700W, 2020MNRAS.492.1279S, 2022MNRAS.514.2315C, 2023MNRAS.522.1939Q, 2024A&A...681A..84G} and observationally \citep[e.g.][]{2018ApJ...860...77E, 2023A&A...679A..82M, 2024ApJ...976..132H, 2025A&A...701A.139R}. Observational efforts have primarily focused on the Orion star-forming region \citep[e.g.][]{2017AJ....153..240A, 2023A&A...673L...2V, 2024A&A...687A..93A}, where disks are strongly irradiated by nearby massive stars. Star-forming regions containing a substantial population of massive stars are expected to have enhanced radiation levels and stellar densities. When both physical processes are present, theoretical studies suggest that external photoevaporation usually dominates over tidal truncation unless the region is very dense \citep{2018MNRAS.478.2700W}. To study the effect of dynamical encounters on disc evolution alone, we therefore have to shift our focus to regions with high stellar densities but comparatively weak external irradiation \citep[e.g.][]{2021ApJ...923..221O}. The scarcity of O/B stars and the relatively higher stellar surface density in Serpens \citep{2007ApJ...663.1149H,2009ApJS..181..321E,2016AJ....151....5M} therefore make it a unique testbed.

The Serpens star-forming region is centered around $\alpha\sim278^{\circ}$ in right ascension $\delta \sim 0^{\circ}$ in declination, at a distance of $370$--$500~$pc \citep{2019ApJ...878..111H}. The optical extinction map shows the entire cloud spans $>10~\mathrm{pc^{2}}$ \citep{1999A&A...345..965C} and the active star formation region covers $\sim1.5~\mathrm{pc^{2}}$ \citep{2007ApJ...666..982E}. The star formation in this region is proposed as a result of cloud-cloud collisions \citep{2011A&A...528A..50D} and was first recognized as an active star-forming region by \citet{1974ApJ...191..111S}. Serpens has been observed in millimeter wavelengths across various length scales, ranging from the entire cloud \citep[e.g.][]{2010MNRAS.409.1412G}, to filaments \citep[e.g.][]{2014ApJ...797...76L, 2018ApJ...853..169D}, cloud cores \citep[e.g.][]{2010A&A...523A..29D}, condensations \citep[e.g.][]{2019ApJ...887..209A} and down to individual protoplanetary systems \citep[e.g.][]{2017AJ....154..255L, 2022ApJ93855A, 2024ApJ...973..138H}. 

Previous millimeter disk surveys dedicated to Serpens show that the disk dust masses in Serpens are comparable to those in Lupus and Taurus \citep{2017AJ....154..255L, 2022ApJ93855A}, and appear insensitive to the proposed high stellar density. However, these studies did not dive into the underlying physics of the similarity between regions. Also, since they were done at low angular resolution, it is possible that continuum fluxes were confused by nearby objects. In this paper, we analyze ALMA Band 6 ($\lambda=1.3~
\mathrm{mm}$) observations of 321 disks in Serpens with a resolution of $\sim$0.\arcsec25 and systematically study their disk dust masses, assisted with the 3D stellar density computed from Gaia data. We compare Serpens with other nearby star-forming regions on these properties, and explore the potential underlying physics between disk dust masses and stellar densities. This paper is organized as follows: we show the sample selection for this survey in Section \ref{sec:sample}. We present the observation and results in Sections \ref{sec:obs} and \ref{sec:results}, respectively. We discuss the environmental effects on disk evolution in Section \ref{sec:discussion} and conclude our findings in Section \ref{sec:conclusion}. 

\section{Sample Selection}\label{sec:sample}
We selected young stellar object candidates in the Serpens star-forming region from the Spitzer c2d and Gould Belt surveys \citep{2009ApJS..181..321E, 2015ApJS..220...11D}. We apply three criteria to (a) exclude very low-mass stars ($\lesssim 0.2M_\odot$); (b) exclude objects that are far beyond Serpens ($d>650~$pc); and (c) preferentially include Class I--II disks \citep[through the infrared spectral slope $\alpha_\mathrm{IR}=d\log(\lambda F_\lambda)/d\log\lambda>-1.6$,][]{2011ARA&A..49...67W}. This yields 321 disks in total. These samples are re-classified as discussed in \cite{2022ApJ93855A} to account for the high optical extinction in Serpens due to its proximity to the Galactic plane. \cite{2022ApJ93855A} show the high optical extinction can bias the SED-based classification by ``rejuvenating'' targets, i.e. shifting Class III to Class II disks, and Class II to Flat Spectrum. To mitigate this, the spectral slope used for classification is computed from 3.6 to 24$~\mu m$ rather than from 2 to 24$~\mu m$. This reclassification results in 16 Class I, 39 Flat Spectrum, 250 Class II and 16 Class III disks. 

\section{Observations}\label{sec:obs}
\subsection{Data and data reduction}
The sample was observed with Band 6 receivers during ALMA Cycle 7 (2019.1.00218.S, PI: van der Marel) in both compact and extended configurations. This paper mainly focuses on the higher-resolution dataset; analysis of the lower-resolution data is presented in \cite{2022ApJ93855A}, which achieves a typical resolution of $\sim 1.^{\prime\prime}2$ and a typical root-mean-square noise (rms) of $0.45~\mathrm{mJy~beam^{-1}}$ (see Table \ref{table:obs_info}). The detection rate for the lower-resolution observations is discussed in Section \ref{subsec:detections} and summarized in Table \ref{tb:detection_rate}. 

The higher-resolution observations consist of three execution blocks observed between May and June 2021 with baselines of $15-2517~$m. The spectral setup includes three broad continuum spectral windows, centered at $232.810$, $245.009$ and $247.508$~GHz, and one narrow window covering the $\mathrm{{}^{12}CO}$ 2-1 line, centered at $240.16$~GHz with a resolution of $0.32~\mathrm{km~s^{-1}}$. Details of each execution block are summarized in Table \ref{table:obs_info}.

\begin{table*}
\renewcommand{\arraystretch}{1.1}
\centering
\caption[]{Observation information.}
\begin{tabular}{ccccccccccc}

\hline\hline
   & Date & \multicolumn{2}{c}{Baselines [m]} & Int. time [s] & \multicolumn{3}{c}{Beam [${\arcsec}\times{\arcsec}]~$} & \multicolumn{3}{c}{rms [$\mathrm{mJy~beam^{-1}}$]} \\
  &  &max  &min &  & max & min& mean& max& min& mean   \\
   \hline
 \multirow{3}{*}{HR} &May 6-9, 2021 & 2517.3& 15 & 100 & $\mathrm{0.26\times0.21}$ & $\mathrm{0.22\times 0.21}$ & $\mathrm{0.24\times0.21}$ & $0.18$ & $0.08$  &  $0.10$\\
& Jun. 26, 2021& 2009.9& 23 & 25& $\mathrm{0.26\times0.22}$ & $\mathrm{0.26\times 0.21}$& $\mathrm{0.26\times0.22}$& $0.24$ & $0.14$ & $0.16$\\
& Jun. 30, 2021 & 2114.1 & 15  & 25 & $\mathrm{0.46\times 0.15}$& $\mathrm{0.36\times0.15}$& $\mathrm{0.40\times 0.15}$ & $0.17$ & $0.12$ & $0.14$ \\
\hline
 \multirow{3}{*}{LR}&  Dec. 19, 2019 & \multirow{3}{*}{313.7}  & \multirow{3}{*}{15.1} & \multirow{3}{*}{20}& \multirow{3}{*}{$1.4\times 1.0$}&  \multirow{3}{*}{$1.2\times 1.0$}& \multirow{3}{*}{--}& \multirow{3}{*}{3.2}& \multirow{3}{*}{0.13} & \multirow{3}{*}{0.45} \\
&Dec. 30, 2019&&&&        &        &        &        &       &        \\
&Dec. 30, 2019&&&&  &  & &  & &\\
\hline
\end{tabular}
\vspace{3pt}
\tablefoot{${}^\mathrm{a}$ HR and LR are abbreviations for high- and low-resolution observations, respectively. ${}^\mathrm{b}$ Information for the low-resolution observations is obtained from \cite{2022ApJ93855A}.}
\label{table:obs_info}
\end{table*}

Data reduction and imaging are performed with the Common Astronomy Software Applications package \citep[CASA][]{2007ASPC..376..127M, 2022PASP..134k4501C}. For the continuum, we use CASA task \texttt{tclean} with Briggs weighting (robust=0.5) to compromise between resolution and sensitivity. This yields a mean rms of $\sim 0.13~\mathrm{mJy~beam^{-1}}$, ranging from $0.08$ to $0.24~\mathrm{mJy~beam^{-1}}$ (Table \ref{table:obs_info}). For the $\mathrm{{}^{12}CO}$ 2-1 line, we subtract the continuum baseline using CASA task \texttt{uvcontsub}, and concatenate high- and low-resolution observations with CASA task \texttt{concatenate} to improve their uv coverage and the signal-to-noise ratio. This processing is applied to targets showing line emission other than cloud features in the higher resolution dataset. The concatenated visibilities are then imaged with a velocity resolution of $1~\mathrm{km/s}$. We generate the zeroth and first moment maps using \texttt{bettermoments} \citep{2018RNAAS...2c.173T}. Only pixels with values $>3\sigma$ are considered.

\subsection{Detections}\label{subsec:detections}
\begin{figure*}
    \centering
    \includegraphics[width=0.92\textwidth]{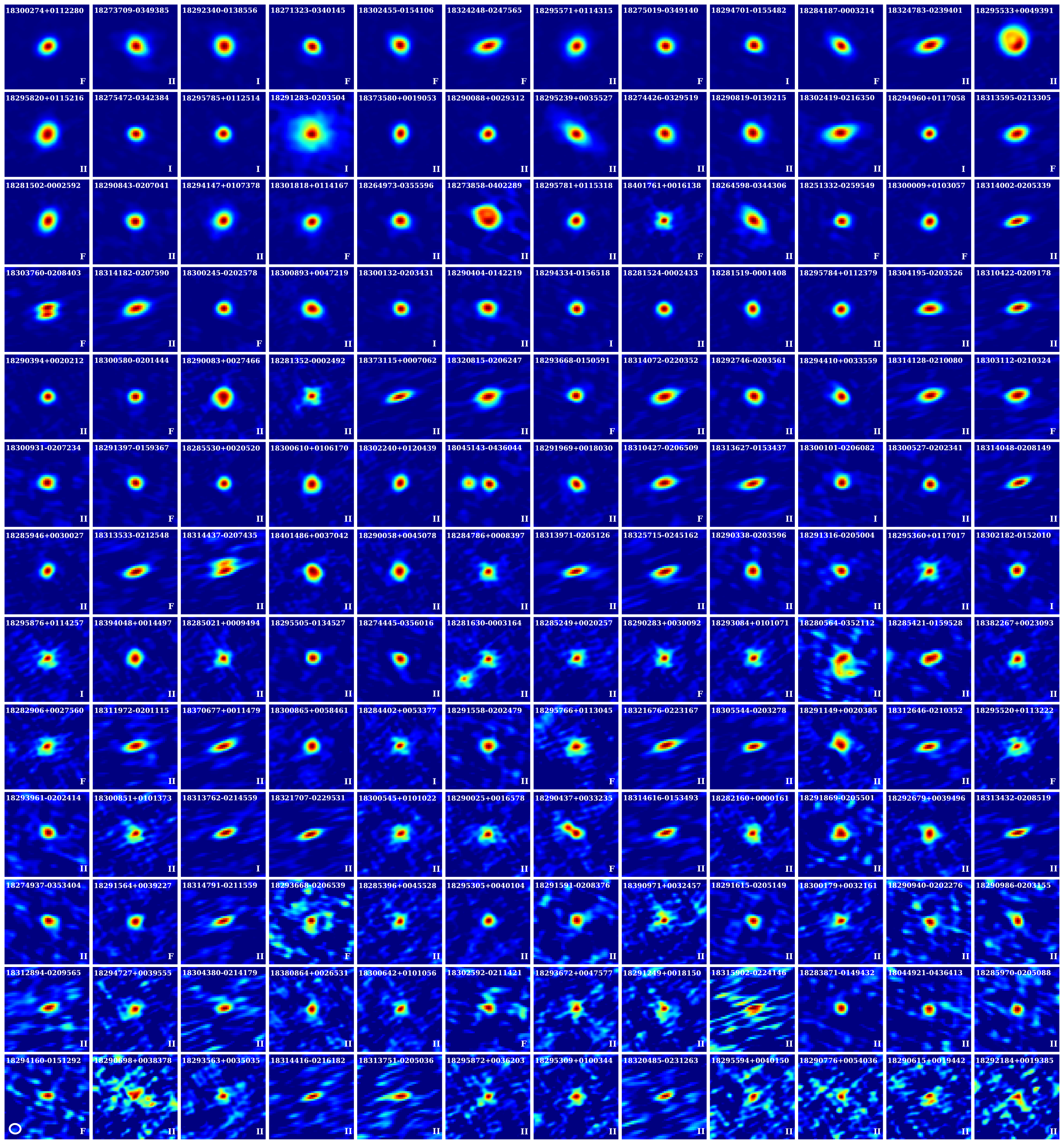}
    \caption{1.3-mm continuum detections from the higher-resolution dataset. The gallery is arranged by the continuum fluxes $F_\mathrm{1.3mm}$ measured through the aperture analysis (see more in Section \ref{subsec:disk_mass} and Table \ref{table:disk_info}). Each panel displays a detected young stellar object in a $2^{\prime \prime}\times2^{\prime \prime}$ box, scaling from $10^{-5}~\mathrm{mJy~beam^{-1}}$ to the maximum pixel value of the target. The spectral classification and the 2MASS ID are labelled in the lower left and upper middle corners of each panel, respectively. A representative beam of $0.^{\prime\prime}24 \times 0.^{\prime\prime}21$ ($-62.^{\circ}73$) is shown at the lower left corner of the figure.}
    \label{fig:detect_gallery}
\end{figure*}

We define a target as "detected" if the continuum emission is present within a $1\arcsec \times1\arcsec$ box centered on the field of view (FoV) and has an aperture-integrated signal-to-noise ratio (SNR) exceeding 5 (see Section \ref{subsec:disk_mass}). Targets with SNR$>5$ but lacking a clearly elliptical disk morphology on visual inspection are labelled as ``tentative detections''. We show detections for 156 disks in Fig. \ref{fig:detect_gallery}. The comparison for detection rates between the lower-resolution \citep{2022ApJ93855A} and the higher-resolution observations is shown in Table \ref{tb:detection_rate}. The higher-resolution observations have generally higher detection rates due to enhanced sensitivity (see Table \ref{table:obs_info}). 

\begin{table}[]
\caption{Detection rates of objects by class for lower- and higher-resolution observations.} \label{tb:detection_rate}
\centering
\renewcommand{\arraystretch}{1.15}
\begin{tabular}{cccccc}
\hline
\hline
   & Class I & Flat  & Class II & Class III & Ref.\\
\hline
low-res &  $87\%$ & $63\%$ & $37\%$  &  $0\%$ & 1 \\
high-res & $81\%$ & $74\%$& $45\%$ & $0\%$ & 2 \\
\hline
\end{tabular}
\tablefoot{1. \citet{2022ApJ93855A}; 2. this work.}
\end{table}
 
\begin{figure}
    \centering
    \includegraphics[width=0.46\textwidth]{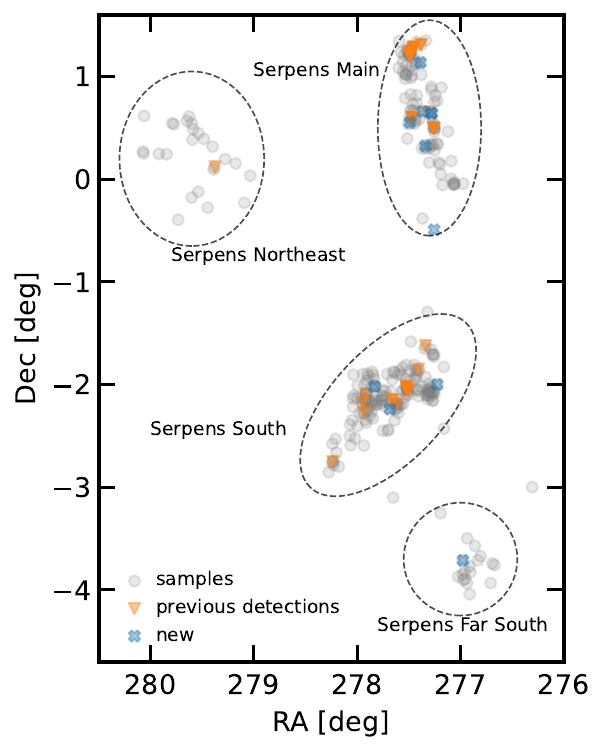}
    \caption{Sky scatter map of center and off-center sources that have been observed in this survey. Grey dots indicate young stellar objects of the original survey that fulfill the distance criteria (Section \ref{sec:sample}). Orange triangles are for young stellar objects that have been observed before but are not part of the original survey (Fig. \ref{fig:nd_gallery}). Blue crosses are newly identified millimeter sources (Fig. \ref{fig:ni_gallery}).}
    \label{fig:scatter}
\end{figure}

In addition to the central targets, we identify 40 disk candidates (e.g., from Spitzer surveys) that have been reported in previous studies but that are not part of our primary target list (Section~\ref{sec:sample}), and 11 new disk candidates that have not been reported before. These disk candidates typically lie away from the center of FoV and several newly found candidates are located close to previously known disks. The observed proximity can be attributed to projection effects and does not guarantee that these close systems are physically bound binary/multiple systems. We show these previously known and newly identified candidates in Fig. \ref{fig:nd_gallery} and Fig. \ref{fig:ni_gallery}, respectively, and summarize them at the end of Table \ref{table:disk_info}. Previously known disc candidates are also summarized in Table \ref{tb:nd_table}. These sources are excluded from the dust mass distribution analysis in Section \ref{subsec:disk_mass}, and we discussed the contributions from disks in Fig. \ref{fig:nd_gallery} to the cumulative disk dust mass distributions in Appendix \ref{appendix:tent_det} (Fig. \ref{fig:cumulative_distr_wnt})\footnote{Only 9 out of 40 previously known disks in Fig. \ref{fig:nd_gallery} have Gaia memberships. Of these, two are identified as ``distributed'' members of Serpens, which are not associated with the four sub-clusters shown in Fig.\ref{fig:scatter}, and the remaining seven are not identified as Serpens members \citep{2019ApJ...878..111H}. We note that Serpens South has high extinction in optical and few disks there have optical counterparts \citep{2019ApJ...878..111H}.}.

Fig. \ref{fig:scatter} shows a sky map of disks/disk candidates in this survey. Disks are distributed in four sub-clusters--Serpens main, Serpens Northeast, Serpens South and Serpens Far South--based on \cite{2019ApJ...878..111H}, appearing from upper left to lower right in the figure. A similar plot with color-encoded spectral classification is provided in Figure~2 of \citet{2022ApJ93855A}.

\section{Results}\label{sec:results}
\subsection{Disk dust mass}\label{subsec:disk_mass}
\begin{figure*}
    \centering
    \includegraphics[width=0.78\textwidth]{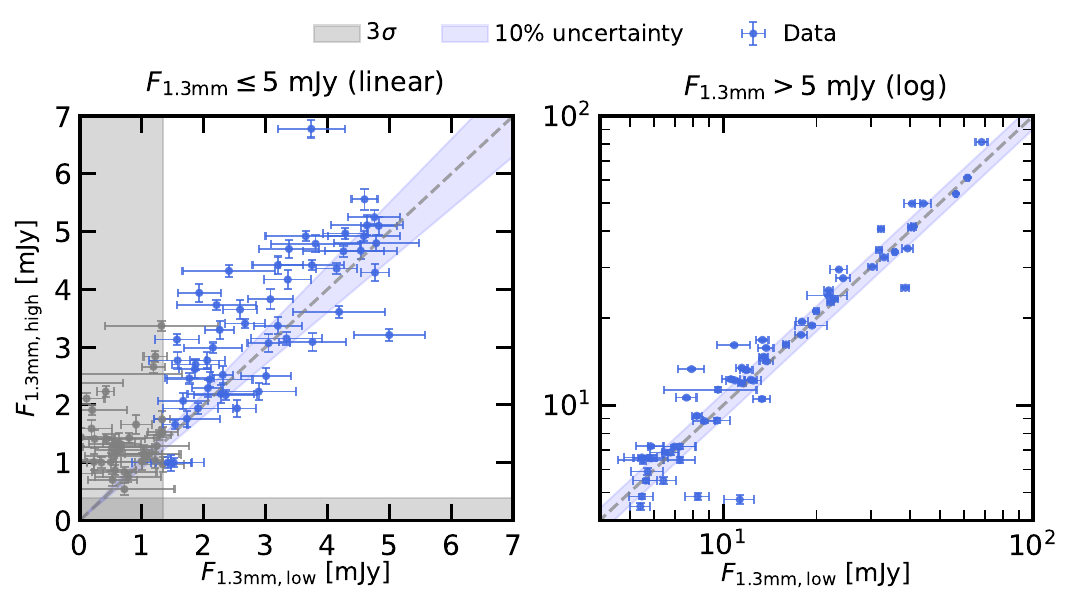}
    \caption{Comparison of continuum fluxes $F_\mathrm{1.3mm}$ between lower- \citep[horizontal axes,][]{2022ApJ93855A} and higher-resolution observations from this work (vertical axes). The left panel shows disks with $F_\mathrm{1.3mm}\leq5$ mJy in the linear scale, and the right panel shows disks with $F_\mathrm{1.3mm}>5$ mJy in the log scale. The log-scale panel better shows disks with continuum fluxes over a wide range and also disks that are sparsely distributed in the high-flux end. The error bars indicate the rms level. The vertical and horizontal gray shaded regions in the left panel indicate the $3\sigma$ limits of the lower- and higher-resolution observations, respectively. The gray dashed line marks equal continuum fluxes between the two measurements. The blue shaded regions show the $10\%$ calibration uncertainty.}
    \label{fig:flux_comp}
\end{figure*}

We measure continuum fluxes $F_\mathrm{1.3mm}$ and rms noise $\sigma$ using the aperture analysis described in \cite{2016ApJ...828...46A} and \cite{2022ApJ93855A}; their values are summarized in Table \ref{table:disk_info}. Briefly, a circular aperture is drawn around the maximum-valued pixel associated with the targets to measure fluxes. An aperture of $0.\arcsec5$ is adopted for targets with $F_\mathrm{1.3mm}\geq10~\mathrm{mJy}$ and a smaller aperture of $0.\arcsec35$ for fainter targets with $F_\mathrm{1.3mm}<10~\mathrm{mJy}$. These two radii are selected to ensure that the total emission from most targets is covered. The rms uncertainty $\sigma$, is measured by placing ten apertures identical to the one for flux measurement uniformly around the target at a radius of five times the flux-aperture radius. If an rms aperture overlaps an off-center source, we slightly adjust its location or the radius to avoid the contamination. For potential binary systems, we adjust the aperture radius so that only one component is measured each time. The continuum fluxes measured from the low- and high-resolution observations are compared in Fig. \ref{fig:flux_comp}. The differences in continuum fluxes between the lower- and higher- resolution observations are within $10\%$ calibration uncertainties for most targets. Among targets with continuum flux differences $>10\%$, most have higher fluxes measured from the higher-resolution data than from the lower-resolution data \citep{2022ApJ93855A}. This is because these disks have faint and radially extended emission that falls below the sensitivity of the low-resolution observations, and is therefore not fully accounted for. Targets with lower continuum fluxes in higher-resolution data can be explained by either (a) a source identified as a single object in \citet{2022ApJ93855A} being spatially resolved into 2 close Gaussian disks, possibly corresponding to a close binary; or (b) some large-scale noisy structures being included as disk emission in the lower-resolution observations.

Distances are derived from a simple inverse of parallaxes from Gaia Data Release 3 \citep[Gaia DR3,][]{2023A&A...674A...1G}. We adopt parallaxes if the parameter assessing the goodness of fit RUWE$\leq1.4$ \citep[renormalized unit weight error,][]{2018A&A...616A...2L}\footnote{A more relaxed threshold for RUWE \citep[e.g. 1.6 in][]{2020AJ....160..186L, 2020AJ....160...44L} is adopted in several membership censuses or no threshold is set \citep[e.g.][]{2020AJ....159..282E}. Applying the same criteria to samples here results in 3 more targets with RUWE$\geq1.6$ and 10 more targets when the RUWE threshold is ignored.} and if the parallax has a signal-to-noise ratio $\overline{\omega}/\sigma(\overline{\omega})\geq5$ \citep{2019ApJ...878..111H}. This yields 102 targets with valid Gaia-based distances. For targets without (valid) Gaia parallaxes, the median distance of the sub-clusters they are closest to in the projected sky is assigned. In the following analysis, we exclude two targets (2MASS J18045143-0436044 and J18044921-0436413) with right ascension away from the Serpens four sub-clusters (RA$\sim271^{\circ}$) and 2 targets (2MASS J18293672+0047577 and J18294122+0049020) that has $\overline{\omega}/\sigma(\overline{\omega})\geq5$ (RUWE$<2$) and is beyond the Serpens distance range accounting for the uncertainties. This results in a total of 317 disks that we consider as members of the Serpens star-forming region \citep{2019ApJ...878..111H, 2022ApJ93855A}\footnote{44 disks considered here are not identified as part of Serpens after crossing-match with Gaia memberships in \citet{2019ApJ...878..111H}. We compare disk dust mass distributions for Class II disks between these likely Serpens members and non-members, and no significant difference has been found.}. 

The disk dust masses are computed from the continuum fluxes and Gaia-based distances under three assumptions: (a) dust emission is optically thin at the observing wavelengths; (b) the dust opacity depends only on the frequency and the opacity coefficient follows $\kappa_\nu=(\nu/100~\mathrm{GHz})~\mathrm{cm^2~g^{-1}}$, which is $2.41~\mathrm{cm^2~g^{-1}}$ at our observing frequency; and (c) the dust temperature is spatially uniform and depends only on the classification of young stellar objects. We assume $T_\mathrm{dust}=20~$K for Class II disks \citep{2005ApJ...631.1134A}, $30~$K for Flat Spectrum disks \citep{2020A&A...640A..19T}, and $40~$K for Class I disks \citep{2020ApJ...905..162T}, as the protostellar envelopes can keep disks at earlier evolutionary stages warmer. We use
\begin{equation}\label{eq:dust_mass}
    M_\mathrm{dust,Class\,II}=\frac{F_\nu d^2}{\kappa_\nu B_\nu (T_\mathrm{dust})}=4.00 M_\oplus \bigg(\frac{F_\mathrm{1.3mm,Class\,II}}{1~\mathrm{mJy}}\bigg)\bigg(\frac{d}{400~\mathrm{pc}}\bigg)^2,
\end{equation}
where $B_\nu$ is the Planck function, $d$ is the distance and $F_\nu$ is the continuum flux at the observing frequency. The pre-factor in Eq. \ref{eq:dust_mass} is modified to 1.71 and 2.40 to adapt to Class I and Flat spectrum disks, respectively, to account for the change in $T_\mathrm{dust}$. For non-detections with $F_\mathrm{1.3mm}<5\sigma$, we report a $5$-$\sigma$ upper limit on the continuum fluxes.

\begin{table*}[ht]
\caption{Observed disk properties.}
\renewcommand{\arraystretch}{1.2}
\centering
\begin{threeparttable}
\scalebox{0.9}{
\begin{tabular}{cccccccccccc}
    \hline
    \hline
    (1) & (2) & (3) & (4) & (5) & (6) & (7) & (8) & (9) & (10) & (11) & (12)\\
    2MASS ID & RA & Dec & Class & $d$ & $\sigma_d$ & $F_\mathrm{1.3mm}$ & rms & $\sigma_\mathrm{tot}$ &  $M_\mathrm{dust}$ & $\sigma_M$ & Flag$^{*}$ \\
    & [$^{\circ}$] & [$^{\circ}$] & & [pc] & [pc] & [mJy] & [$\mathrm{mJy~beam^{-1}}$] & [mJy]  & [$M_\oplus$] & [$M_\oplus$] &  \\
    \hline
18300274+0112280&277.51138&1.20784&F&430.85&46.13&81.37&0.17&8.14&226.57&53.55&--\\
18273709-0349385&276.90457&-3.8274&II&379.74&16.45&61.24&0.24&6.13&220.77&29.23&--\\
18292340-0138556&277.34753&-1.64879&I&488.05&65.06&53.83&0.18&5.39&137.03&39.03&O\\
18271323-0340145&276.80514&-3.67071&F&476.47&23.93&49.89&0.2&4.99&169.89&24.08&--\\
18302455-0154106&277.60236&-1.90295&F&488.05&65.06&49.82&0.18&4.99&178.0&50.7&O\\
    \hline
\end{tabular}}
\vspace{3pt}
\tablefoot{The full Table is available at the CDS. Column 1: 2MASS Source ID; column 2: right ascension; column 3: declination; column 4: the infrared spectral classification of young stellar objects from \citet{2022ApJ93855A}; columns 5 and 6: Gaia-based distances and uncertainties; columns 7 and 8: continuum fluxes and rms noise measured from the aperture analysis (see Section \ref{subsec:disk_mass}); column 9: combined uncertainties computed as $\sigma_\mathrm{tot}=\sqrt{\sigma_\mathrm{1.3mm}^2+\mathrm{rms}^2}$, where $\sigma_\mathrm{1.3mm}$ is the 10\% uncertainty of the fluxes contributed by the calibrator; columns 10 and 11: the disk dust mass computed from Eq. \ref{eq:dust_mass} and the uncertainty computed using the propagation of error; column 12: flags identifying a subset of disk properties.\\
    *Abbreviations in column 12:
    \\T: Transition disk candidates identified from SED modelling in \cite{2016A&A...592A.126V}.
    \\O: Targets with $\mathrm{^{12}CO}$ outflows detected in our survey (Section \ref{subsec:outflows}).
    \\G: Targets that have a distance of $<275~$pc or $>675~$pc.
    \\E: Targets that are not part of the original sample but that are detected off the center of the field of view.
    \\N: Targets that have not been reported previously.}
\end{threeparttable}
\label{table:disk_info}

\end{table*}

\begin{figure}
    \centering
    \includegraphics[width=0.47\textwidth]{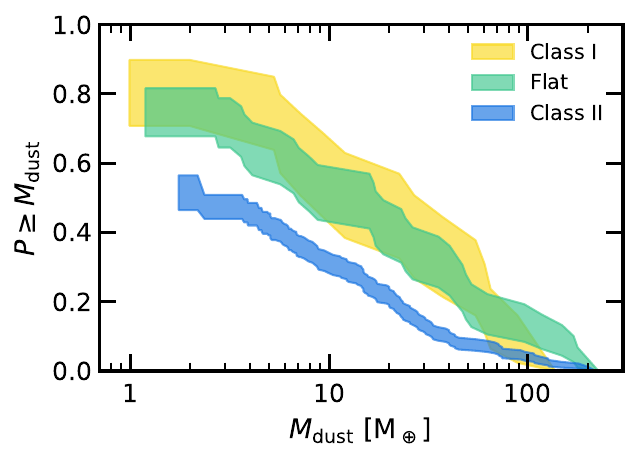}
    \caption{The cumulative disk dust mass distributions for Class I, Flat and Class II YSOs in Serpens. The width of the ribbons give the 68\% confidence interval.}
    \label{fig:mass_distri}
\end{figure}

Figure \ref{fig:mass_distri} shows the cumulative distributions of disk dust masses for Class I, Flat Spectrum, and Class II disks, computed with the Kaplan-Meier estimator implemented in \texttt{lifelines} \citep{Davidson-Pilon2019}, which accounts for non-detections. Tentative detections  (Section \ref{sec:obs}) are treated as non-detections in this analysis. It is clear that disk dust masses decrease from earlier to later evolutionary stages: the median dust masses for Class I, Flat Spectrum and Class II disks are $22.55_{-0.37}^{+0.62}$, $15.97$$_{-0.41}^{+0.56}$, and $2.36_{-0.44}^{+0.54}$ $M_\oplus$, respectively. 

We compare the cumulative disk dust mass distributions between higher-resolution data presented in this work and lower-resolution data from \citet{2022ApJ93855A}. We perform log-rank tests for each young stellar object classification. The p-values of $\gtrsim 0.7$ for Class I, Flat Spectrum and Class II objects indicate that there are no statistically significant differences between the cumulative distributions derived from the lower- and higher-resolution observations. This again validates that the disk dust mass distribution in Serpens is indistinguishable from those in Lupus and Taurus \citep{2017AJ....154..255L, 2022ApJ93855A}.

\subsection{Transition disks}\label{sec:td}
In the survey, 21 out of 321 disks are identified as transition disk candidates based on either the spectral energy distribution (SED) analysis in \citet{2016A&A...592A.126V} or visual inspection of the continuum images. These disks are summarized in Table \ref{tb:TD_candidate}.

We parametrically model the ALMA visibility of selected objects to reassess their candidacy and measure their cavity sizes, using \texttt{galario} \citep{2013PASP..125..306F, 2018MNRAS.476.4527T}. A target is modeled if it: (i) has a Gaia-based distance $275$--$675$ pc; (ii) is detected in continuum with $F_\mathrm{1.3mm}>3~$mJy (see Section \ref{subsec:disk_mass}); and (iii) has an estimated cavity size of $r_\mathrm{cav}>10~$au, identified by the SED analysis \citep{2016A&A...592A.126V}, since smaller cavities cannot be resolved in this survey. Criterion (iii) is relaxed when the spectral type of the host star is unknown, in which case disks with $r_\mathrm{cav}<10~$au are also taken into account. Applying these criteria results in 9 transition disk candidates for modeling, as detailed below and in Table \ref{tb:TD_candidate}. 

\begin{table}
\caption{Summary of 21 transition disk candidates.}\label{tb:TD_candidate}
\centering
\begin{tabular}{ccccc}
\hline\hline
(1)& (2) & (3) & (4) & (5)\\
\# & 2MASS ID & $F_\mathrm{1.3mm}$& origin(s) & modeled?\\
 & & [mJy]& &\\
 \hline 
1 & J18273858-0402289  &14.2& S+V & \cmark\\ 
2 & J18280564-0352112${}^{a}$ & 3.09 & V & \cmark \\
3 & J18281524-0002433     &8.82& S(u) & \cmark\\ 
4 & J18282906+0027560${}^{a}$ &2.84 & S(u)& \\ 
5 & J18285021+0009494 &3.37& S, \xmark&\\ 
6 & J18290437+0033235${}^{a}$ & 2.11  & V &  \\
7 & J18290819-0139215${}^{a}$ &21.15& S(u)& \cmark\\ 
8 & J18291149+0020385 & 2.46  & S& \\ 
9 & J18293563+0035035 & 1 & S& \\ 
10 & J18293961-0202414 & 2.29 & S(u)& \\ 
11 & J18294410+0033559 & 5.49 & S&\cmark\\
12 & J18294721-0148301 & 0.78${}^{b}$&S(u)& \\
13 & J18295533+0049391 &32.47&S+V& \cmark\\
14 & J18295741-0151541 & 1.47${}^{c}$ &S & \\ 
15 & J18303321-0152563 & 0.3${}^{b}$ &S&  \\ 
16  & J18303760-0208403${}^{a}$ & 12.17 & V & \cmark\\
17 & J18313657-0157320 &  0.04${}^{b}$ &S(u) &  \\ 
18 & J18324783-0239401 &33.9&S(u)& \cmark\\ 
19 & 18394048+0014497 &3.41&S, \xmark & \\ 
20 & 18401486+0037042 &4.42&S & \cmark\\ 
21 & 18044921-0436413 &1.02&S & \\   
\hline
\end{tabular}
\tablefoot{All disks fulfill the distance criterion ($d>275$ pc and $d<675$ pc). Column 2: 2MASS Source ID; column 3: 1.3-mm continuum fluxes; Only disks with $F_\mathrm{1.3mm}>3~$mJy are considered for visibility modeling; column 4: methods for candidate identification. ``S'' is for candidates identified from the SED analysis \citep{2016A&A...592A.126V}; ``u'' in parentheses denotes targets with unknown stellar spectral types. ``V'' is for candidates identified through visual inspection. For targets with known stellar spectral types, an additional \xmark~is shown if the inferred cavity size $r_\mathrm{cav}<10~\mathrm{au}$, which is not expected to be spatially resolved in this survey; column 5: disks that are modeled in the visibility plane are marked as \cmark.\\
${}^\mathrm{a}$ Targets with no Gaia-based distances. A median distance to the sub-cluster they are closest to is assigned as the distance instead.\\
${}^\mathrm{b}$ Non-detections.\\
${}^\mathrm{c}$ Tentative detections.}

\end{table}
We model each disk twice with a ring model and a Gaussian model, with the exception of two disks--J18273858-0402289 and J18295533+0049391--that show clear continuum inner cavities, which can be seen in Fig. \ref{fig:detect_gallery}. We model them with only the ring model. The ring model is described by $I = I_0\exp{\big[-(r-r_c)^2/r_w^2\big]}$, where $I_0$ is the intensity, $r$ is the radius, $r_c$ is the inner radius of the ring and $r_w$ is the width of the ring. The Gaussian model is described by $I = I_0\exp{\big(-r^2/r_{wc}^2\big)}$, where $r_{wc}$ is the radius of the disk. These two models include 7 and 6 parameters, respectively, along with four addition parameters including the inclination angle ($i$), position angle (PA), offset in declination and offset in right ascension ($\Delta$ DEC, $\Delta$ RA). We run a Markov-Chain Monte Carlo (MCMC) analysis to infer the best-fit parameters for each model. We set $40$ walkers for $2000-3000$ iterations each run and then use the last $500$ samples to identify best-fit parameters and create corner plots for the posteriors. We report the $68\%$ confidence interval of each parameter in Table \ref{tb:mcmc_params}. 

The visibility model of the two disks with clear inner cavities is shown in Fig. \ref{fig:2td_vis}. Their posteriors are shown in Fig. \ref{fig:TD1827} and Fig. \ref{fig:TD1829}. The best fit of both disks converges to the ring model, suggesting a cavity radius of 0.\arcsec19 for J18273858-0402289, which corresponds to $r_\mathrm{cav}=87$~au and 0\arcsec.20 for J18295533+0049391, which corresponds to $r_\mathrm{cav}=94$~au. The best-fit parameters are applied to generate the modeled continuum images, which are compared with the observed continua in Fig \ref{fig:2td_image}, along with residuals. An asymmetry ($>11\sigma$) is found in the residual of J18295533+0049391 (the upper panel of Fig \ref{fig:2td_image}). Follow-up higher sensitivity and resolution observations are required to investigate its origin. 

The same analysis is applied to the other 7 candidates. Most candidates converge neither to the ring model nor {to the Gaussian model. We attribute this to the low signal-to-noise ratio and low-resolution of the observations, and possibly inadequate description of the simple models. J18290819-013921 converges to both ring and Gaussian models, and we take it as a Gaussian disk as the best-fit inner radius in the ring model approximates to $0\arcsec$. We show the best-fit visibility models of these 7 disks in Fig. \ref{fig:TD_cand_vis} and the modeled continuum imaging in Fig. \ref{fig:TD_cand_model}. 

We note that disks in Serpens are at relatively large distances and our survey resolution is moderate. These factors jointly require the disk cavities to be sufficiently large to be well resolved ($\gtrsim 90~$au, for a target located at 450~pc away when observed with a resolution of 0.\arcsec2, which is the typical beam minor axis size). Consequently, our visibility modeling does not rule out that targets that do not converge to either model are transition disks with smaller cavities. Follow-up, higher-resolution observations with improved sensitivity are required to confirm whether they have inner cavities, and if confirmed, to measure the exact cavity size, which may differ from those derived from the SED analysis \citep[e.g.][and references therein]{2023EPJP..138..225V}.

\begin{figure}
    \centering
    \includegraphics[width=0.45\textwidth]{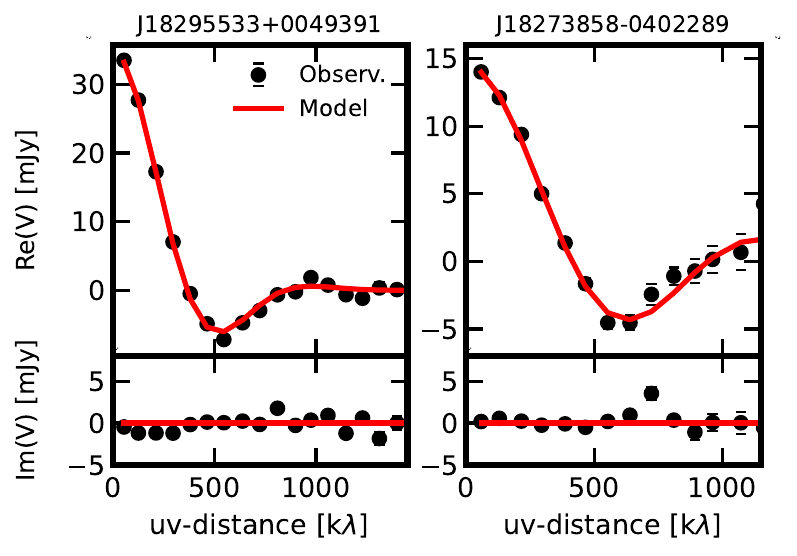}
    \caption{The best-fit visibility of a ring model (red lines) overlapped on the observed visibility (black dots) for two transition disk candidates J18295533+0049391 (left) and J18273858-0402289 (right).}
    \label{fig:2td_vis}
\end{figure}

\begin{figure}
    \centering
    \includegraphics[width=0.47\textwidth]{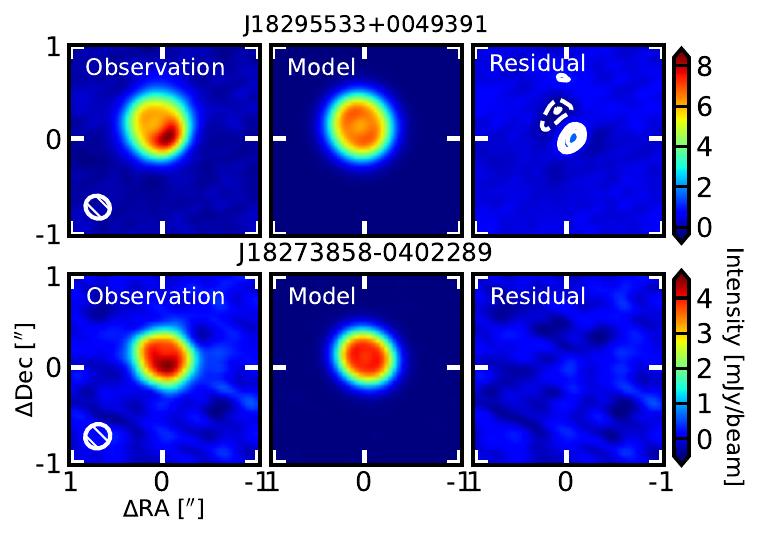}
    \caption{Comparison between 1.3-mm observations (left), best-fit models (middle) and residuals (right) for transition disk candidates J18295533+0049391 (upper panels) and J18273858-0402289 (lower panels). The over-plotted contours in the right panels indicate [5, 8, 11] $\times$ rms (dashed lines for negatives and solid lines for positives). The beam sizes are shown in the bottom left corner of the observation panels.}
    \label{fig:2td_image}
\end{figure}

\subsection{Outflows}\label{subsec:outflows}
We detect 12 outflow-shaped features traced by ${}^{12}\mathrm{CO}$ $\mathrm{2-1}$ in our Serpens survey. Their zeroth and first moment maps are shown in Fig. \ref{fig:outflow}. Not all features are associated with the sources centered in the FoV. In such cases, the white cross marking the central source is offset from the image center in Fig. \ref{fig:outflow}. 

In the lower-resolution dataset, \citet{2022ApJ93855A} reported 15 large-scale ${}^{12}\mathrm{CO}$ 2-1 structures potentially associated with the central targets. Of these, 6 overlap with the outflow-shaped features in Fig. \ref{fig:outflow}. The morphologies of these 6 outflows are consistent between the lower resolution observations and the concatenated data used in this work (see Section \ref{sec:obs}). The remaining 9 targets are not reported here for the following reasons: (i) two are not detected in the higher-resolution observations (J18293766-0152049, J18320048-0207440); (ii) seven show ambiguous features, which appear in a single channel or are not cone-shaped  (J18290843-0207041, J18292072-0137173, J18300132-0203431, J18302592-0211421, J18302712-0210564, J18303760-0208403,  J18401761+0016138). Nevertheless, we does not exclude the possibility that they are molecular outflows associated with the central target.

\begin{figure*}
    \centering
    \includegraphics[width=0.88\textwidth]{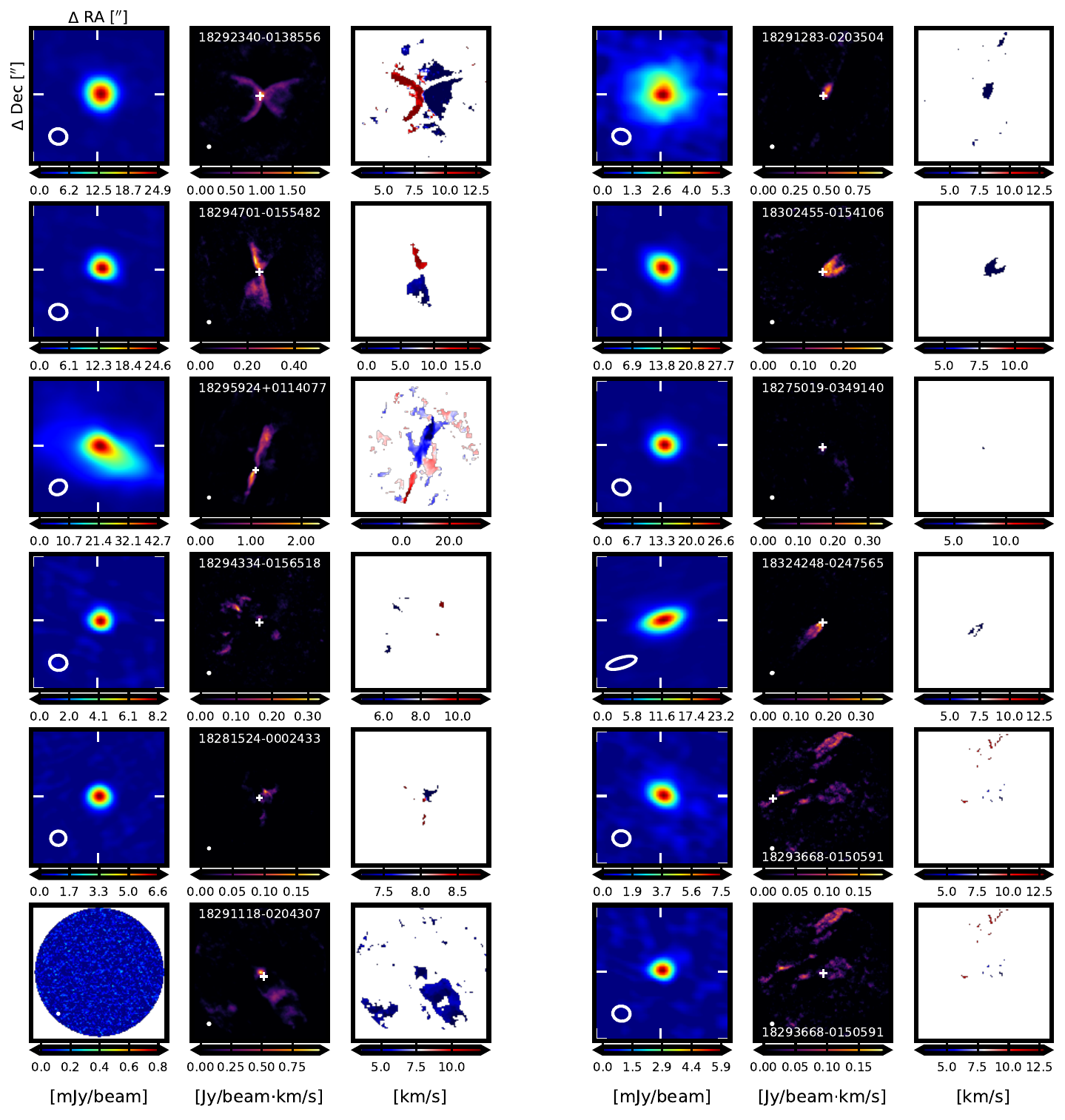}
    \caption{Gallery of large-scale structures traced by ${}^{12}\mathrm{CO}$ in the Serpens survey. The three columns show, from left to right, the continuum emission with which the large-scale structures are possibly associated (left, in boxes of 2\arcsec$\times$2\arcsec~if detected), the zeroth moment maps (middle, in boxes with a side length of $37.92$\arcsec on one side, except for 18295924+0114077 and 18281524-0002433, which are shown in boxes with a side length of $42.5$\arcsec), and the first moment maps (right, box sizes are the same as the middle columns). We assume a systemic velocity of $v_\mathrm{LSR}=8~\mathrm{km~s^{-1}}$ for objects in Serpens \citep{2019ApJ...887..209A} when creating the color scale for the first moment map. White crosses indicated the location of the continuum the large-scale structures are likely associated with. The 2MASS ID of the central source is labeled in the middle panels.}
    \label{fig:outflow}
\end{figure*}

\section{Discussion}\label{sec:discussion}
\subsection{Stellar density}\label{subsec:stellar_density}
\subsubsection{3D stellar density measurements}\label{subsubsec:3d_density}
The stellar density, defined as the number of stars in a given volume, reflects the likelihood of stellar encounters. When the local stellar density is high, frequent encounters can truncate disks and reduce continuum fluxes, shifting the cumulative distributions of dust disk sizes and masses to lower ends \citep[e.g.][]{1993MNRAS.261..190C, 2014MNRAS.441.2094R, 2021ApJ...923..221O}.

The stellar density has been previously measured in 2D. \citet{2009ApJS..181..321E} measured global 2D stellar surface densities for nearby star-forming regions ($<500$ pc) with Spitzer data. By defining it as the total number of young stellar objects (YSOs) per projected two-dimensional area, they found that Serpens, followed by Ophiuchus, is relatively denser than other star-forming regions, such as Lupus and Chamaeleon II (their Fig. 3). \citet{2016AJ....151....5M} measured the local 2D stellar surface density with the 10 nearest stellar neighbors. By computing the medians and cumulative distributions of 10-nearest neighbors for each star-forming region, they reached a similar conclusion that Serpens and Ophiuchus are denser than Chameleon, Lupus and Taurus (their Fig. 15).

However, the cumulative disk dust mass distribution of Serpens, measured in both this work and in \citet{2022ApJ93855A}, is similar to those of Lupus and Taurus \citep{2017AJ....154..255L, 2022ApJ93855A}. This suggests that the 2D surface stellar density in Serpens does not affect the disk dust mass as strongly as expected. This could indicate that Serpens is not as dense in 3D as inferred previously from the 2D distribution, or that its stellar density, though relatively higher, does not reach the threshold required for significant truncation. This motivates us to re-assess the stellar density of nearby star-forming regions in 3D with the revolutionary Gaia data. This approach can mitigate the contamination from infrared background objects and factor in the distance variations along the line-of-sight within each region.

Leveraging Gaia phase-space information \citep{2016A&A...595A...1G, 2018A&A...616A...1G, 2023A&A...674A...1G} and ground-based optical spectroscopy, recent studies have conducted membership censuses of nearby star-forming regions and their sub-clusters, including
\begin{itemize}
    \item Chamaeleon I \& II \citep[Cha I North and South,][]{2021A&A...646A..46G}
    \item Corona Australis \citep[Main, North and the core,][]{2022AJ....163...64E, 2023A&A...677A..59R}
    \item Lupus \citep[Lupus I, II, III and IV,][]{2020AJ....160..186L}
    \item Ophiuchus \citep[L1688, L1689, L1709, $\rho$ Oph,][]{2020AJ....159..282E}
    \item Serpens \citep[Serpens Main, Northeast and Far South,][]{2019ApJ...878..111H}
    \item Taurus \citep[B209N, L1489/L1498, L1495/B209, L1517, L1521/B213, L1524/L1529/B215, L1527, L1536, L1544, L1551, L1558 and TTau,][]{2023AJ....165...37L}, and
    \item Upper Scorpius \citep[$\nu$ Sco, $\delta$ Sco, $\beta$ Sco, $\sigma$ Sco, $\rho$ Sco and Antares,][]{2020AJ....160...44L}.
\end{itemize}

These star-forming regions have been found to include multiple sub-clusters, and we compute the stellar density for each sub-cluster separately. The spatial distributions of stars in these star-forming regions and in their sub-clusters are shown in Fig. \ref{fig:star_dens} (panels a-h). Stars in Corona Australis have been divided into three sub-clusters using their Galactic coordinates and the J-band extinction, $A_J$, with one sub-cluster being much younger than the other two \citep{2022AJ....163...64E}. Following the nomenclature of \citet{2023A&A...678A..71R}, we refer to the more sparse and older sub-cluster with galactic latitude $b\geq -17.^{\circ}3$ as Corona Australis North, and the less sparse but still old sub-cluster with galactic latitude $b<-17.^{\circ}3$ and $A_J\leq1$ as Corona Australis Main. The youngest sub-cluster with galactic latitude $b<-17.^{\circ}3$ and $A_J>1$ \citep{2022AJ....163...64E}, which is not identified separately in \citet{2023A&A...678A..71R}, is referred to as the Corona Australis core. For Upper Sco, where sub-clusters are not identified in the corresponding membership studies listed above, we adopt the sub-clusters identified by \citet{2023A&A...677A..59R}. Since this study uses only Gaia coordinates, parallaxes and proper motion, we conservatively include only members that appear in both \citet{2023A&A...677A..59R} and \citet{2020AJ....160...44L} for Upper Sco ($\nu$ Sco, $\delta$ Sco, $\beta$ Sco, $\sigma$ Sco, $\rho$ Sco and Antares)\footnote{We do not consider two more extended sub-clusters--US-foreground and Scorpio-Body--which are more evolved than the other 6 sub-clusters discussed here \citep{2023A&A...678A..71R}.}. By doing this, we do not include members that might be more embedded and have no Gaia membership, providing only a lower-limit on the stellar density. 

Combining these membership catalogs with Gaia stellar coordinates and parallaxes, we compute the 3D stellar density of each region. We follow the method described in \citet{2016AJ....151....5M} but extend it to 3D by considering distance variations among stars within a (sub-)cluster instead of assuming a single distance for stars in a star-forming region.

The stellar distances are computed by sampling Gaussian distributions of Gaia DR3 parallaxes and their uncertainties using an MCMC sampler \citep[here we use \textsc{emcee},][]{2013PASP..125..306F}, and then inverting the sampled parallaxes. With the derived distances and their sky positions (RA and Dec), we compute the 3D separations between individual stars in Cartesian coordinates. For each star, the distance to its $10^\mathrm{th}$ nearest neighbors $d_{10}$ is then applied to compute the local stellar density 
\begin{equation}\label{eq:local_stellar_density}
    \rho_{10}=\frac{10-1}{\frac{4}{3}\pi d_{10}^3}.
\end{equation}

For each (sub-)cluster, we run the MCMC for 2,000 steps with the number of walkers set to twice the number of members. At each step, we compute the local stellar density for each star, following Eq. \ref{eq:local_stellar_density}. We then use the local stellar densities computed from the last 1,500 steps to derive the two metrics introduced below and their uncertainties. 

Not all members included in previous membership censuses are considered here, since some censuses adopted parallaxes from Gaia DR 2, and some members appear to lie beyond the (sub-)clusters when Gaia DR3 parallaxes are used. We therefore exclude sources whose Gaia DR3 parallaxes differ from the median parallax of members with known parallaxes by more than 30\%, taking the parallax uncertainties into account. This removes only previously identified members with extreme Gaia DR3 parallaxes.

For members without Gaia parallaxes and for those with large uncertainties in parallaxes ($\overline{\omega}/\sigma(\overline{\omega})<5$), we sample parallaxes from the distribution of other members in the same (sub-)cluster with known parallaxes, and adopt the median parallax uncertainty of that sample. This approach may overestimate or underestimate the distances of stars located in the outskirts of a cluster, possibly leading to slightly overestimated local stellar densities. However, we do not expect this to substantially affect our final results when the number of members with unknown parallaxes and with large uncertainties in parallaxes is small compared to the total membership. In cases where the fraction of members without reliable parallaxes is large, such as in the Corona Australis core ($\sim 53\%$) and Ophiuchus ($\sim 32\%$ in the sub-cluster L1688), this likely reflects the youth, high gas density, embedded nature, and strong extinction of these regions. Such (sub)-clusters are likely to be more spatially concentrated, with some members not detected by Gaia. Therefore, sampling parallaxes for members without reliable measurements using the method described above is likely to provide a lower limit on the stellar density in these sub-clusters. The numbers of members with known and unknown parallaxes, and that of members with $\overline{\omega}/\sigma(\overline{\omega})<5$ are listed in Table \ref{tb:known_distance} for each sub-clusters.

We characterize the stellar density of each region by (a) the median local stellar density; and (b) the fraction of groups, consisting of the nearest 10 stars, with local stellar densities $>0.1~\mathrm{pc^{-3}}$. We compute these two quantities and corresponding standard deviations from MCMC sampling and show them in Fig. \ref{fig:star_dens}. As the local stellar density is computed based on the ten nearest stellar neighbors, we do not consider sub-clusters with members fewer than 30. This excludes $\rho$ Oph, Oph L1709, Oph L1689, Lupus I, II, IV, Tau B209N, Tau HD28354, Tau L1489/L1498, Tau L1521/B213, Tau L1527, Tau L1544, Tau L1558, T Tau, $\rho$ Sco and the Corona Australis core in Table \ref{tb:known_distance}. Though the Corona Australis core contains slightly more than 30 disks, more than half of them lack Gaia parallaxes. We, therefore, do not compute its stellar density. The comparison among regions is shown in the two panels in the bottom row of Fig. \ref{fig:star_dens} (panels i and j).

\begin{figure*}
    \centering
    \includegraphics[width=0.95\textwidth]{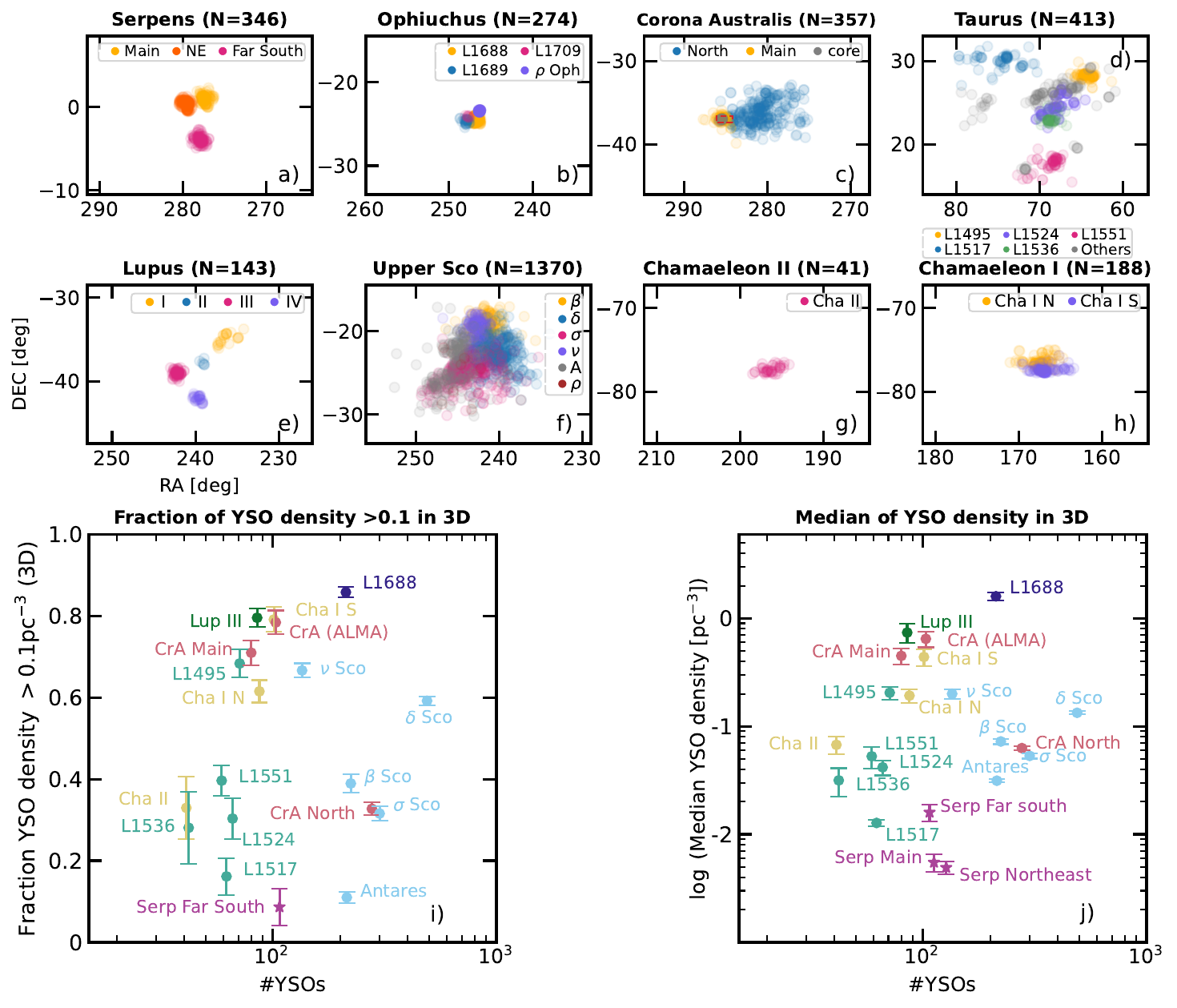}
    \caption{Spatial distributions of young stellar objects identified in Gaia-based membership censuses are shown for (a) Serpens \citep{2019ApJ...878..111H}, (b) Ophiuchus \citep{2020AJ....159..282E}, (c) Corona Australis \citep[][]{2022AJ....163...64E}, (d) Taurus \citep{2023AJ....165...37L}, (e) Lupus \citep{2020AJ....160..186L}, (f) Upper Scorpius \citep{2020AJ....160...44L, 2023A&A...677A..59R} and (g \& h) Chamaeleon \citep[I (h) \& II (g),][]{2021A&A...646A..46G}. Numbers in parentheses indicate the number of members in each region. When a star-forming region contains multiple sub-clusters, young stellar objects in different sub-clusters are shown in different colors. The bottom two panels compare the stellar density by the fractions above $0.1~\mathrm{pc^{-3}}$ (i) and by the median density (j). The stellar density for Corona Australis is also measured for the sub-region that has been surveyed by ALMA, which is circled by a red box. Only the five sub-clusters in Taurus with $\geq30$ members are shown here for visual clarity. Sub-clusters with less than $30$ members, including B209N, HD28354, L1489/L1498, L1521/B213, L1527, L1544, L1558 and T Tau, are shown in gray as a whole in panel d, and their stellar densities are not computed for the reason explained in the final paragraph of Section \ref{subsubsec:3d_density}. Sub-cluster A in Upper Sco (panel f) indicates Antares; In panels i and j, CrA indicates Corona Australis, Cha indicates Chamaeleon, Lup indicates Lupus, Sco indicates Upper Sco, Serp indicates Serpens.}
    \label{fig:star_dens}
\end{figure*}

\subsubsection{Comparison across star-forming regions}\label{subsec:comparison_with_stellar_density}
The stellar densities characterized by two metrics in Fig. \ref{fig:star_dens} are consistent. Interestingly, irrespective of the metric used, Serpens Far South is among the least dense regions studied here--even though it is the densest sub-cluster in Serpens. We attribute the unexpected low stellar density in Serpens to two factors. First, Serpens is more elongated along the line of sight than other regions (Fig. \ref{fig:distance}). Therefore, expanding it from 2D to 3D leads to a significant drop in the stellar density. This effect is more significant in \citet{2009ApJS..181..321E}, where they underestimate the distance to Serpens ($260\pm10~\mathrm{pc}$). Second, the high optical extinction due to Serpens's proximity to the Galactic plane induces extra challenges in identifying optically faint objects, leading to incomplete memberships. The incomplete membership problem may also exist for regions that are young ($\lesssim 3~$Myr) but that have stellar densities much lower than that of $\sigma$ Sco. This is because clusters like $\sigma$ Sco with an age of $5$--$10$ Myr, irrespective of isochronal models used in \citet{2023A&A...678A..71R}, are expected to have lower stellar densities than much younger regions as the results of proper motion of gravitationally unbound stellar populations \citep{2019ApJ...870...32K}, following the dispersal of natal molecular gas \citep{2008hsf2.book..235P}. Such processes are well demonstrated by the sub-clusters of Upper Sco. The stellar density generally decreases from $\nu$ Sco to Antares with the increasing ages of the sub-clusters \citep{2023A&A...678A..71R}\footnote{In \citet{2023A&A...678A..71R}, $\delta$ Sco is slightly older than $\beta$ Sco, while in some other studies \citep[e.g.][]{2021ApJ...917...23K}, $\delta$ Sco is younger than $\beta$ Sco. The local stellar density in Fig. 8 is more consistent with the latter.} in both panels (panels i and j) of Fig. \ref{fig:star_dens}. Although some star-forming regions are intrinsically sparse, such as Taurus, where 13 sub-clusters are identified in \citet{2023AJ....165...37L}, but none of them contains more than 100 members, and they all show relatively low stellar densities in Fig \ref{fig:star_dens} given their young ages.

In contrast to the sub-clusters in Serpens, some sub-clusters are much denser. The most prominent region is L1688 in Ophiuchus, regardless of which metric is used to characterize the local stellar density. Since $\sim 32$\% of the members in L1688 do not have known Gaia parallaxes (see Table \ref{tb:known_distance}), the values of these two metrics should be regarded as lower limits, as discussed in Section \ref{subsubsec:3d_density}. A similar case is the sub-region named as Corona Australis (ALMA) in Fig. \ref{fig:star_dens}, where $\sim 20\%$ of its members have no known Gaia parallaxes. This is the sub-region of Corona Australis that has been studied by ALMA \citep{2019A&A...626A..11C} and includes members from both Corona Australis Main and the core. Since only $\lesssim8\%$ members in Lupus III and Chamaeleon I South have no known parallaxes, we expect L1688 and Corona Australis (ALMA) are likely to be denser than other sub-clusters shown here. These high stellar densities coincide with the lower cumulative disk dust mass distributions found in the ALMA disk surveys of Ophiuchus \citep{2017ApJ...851...83C, 2019ApJ...875L...9W, 2021ApJ...913..149E} and Corona Australis \citep[][ see also Fig. \ref{fig:cum_mass_distri_multi} for a comparison of the cumulative distributions]{2019A&A...626A..11C}. This motivates us to hypothesize that the lower disk dust masses in these two sub-clusters may result from more frequent star-disk interactions in an even earlier evolutionary stages, when the stellar densities were likely even higher. We explore this possibility in the next subsection. 

\subsubsection{Local stellar densities and cumulative distributions of disk dust masses}\label{subsubsec:stellar_density_culmulative distr}

To test whether the high stellar densities of L1688 in Ophiuchus and Corona Australis (ALMA), identified in Section \ref{subsec:comparison_with_stellar_density} could explain the lower disk dust masses in these two regions, we re-examine the membership of disk samples in ALMA surveys. These samples are typically selected based on infrared excess and they are not separated by sub-clusters, which usually at least requires knowledge of parallaxes and proper motion. As a result, these ALMA surveys may include disks that are associated with nearby older sub-clusters. For example, disks from the older Corona Australis North might be considered as disks from the Corona Australis core in the Corona Australis ALMA survey. Therefore, we cross-match these samples with Gaia-identified members to exclude contamination from field objects or objects associated with nearby sub-clusters, and assign each disk to its corresponding sub-cluster.

For the ALMA survey in Ophiuchus, we adopt data from \citet{2019ApJ...875L...9W}, which includes a total of 289 disks. We limit our discussion to 172 Class II disks only, and cross-matched these disks with the Gaia membership census by \citet{2020AJ....159..282E} and by \citet{2023A&A...677A..59R}, respectively. The former classifies disks into four sub-clusters--L1688, L1689, L1709 and $\rho$ Oph (as panel b in Fig. \ref{fig:star_dens}), and the latter only considers one combined sub-clusters--L1688/$\rho$ Oph. The cross-match results are shown in Table \ref{tb:alma_gaia_oph}, and there are 55 and 4 disks in L1688 and $\rho$ Oph, respectively, from \citet{2020AJ....159..282E}, and 97 disks in L1688/$\rho$ Oph from \citep{2023A&A...677A..59R}. 72 disks from the Ophiuchus ALMA survey have no Gaia membership.

We then plot the cumulative disk dust mass distributions for (a) the entire sample of the Ophiuchus survey, (b) disks identified as members of L1688 and $\rho$ Oph, and (c) disks without Gaia information in the upper left panel of Fig. \ref{fig:oph_cra_distribution}, respectively. Other Ophiuchus sub-clusters are not considered here as they contain $\lesssim 10$ members. The cumulative distribution of disks in $\rho$ Oph and L1688 is generally above that of the full Ophiuchus ALMA survey sample \citep{2019ApJ...875L...9W}. It is closer to that of Lupus but have a higher fraction of disks with lower disk dust masses.

Instead, disks in Ophiuchus having lower disk dust masses is driven by disks that have no Gaia information and that are spatially clustered in the 2D projected sky plane (see the upper right panel of Fig. \ref{fig:oph_cra_distribution}). These disks are unlikely to be associated with Upper Sco, part of which spatially overlaps with Ophiuchus but is more dispersed and less embedded as it is more evolved. The lack of Gaia identifications may have two possible explanations. First, the host stars may be too low in mass and luminosity to be detected by Gaia. Such stars typically host disks with lower dust masses \citep[e.g.][]{2013ApJ...771..129A}. Second, the lack of Gaia information may reflect the high optical depths and embedded nature of these young systems. Combined with their spatial clustering in the 2D projected sky plane, this may suggest that these disks were born in a dense environment, where early tidal truncation reduced their disk sizes and masses.

To test whether the stellar mass dependence alone could explain the lower disk dust mass of the no Gaia systems, we retrieve spectral types from \textsc{SIMBAD} for members of sub-clusters L1688 and $\rho$ Oph identified in \citet{2020AJ....159..282E}, and for systems in the ALMA survey in Ophiuchus without Gaia data. We exclude spectroscopic binaries (one in each sample) and restrict the comparison to systems with spectral types from M4 to M8. Applying this criteria to the two groups yields 35 out of 48 (81\%) systems that lack Gaia memberships but have known spectral types, and 23 out of 49 (47\%) systems that have both Gaia memberships and known spectral types. We compute the cumulative disk dust mass distributions for these 35 and 23 systems, respectively, and show their disk dust mass distributions for each spectral type from M4 to M8 in Fig. \ref{fig:oph_stellar_mass_dependence}. It is clear that the stellar mass dependence alone cannot account for the low disk dust masses of disks without Gaia information.

For these discs without Gaia memberships, we cannot estimate their local stellar densities or directly test the second scenario that earlier tidal truncation leading to the lower disk dust mass. Nevertheless, their low disk dust masses, lack of Gaia information, and their 2D clustering are consistent with the hypothesis that disks born in crowded local environments may experience tidal truncation at early stages, leading to less massive dust disks.

\begin{table}[]
\centering
\renewcommand{\arraystretch}{1.15}
\caption{Cross-matching results between the ALMA disk survey and the Gaia member census for 172 Class II disks in Ophiuchus.}\label{tb:alma_gaia_oph}
\scalebox{0.93}{
\begin{tabular}{ccc}
\hline\hline
sub-clusters & Esplin et al. 2020 & Ratzenböck et al. 2023\\
\hline
$\rho$ Oph   & 4   & \multirow{2}{*}{81} \\
L1688        & 48  &   \\
L1689        & 8   & --  \\
L1709        & 2   & --  \\
$\sigma$ Sco & --  & 1  \\
$\rho$ Sco   & --  & 1  \\
$\beta$ Sco  & --  & 2  \\
Antares      & --  & 5  \\
unknown      & 38  & 10 \\
No Gaia      & 72  & 72 \\
\hline
\end{tabular}}
\tablefoot{The ALMA disk survey is adopted from \citet{2019ApJ...875L...9W} and the Gaia member census is adopted from \citet{2020AJ....159..282E, 2023A&A...677A..59R}. ``Unknown'' indicates ALMA disks that have Gaia IDs but are not classified into any sub-clusters considered in the member census. ``No Gaia'' indicates ALMA disks without Gaia identifications. 9 ALMA disks are classified as members of Upper Sco sub-clusters in \citet{2023A&A...677A..59R}, as that study focuses on the Sco-Cen association and Ophiuchus lies close to Upper Sco in the 2D projected sky plane (see panels b and f in Fig. \ref{fig:star_dens}). We note that \citet{2023A&A...677A..59R} considers members of $\rho$ Oph and L1688 together, comprising a sample of 81 sources observed in the ALMA disk survey.}
\end{table}

We apply a similar analysis to the sub-clusters in Corona Australis. The cross-matching results are shown in Table \ref{tb:alma_gaia_cra}. Out of 43 Class II disks in the ALMA survey \citep{2019A&A...626A..11C}, 22 are assigned to Corona Australis Main. When disks with no reported $A_J$, galactic latitude $b<-17.^{\circ}3$, and no Gaia membership are excluded, 11 disks are assigned to the youngest sub-cluster, the Corona Australis core. Including these disks increases the core sample to 18 disks. We plot the cumulative distributions of disk dust masses for (a) the entire ALMA samples, (b) Corona Australis Main and (c) the Corona Australis core, respectively, in the lower left panel of Fig. \ref{fig:oph_cra_distribution}. The relatively small sample size results in large uncertainties for the three distributions.

Disks from Corona Australis Main and the core contribute differently to the cumulative disk dust mass distributions of the whole ALMA sample. Disks from the Corona Australis core generally have higher disk dust masses than disks from Corona Australis Main, and this is consistent with the increasing ages from the core (1-2 Myr) to Main \citep[>5 Myr, e.g.][]{ 2022AJ....163...64E, 2025A&A...701A.242R}.

Compared with other previously studied star-forming regions and considering only disks with $M_\mathrm{dust}>0.5~M_\oplus$, which is approximately the lowest sensitivity achieved by disk surveys shown in the lower left panel of Fig. \ref{fig:oph_cra_distribution}, the Corona Australis core has a similar cumulative distribution to that of similarly aged Lupus\footnote{The Lupus sample considered here is adopted from \citet{2018ApJ...859...21A}, after excluding 3 of the 95 disks that lie beyond 200 pc when parallax uncertainties are taken into account. Restricting the sample to the 61 of 95 disks that are explicitly associated with one of the four Lupus clouds \citep{2020AJ....160..186L} does not alter the results reported in Sections \ref{subsubsec:stellar_density_culmulative distr} and \ref{subsec:external_pe}.}, with a p-value of $0.72$ in a log-rank test, when 7 disks shown as green filled circles with purple edges in the lower right panel of Fig. \ref{fig:oph_cra_distribution} are excluded. These disks have galactic latitude $b<-17^{\circ}.3$ and lack both Gaia identifications and extinction $A_J$. When these 7 disks are included, the p-value becomes $0.69$, indicating that the 7 disks, which are clustered in the projected sky plane in the right lower panel of Fig. \ref{fig:oph_cra_distribution}, decrease the cumulative disk dust mass distribution of the Corona Australis core, and make it slightly more different from the distribution of Lupus. This is similar to Ophiuchus, where disks without Gaia information are also spatially clustered and preferentially contribute to the low end of the cumulative distribution. We do not test the stellar mass dependence for these 7 disk as we did for Ophiuchus, since only 2 of them have known spectral type (M2 and F5). Nevertheless, this again may suggest these lower disk dust masses in the Corona Australis core result from tidal truncation at an earlier evolutionary stage.   

We also compare the distribution of disks in Corona Australis Main with that of Upper Sco, whose six sub-clusters considered here have ages of 4–13 Myr \citep{2023A&A...678A..71R}, in the lower left panel of Fig. \ref{fig:oph_cra_distribution}. A log-rank test suggests that the two distributions are statistically similar, with a $p$-value of $0.81$ when only disks with $M_\mathrm{dust}>0.5~M_\oplus$ are considered. This result is consistent with recent age estimates for Corona Australis Main. However, although both Corona Australis Main and Upper Sco are relatively evolved, Corona Australis Main lacks disks with dust masses $\gtrsim 20~M_\oplus$. This may simply reflect the fact that more massive disks were not included in existing ALMA surveys when similar initial mass functions and similar disk mass-stellar mass correlations are assumed for these two star-forming regions. Alternatively, more massive disks may simply not exist in the sub-cluster with relatively few members by coincidence. Another possibility is that Corona Australis Main experienced higher stellar densities at earlier evolutionary stages, leading to more frequent dynamical interactions and lower dust masses through tidal truncation.

The possibility of a once higher stellar density in Corona Australis has been discussed and ruled out in \citet{2019A&A...626A..11C}, based on its low velocity dispersion \citep{2000A&AS..146..323N}, which indicates that stars are still well gravitationally bound. However, more recent studies consistently show that Corona Australis Main is more evolved than typical young star-forming regions (1-3 Myr). Its median/mean age has been estimated to be 5-6 \citep{2021A&A...646A..46G, 2025A&A...701A.242R}\footnote{\citet{2021A&A...646A..46G} classifies Corona Australis into on-cloud and off-cloud populations. Cross-matching between members considered for stellar densities in this study with those in \cite{2021A&A...646A..46G} shows that most members in Corona Australis Main from \citep{2023A&A...677A..59R} are on-cloud sources, and most members in Corona Australis north are off-cloud sources.}, 6-11 Myr \citep{2023A&A...678A..71R} or even $\sim15$ Myr \citep{2022AJ....163...64E}. The relatively high stellar density of Corona Australis Main for its old age, combined with its lack of massive disks ($>20~M_\oplus$), may suggest its low disk dust mass distribution results from the combined effects of aging and a once denser environment. A more complete ALMA survey of Corona Australis Main (Y. Guo et al., in prep.), combined with gas disk size measurements, would help distinguish between these possibilities.

\begin{table}[]
\centering
\renewcommand{\arraystretch}{1.15}
\caption{Cross-matching results between the ALMA disk survey and the Gaia member census in Corona Australis.} \label{tb:alma_gaia_cra}
\scalebox{0.93}{
\begin{tabular}{cc}
\hline\hline
sub-clusters & Esplin\& Luhman 2022\\
\hline
Main   & 22 \\
North  & 3 \\
core (excl.)& 11\\
core (incl.)&18\\
\hline
\end{tabular}}
\tablefoot{Cross-matching results between the ALMA disk survey \citep{2019A&A...626A..11C} and the Gaia member census \citep{2022AJ....163...64E} in Corona Australis, which includes sub-clusters Main, North and core following the nomenclature in \citet{2023A&A...677A..59R}. We assign disks from \citet{2019A&A...626A..11C} with no reported $A_J$, galactic latitude $b<-17^{\circ}.3$ and Gaia memberships as members of ``Main''. Disks with no reported $A_J$, $b<-17^{\circ}.3$, and no Gaia memberships are excluded from ``core (excl.)'' but included in ``core (incl.)’’.}
\end{table}

\begin{figure*}
    \centering
    \includegraphics[width=0.83\linewidth]{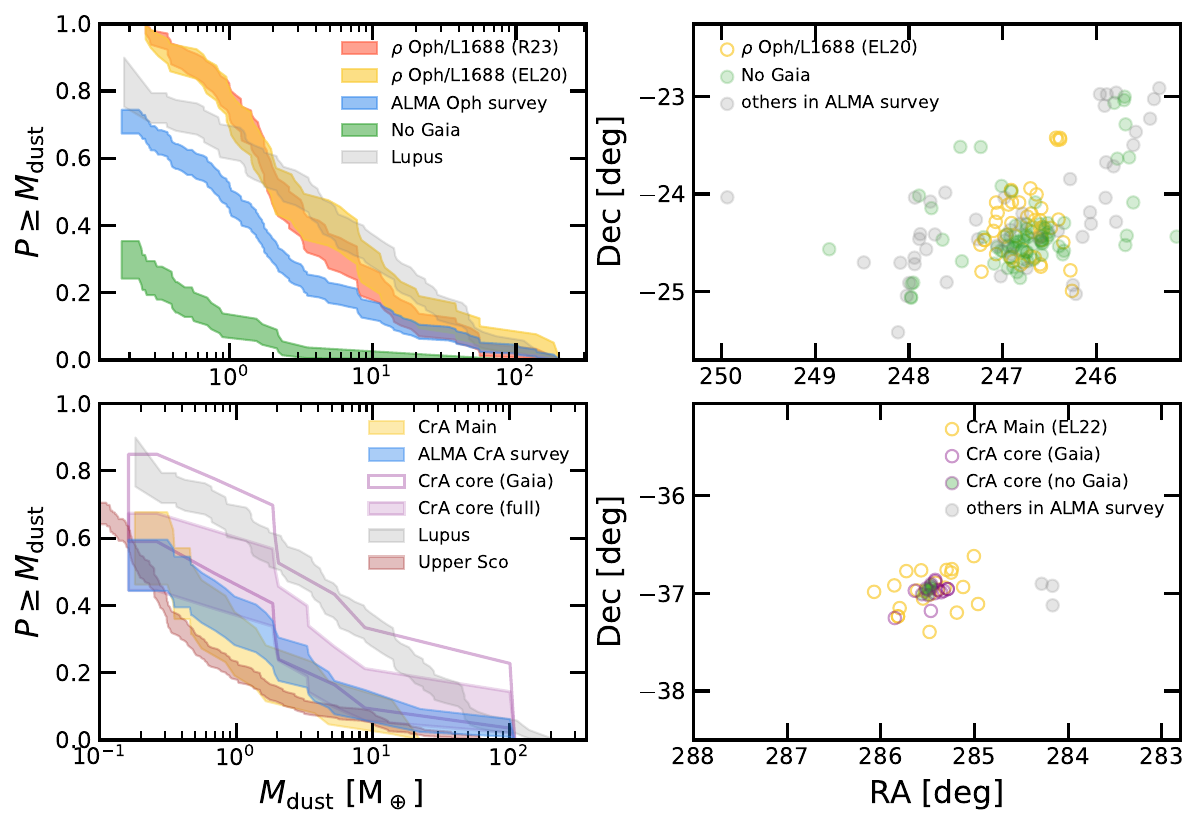}
    \caption{Cumulative disk dust mass distributions (left panels) and spatial distributions (right panels) of ALMA disks in Ophiuchus (upper panels) and Corona Australis (lower panels). Upper left panel: cumulative disk dust mass distributions for (a) the full sample from the ALMA Ophiuchus disk survey \citet{2019ApJ...875L...9W}, (b) ALMA disks identified as members of $\rho$ Oph/L1688 by \citet{2020AJ....159..282E} (denoted as EL20) and by \citet{2023A&A...677A..59R} (denoted as R23), and (c) disks without Gaia identifications. These distributions are in comparison to that of Lupus \citep{2018ApJ...859...21A}. Upper right panel: spatial distributions of disks identified as members in $\rho$ Oph/L1688 by \citet{2020AJ....159..282E} and of disks without Gaia identifications. Disks that have been studied by ALMA but that are not one of these two groups are plotted in grey. Lower left panel: cumulative disk dust mass distributions for (a) the full sample from the ALMA Corona Australis disk survey \citet{2019A&A...626A..11C}, (b) disks identified as members of Corona Australis Main \citep{2022AJ....163...64E}, and (c) disks identified as members of the Corona Australis core, which has an age of 1-2 Myr \citep{2022AJ....163...64E}. The core sample is plotted without (purple empty band) and with (purple filled band) disks satisfying the following criteria: i. galactic latitude $b<-17.^{\circ}3$; ii. no reported extinction $A_J$, and iii. no Gaia identifications. These distributions are compared with that of Lupus \citep{2018ApJ...859...21A} and that of Upper Sco \citep{2025ApJ...978..117C}, for which we consider only the six sub-clusters shown in Fig. \ref{fig:star_dens}). Lower right panel: spatial distributions of disks identified as members of Corona Australis Main (yellow) and the core \citep[purple][]{2022AJ....163...64E}. Disks in the Corona Australis core with galactic latitude $b<-17.^{\circ}3$, no reported extinction, and no Gaia identification are shown as green filled circles with purple edges, and disks that are not part of the above three populations are plotted in grey.}
    \label{fig:oph_cra_distribution}
\end{figure*}

\begin{figure}
    \centering
    \includegraphics[width=0.83\linewidth]{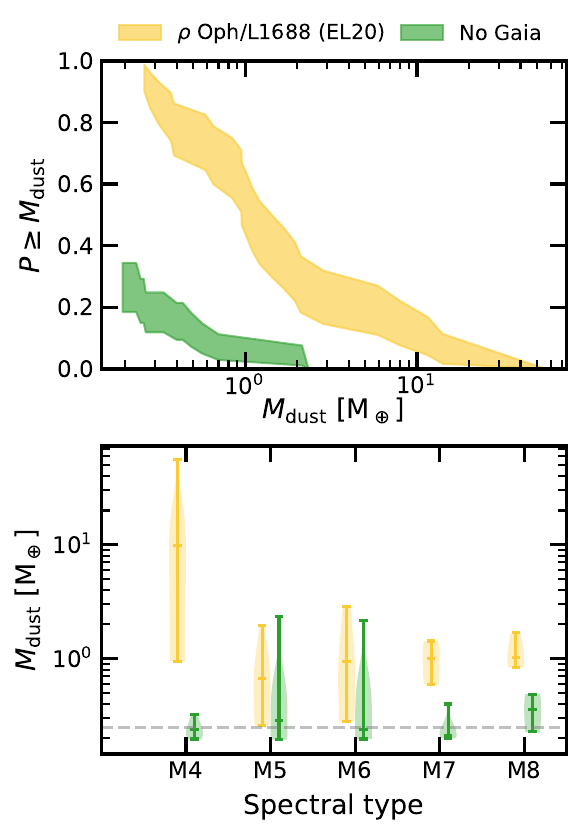}
    \caption{Upper panel: Cumulative disk dust mass distributions for disks around host stars with spectral types M4-M8. The yellow band shows disks in sub-clusters $\rho$ Oph and L1688, identified by \citet{2022AJ....163...64E}, and the green band shows disks without Gaia identifications. Lower panel: The disk dust mass distributions for the two samples in each spectral type. In each violin plot, the whiskers, from top to bottom, indicates the maximum, median and minimum disk dust masses. The dashed gray line shows the detection limit, below which most disks are not detected.}
    \label{fig:oph_stellar_mass_dependence}
\end{figure}

In summary, star-forming regions, such as Lupus and Chamaeleon I (South and North), are more typical regions in terms of their stellar densities, and are therefore expected to be less strongly affected by tidal truncation than more clustered regions. Such clustered populations include the disks without Gaia identifications that are spatially concentrated in L1688 (green dots in the upper right panel of Fig. \ref{fig:oph_cra_distribution}) and in the Corona Australis core (green filled circles with purple edges in the lower right panel of Fig. \ref{fig:oph_cra_distribution}). The similar disk dust mass distributions of Serpens, Taurus and Lupus \citep{2022ApJ93855A} may suggest that Serpens is not as dense as some sub-regions of Ophiuchus or Corona Australis even when we account for the incomplete memberships in Gaia due to high extinction. 

We note that although tidal truncation is proposed as a possible explanation for a sub-sample of disks in Ophiuchus and Corona Australis that have (a) low disk dust masses; (b) high spatial concentration in the projected sky plane (Fig. \ref{fig:oph_cra_distribution}); and (c) no Gaia memberships, we do not exclude other possibilities that could also produce disks with low disk dust masses or a small disk radii, such as stronger magnetic braking caused by higher ionization levels \citep{2020A&A...639A..86K} and more efficient dust drift \citep{1972fpp..conf..211W, 1977MNRAS.180...57W} driven by mechanisms other than tidal truncation. However, any plausible mechanisms invoked to explain the low disk dust masses of the sub-sample in Corona Australis and Ophiuchus should also account for their 2D spatial concentration and lack of Gaia memberships.

\subsection{External photoevaporation in Serpens}\label{subsec:external_pe}
External photoevaporation is another physical process that can truncate disk sizes by heating and stripping gas from the potential well of the central star. It usually dominates the disk evolution in highly irradiated regions, such as in the Orion Nebula Cluster \citep[e.g.][]{1998A&A...330..696M, 2018ApJ...860...77E, 2024A&A...687A..93A} and NGC 2024 \citep[e.g.][]{2020A&A...640A..27V, 2021MNRAS.501.3502H}, but recent studies show that moderate FUV (far ultra-violet) radiation down to tens of $G_0$\footnote{The unit to quantify the intensity of the FUV radiation field, normalized to the solar neighbourhood level \citep{1968BAN....19..421H}. 1 $G_0=1.6\times 10^{-3}~\mathrm{erg~s^{-1}~cm^{-2}}$} can also truncate disk sizes, resulting in observable less massive disks \citep{2023A&A...673L...2V} and smaller disk sizes \citep{2025ApJ...989....8A}. Two sub-clusters in Serpens--Serpens South ($44.27_{-26.61}^{+190.37}$ $\mathrm{G_0}$) and Serpens Far South ($\sim 40~\mathrm{G_0}$)--are also found to have moderate median FUV radiation \citep{2025A&A...695A..74A}, raising the question of whether disks in these two sub-clusters are affected by external irradiation and therefore become smaller and less massive. As most disks are not spatially resolved, our discussion in the following paragraphs focuses on the disk dust mass, which is positively correlated to the dust disk sizes \citep[e.g.][]{2010ApJ...723.1241A}.

We take Class II disks in Lupus as a reference (see footnote 6). These Lupus disks are exposed to a median FUV level of $3.48_{-0.31}^{+21.03}~\mathrm{G_0}$ \citep{2025A&A...695A..74A}, which is thought to be a typical FUV level of a star-forming region. We compare the Class II disk dust mass cumulative distributions between Lupus and four sub-clusters in Serpens, where two sub-clusters--Serpens South and Far South--have median FUV fluxes an order of magnitude higher than the other sub-clusters--Main and Northeast. The number of ALMA disks from each (sub-)cluster, their median disk dust masses and FUV fluxes are listed in Table \ref{tb:PE_disk_mass}. 

Due to its proximity to the Galactic plane, 75\% of the millimeter-dust disks in Serpens have no Gaia memberships. 44 and 9 disks in our survey are classified as non-members and members of distributed populations in Serpens, which are not associated to any sub-clusters \citep{2019ApJ...878..111H}. After excluding these disks, we assign other disks that have not been classified to any sub-clusters in the Gaia membership census, possibly due to the lack of Gaia memberships, to the nearest sub-cluster by their 2D projected locations. We plot the cumulative dust mass distributions in Fig. \ref{fig:mass_distri_pe} for (sub-)clusters listed in Table \ref{tb:PE_disk_mass}. As the sample sizes in Serpens Northeast and Serpens Far South are small ($<20$), our discussion focuses on Lupus, Serpens Main and Serpens South.

\begin{table}[]
\centering
\renewcommand{\arraystretch}{1.3}
\caption{Summary of the median disk dust masses and the estimated FUV strengths in sub-clusters of Serpens, in comparison to those of Lupus.}\label{tb:PE_disk_mass}
\begin{tabular}{cccc}
\hline
\hline
(1) & (2) &  (3) & (4) \\
 & \# & $M_\mathrm{d,med}$($\geq8M_\oplus$) & FUV$^1$ \\
 &   & [$M_\oplus$] & [$G_0$] \\
\hline
Serp Main & 47 & $23.06_{-0.36}^{+0.58}$ &  $3.93_{-1.42}^{+1.02}$\\
Serp NE & 14 & $40.00_{-0.19}^{+0.60}$ &  $2.79_{-0.49}^{+1.46}$\\
Serp S & 123 & $26.32_{-0.41}^{0.58}$ &  $44.27_{-26.61}^{+190.37}$ \\
Serp FS & 10  & $220.77_{-0.15}^{+0.78}$ &$\sim 40^{a}$\\
Lupus$^2$ &  84$^3$  & $26.23_{-0.40}^{+0.57}$  & $3.48_{-0.31}^{+21.03}$ \\
\hline
\end{tabular}
\vspace{3pt}
\tablefoot{1. \citet{2025A&A...695A..74A}; 
2. \citet{2016ApJ...828...46A}; 3. \citet{2020AJ....160..186L}. Note: a. The exact value of the median FUV strength in Serpens Far South is not explicitly reported in \citet{2025A&A...695A..74A} but it is similar to that in Serpens South (visually estimated from their Fig. 10).}
\end{table}

\begin{figure}
    \centering
    \includegraphics[width=0.83\linewidth]{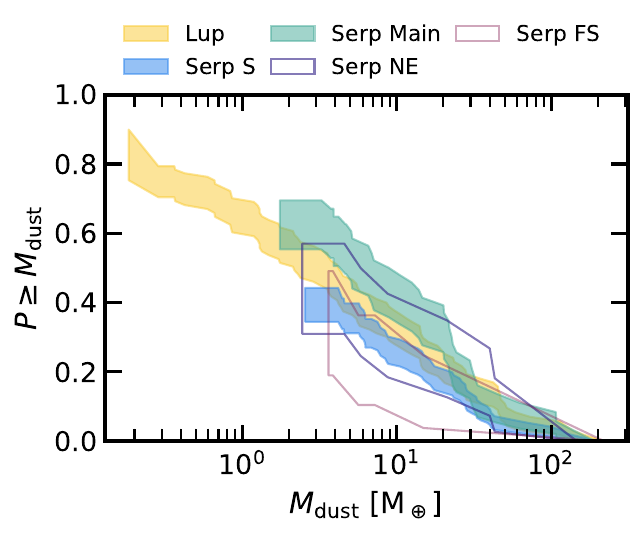}
    \caption{The cumulative disk dust mass distributions for Class II disks from Lupus, Serpens main (Serp Main), Serpens South (Serp S), Serpens Northeast (Serp NE) and Serpens far South (Serp FS).}
    \label{fig:mass_distri_pe}
\end{figure}

No correlations have been found between the cumulative disk dust mass distributions and the FUV strengths. We performed log-rank statistic tests accounting for upper-limits to examine whether the disk dust mass distributions of Lupus, Serpens Main and Serpens South are statistically different. Since the achieved sensitivity for these three regions varies\footnote{Most disks in Serpens Main are observed with an on-target integration time of 100 s, achieving a mean rms noise of $0.1~\mathrm{mJy~beam^{-1}}$, while the majority of disks in Serpens South are observed with an on-target integration time of 20 s, achieving a mean rms noise of $0.15~\mathrm{mJy~beam^{-1}}$, see Table \ref{table:obs_info}.} (as shown in Fig. \ref{fig:mass_distri_pe}), we only limit our comparison to disks massive than $8 M_\oplus$, below which many disk dust masses are detected as (marginally) upper limits in Serpens South. The log-rank tests show that the probability of disks in Lupus, Serpens Main and Serpens South drawing from different distributions is not statistically significant. The p-value for Lupus and Serpens Main is 0.84, for Lupus and Serpens South is 0.44 and for Serpens Main and South is 0.43.

The similar cumulative disk dust mass distributions in Lupus, Serpens Main and Serpens South suggest that external photoevaporation is unlikely to be the dominant process for disk evolution in Serpens South. This is further supported by their comparable median disk dust masses for disks with masses $>8 M_\oplus$ (see Table \ref{tb:PE_disk_mass}). The similarity in median disk dust masses and cumulative distributions among (sub-)clusters with FUV levels differing by an order of magnitude, as predicted by previous studies, may indicate that other mechanisms dominate the disk evolution in Serpens, or that the FUV fluxes have been over-estimated in some sub-clusters of Serpens, possibly because the extinction within sub-clusters is not taken into account, as discussed in \citet{2025A&A...695A..74A}.

Star-disk systems in Serpens are distributed more elongated along the line of sight compared to other star-forming regions (Fig. \ref{fig:distance}). This indicates that systems close in 2D are not necessarily close in 3D. Inferring 3D distances between systems from their 2D projected sky separations, like in \citet{2025A&A...695A..74A}, therefore may yield higher FUV fluxes than reality, especially for clusters like Serpens. Meanwhile, the assumptions of spherical symmetry in \citet{2025A&A...695A..74A} are likely to smooth out the FUV strengths over the entire sub-clusters due to the nature of asymmetries on all length scales of star formation \citep[e.g.,][]{2010A&A...518L.102A,2017ApJ...843...63F}. If the FUV strength is indeed higher in Serpens South and Far South, other mechanisms that can set the initial conditions \citep{2021ApJ...913..149E}, such as the initial stellar density, demonstrated in Section \ref{subsec:stellar_density}, may dominate over the external radiation in determining the disk millimeter continuum. We note, however, that the disk samples in Serpens Northeast and Serpens Far South are small ($<20$). More complete samples from these two sub-clusters are required to confirm which scenario is more applicable.

\subsection{Outflows}
Outflows have been found ubiquitously around low-mass Class 0/I objects in multi-wavelengths by different molecular/atomic tracers \citep{2023ASPC..534..567P}. The low-velocity and wide-angle outflows are usually identified in molecules, such as $\mathrm{CO}$ and its isotopologues \citep[e.g.,][]{2019ApJ...887..209A, 2023ApJ...954..101A, 2023ApJ...947...25H}, $\mathrm{HC_3N}$ \citep[e.g.,][]{2025ApJ...987..197H}, $\mathrm{C{}^{34}S}$ \citep[e.g.,][]{2017A&A...607L...6T}, etc. The high-velocity and highly collimated outflows/jets are usually traced in $\mathrm{SO}$ \citep{2007ApJ...670.1188L}, $\mathrm{SiO}$ \citep{2017NatAs...1E.152L}, $\mathrm{H_2}$ \citep{2002Msngr.109...28M}, $[\mathrm{O~I}]$ $[\mathrm{Fe II}]$ \citep{2007prpl.conf..215B}, etc. Unlike these high-velocity jets, which are typically bipolar and symmetric, a non-negligible number of low-velocity outflows are monopolar \citep[e.g.,][]{2018ApJ...863...19A, 2018A&A...618A.120L,2020A&A...634L..12D}, where the missing component can be either blue-shifted or red-shifted, or asymmetric bipolar \citep[e.g.,][]{2013ApJ...774...39A, 2023ApJ...953..190K}. 

In \ref{subsec:outflows}, we reported 12 outflow-shaped large-scale features detected in ${}^{12}\mathrm{CO}$ $\mathrm{2-1}$. Except for the first three sources on the left columns, which appear like bipolar outflows with relatively high signal-to-noise ratios, other sources only have one-side lobes. These one-side structures are mostly blue-shifted component in our samples with one exception--the last second source on the right column, though it is also possible that the blue-shifted component is beyond the field of view. Due to the limited number of targets with outflows in our survey, we do not conduct any statistical analysis.

Several scenarios have been proposed to explain the monopolar and asymmetric bipolar outflows. The key distinction lies in whether the asymmetries are intrinsic (genuinely asymmetric) or arise from observational biases or interactions with external environments. Intrinsic processes include asymmetries in the magnetic field \citep{2010MNRAS.408.2083L}, spontaneous symmetry breaking in the magnetic field configuration \citep{2017A&A...600A..75B} and spontaneous symmetry breaking caused by vertical stratification in magnetic diffusivities \citep{2024ApJ...972..142W}. In addition, \citet{2024ApJ...963...20T} suggest outflow morphologies depend on the magnetization of the cloud core. If outflows are assumed to be intrinsically symmetric and bipolar, then alternative explanations typically invoke observational biases, such as fore-ground contamination to the red-shifted component, or environmental effects, such as interactions between outflows and ambient medium \citep[e.g.,][]{2011ApJ...743...91O, 2023ApJ...958...98F}. In addition to follow-up observations of large-scale gas emission, a deeper understanding of the system and its local environment is essential to determine the underlying physics of asymmetric outflows.

\section{Conclusion}\label{sec:conclusion}
In this paper, we observe 321 disks in the Serpens star-forming regions with moderate resolution ($\sim$0.\arcsec25) to study disk dust mass distributions in Serpens and investigate why they are not affected by the previously proposed high stellar densities. We measure the disk dust masses in Serpens using ALMA continuum observations, compute the 3D stellar densities for nearby star-forming regions using Gaia data, and then compare these properties across regions. We summarize our results as follows:

\begin{itemize}
    \item The cumulative disk dust mass distribution in Serpens derived from this higher-resolution survey is consistent with the earlier study using lower-resolution data \citep[$1.\arcsec 3$,][]{2022ApJ93855A}, corroborating that disk dust masses in Serpens are similar to those in Lupus and Taurus \citep{2022ApJ93855A}.
    \item We serendipitously detected 2 transition disks with large cavities of $\sim 90$ au and 11 newly detected systems that have not been reported in previous work in the Serpens star-forming region.
    \item The assessment of the 3D stellar density for (sub-)clusters in Serpens and several nearby star-forming regions (Ophiuchus, Corona Australis, Taurus, Lupus, Upper Sco and Chamaeleon I\&II) demonstrates that L1688 in Ophiuchus is the densest sub-cluster studied in this work. Corona Australis Main \citep[$b<-17^{\circ}.3$, $A_J\leq 1$,][]{2022AJ....163...64E} is also unusually dense given its relatively old age (>5 Myr). In contrast, Serpens appears very sparse. We attribute this to incomplete memberships caused by high extinction as a result of its proximity to the Galactic plane. However, given that its cumulative disk dust mass distribution is similar to those of Lupus and Taurus, Serpens is not likely to have stellar densities sufficiently high to induce significant tidal truncation, even after accounting for missing members.
    \item We re-assess the cumulative disk dust mass distributions of Ophiuchus and Corona Australis by assigning disks to sub-clusters identified from Gaia membership censuses \citep{2020AJ....159..282E, 2022AJ....163...64E}. The previously reported tension between the low disk dust masses and young age of Ophiuchus disks is resolved when only disks explicitly identified as members of $\rho$ Oph/L1688 are considered. This is the same for Corona Australis disks when only disks from its youngest sub-cluster (1-2 Myr) are considered.
    \item Disks without Gaia identifications in Ophiuchus and the Corona Australis core tend to be more clustered in the 2D projected sky plane and less massive than disks with Gaia identifications in the same (sub-)clusters. This trend cannot be explained by the dependence of disk dust masses on stellar masses alone for samples in Ophiuchus, and may indicate that their lower disk dust masses result from tidal truncation in earlier denser environments.
    \item We compare the cumulative disk dust mass distributions of two sub-clusters in Serpens whose FUV strengths differ by an order of magnitude, as estimated in previous studies, and find that external photoevaporation is not likely to be the dominant mechanism driving disk evolution in Serpens.
\end{itemize}

\section*{Data availability}
Table 3 is only available in electronic form at the CDS via anonymous ftp to \url{cdsarc.u-strasbg.fr} (130.79.128.5) or via \url{http://cdsweb.u-strasbg.fr/cgi-bin/qcat?J/A+A/.}

\begin{acknowledgements}
We are grateful to Richard Alexander for valuable feedback and helpful discussions. We thank the anonymous referee for helping us to improve the paper. ST acknowledges Min Fang and Gregory Herczeg for fruitful discussion on the stellar density, and Kevin Luhman for helping understand data from Gaia membership censuses. ST acknowledges Leiden Observatory for waiving her international tuition fee during her studies there and the University of Leicester for a Future 100 studentship. This paper makes use of the following ALMA data: 
ADS/JAO.ALMA\#2019.1.00218.S. ALMA is a partnership of ESO (representing its member states), NSF (USA) and NINS (Japan), together with NRC (Canada), NSTC and ASIAA (Taiwan), and KASI (Republic of Korea), in cooperation with the Republic of Chile. The Joint ALMA Observatory is operated by ESO, AUI/NRAO and NAOJ. We acknowledge assistance from Allegro, the European ALMA Regional Center node in the Netherlands. This work has made use of data from the European Space Agency (ESA) mission Gaia (\url{https://www.cosmos.esa.int/gaia}), processed by the Gaia Data Processing and Analysis Consortium (DPAC, \url{https://www.cosmos.esa.int/web/gaia/dpac/consortium}). Funding for the DPAC has been provided by national institutions, in particular the institutions participating in the Gaia Multilateral Agreement. This work made use of \textsc{Jupyter} \citep{jupyter}, \textsc{Matplotlib} \citep{matplotlib}, \textsc{Numpy} \citep{numpy, numpy2}, 
\textsc{SciPy} \citep{2020SciPy-NMeth}, \textsc{Astropy} \citep{astropy2018}, \textsc{emcee} \citep{2013PASP..125..306F}, 
\textsc{galario} \citep{2018MNRAS.476.4527T},
\textsc{Pandas} \citep{pandas2010, pandas2020}, \textsc{CASA} \citep{2022PASP..134k4501C} and the \textsc{SIMBAD} astronomical database \citep{2000A&AS..143....9W}.
\end{acknowledgements}

\bibliographystyle{aa}
\bibliography{reference}

@ARTICLE{2023AJ....165...37L,
       author = {{Luhman}, K.~L.},
        title = "{A Census of the Taurus Star-forming Region and Neighboring Associations with Gaia}",
      journal = {\aj},
         year = 2023,
        month = feb,
       volume = {165},
       number = {2},
          eid = {37},
        pages = {37},
          doi = {10.3847/1538-3881/ac9da3},
archivePrefix = {arXiv},
       eprint = {2211.09785},
 primaryClass = {astro-ph.GA},
       adsurl = {https://ui.adsabs.harvard.edu/abs/2023AJ....165...37L}
}

@INPROCEEDINGS{1972fpp..conf..211W,
       author = {{Whipple}, F.~L.},
        title = "{On certain aerodynamic processes for asteroids and comets}",
    booktitle = {From Plasma to Planet},
         year = 1972,
       editor = {{Elvius}, Aina},
        month = jan,
        pages = {211},
       adsurl = {https://ui.adsabs.harvard.edu/abs/1972fpp..conf..211W}
}

@ARTICLE{1977MNRAS.180...57W,
       author = {{Weidenschilling}, S.~J.},
        title = "{Aerodynamics of solid bodies in the solar nebula.}",
      journal = {\mnras},
         year = 1977,
        month = jul,
       volume = {180},
        pages = {57-70},
          doi = {10.1093/mnras/180.2.57},
       adsurl = {https://ui.adsabs.harvard.edu/abs/1977MNRAS.180...57W}
}

@ARTICLE{2000A&AS..143....9W,
       author = {{Wenger}, M. and {Ochsenbein}, F. and {Egret}, D. and {Dubois}, P. and {Bonnarel}, F. and {Borde}, S. and {Genova}, F. and {Jasniewicz}, G. and {Lalo{\"e}}, S. and {Lesteven}, S. and {Monier}, R.},
        title = "{The SIMBAD astronomical database. The CDS reference database for astronomical objects}",
      journal = {\aaps},
         year = 2000,
        month = apr,
       volume = {143},
        pages = {9-22},
          doi = {10.1051/aas:2000332},
archivePrefix = {arXiv},
       eprint = {astro-ph/0002110},
 primaryClass = {astro-ph},
       adsurl = {https://ui.adsabs.harvard.edu/abs/2000A&AS..143....9W}
}

@ARTICLE{2021ApJ...917...23K,
       author = {{Kerr}, Ronan M.~P. and {Rizzuto}, Aaron C. and {Kraus}, Adam L. and {Offner}, Stella S.~R.},
        title = "{Stars with Photometrically Young Gaia Luminosities Around the Solar System (SPYGLASS). I. Mapping Young Stellar Structures and Their Star Formation Histories}",
      journal = {\apj},
         year = 2021,
        month = aug,
       volume = {917},
       number = {1},
          eid = {23},
        pages = {23},
          doi = {10.3847/1538-4357/ac0251},
archivePrefix = {arXiv},
       eprint = {2105.09338},
 primaryClass = {astro-ph.GA},
       adsurl = {https://ui.adsabs.harvard.edu/abs/2021ApJ...917...23K}
}

@ARTICLE{2025ApJ...978..117C,
       author = {{Carpenter}, John M. and {Esplin}, Taran L. and {Luhman}, Kevin L. and {Mamajek}, Eric E. and {Andrews}, Sean M.},
        title = "{Extending the ALMA Census of Circumstellar Disks in the Upper Scorpius OB Association}",
      journal = {\apj},
         year = 2025,
        month = jan,
       volume = {978},
       number = {1},
          eid = {117},
        pages = {117},
          doi = {10.3847/1538-4357/ad8ebc},
archivePrefix = {arXiv},
       eprint = {2410.21598},
 primaryClass = {astro-ph.SR},
       adsurl = {https://ui.adsabs.harvard.edu/abs/2025ApJ...978..117C}
}

@ARTICLE{2018ApJ...859...21A,
       author = {{Ansdell}, M. and {Williams}, J.~P. and {Trapman}, L. and {van Terwisga}, S.~E. and {Facchini}, S. and {Manara}, C.~F. and {van der Marel}, N. and {Miotello}, A. and {Tazzari}, M. and {Hogerheijde}, M. and {Guidi}, G. and {Testi}, L. and {van Dishoeck}, E.~F.},
        title = "{ALMA Survey of Lupus Protoplanetary Disks. II. Gas Disk Radii}",
      journal = {\apj},
         year = 2018,
        month = may,
       volume = {859},
       number = {1},
          eid = {21},
        pages = {21},
          doi = {10.3847/1538-4357/aab890},
archivePrefix = {arXiv},
       eprint = {1803.05923},
 primaryClass = {astro-ph.EP},
       adsurl = {https://ui.adsabs.harvard.edu/abs/2018ApJ...859...21A}
}

@ARTICLE{2025A&A...701A.242R,
       author = {{Rigliaco}, E. and {Gratton}, R. and {Nascimbeni}, V.},
        title = "{The Corona-Australis star-forming region: New insights on its formation history from detailed stellar and disk analysis}",
      journal = {\aap},
         year = 2025,
        month = sep,
       volume = {701},
          eid = {A242},
        pages = {A242},
          doi = {10.1051/0004-6361/202555158},
archivePrefix = {arXiv},
       eprint = {2508.19757},
 primaryClass = {astro-ph.SR},
       adsurl = {https://ui.adsabs.harvard.edu/abs/2025A&A...701A.242R}
}

@ARTICLE{2013ApJ...771..129A,
       author = {{Andrews}, Sean M. and {Rosenfeld}, Katherine A. and {Kraus}, Adam L. and {Wilner}, David J.},
        title = "{The Mass Dependence between Protoplanetary Disks and their Stellar Hosts}",
      journal = {\apj},
         year = 2013,
        month = jul,
       volume = {771},
       number = {2},
          eid = {129},
        pages = {129},
          doi = {10.1088/0004-637X/771/2/129},
archivePrefix = {arXiv},
       eprint = {1305.5262},
 primaryClass = {astro-ph.SR},
       adsurl = {https://ui.adsabs.harvard.edu/abs/2013ApJ...771..129A}
}

@ARTICLE{2000A&AS..146..323N,
       author = {{Neuh{\"a}user}, R. and {Walter}, F.~M. and {Covino}, E. and {Alcal{\'a}}, J.~M. and {Wolk}, S.~J. and {Frink}, S. and {Guillout}, P. and {Sterzik}, M.~F. and {Comer{\'o}n}, F.},
        title = "{Search for young stars among ROSAT All-Sky Survey X-ray sources in and around the R CrA dark cloud}",
      journal = {\aaps},
         year = 2000,
        month = oct,
       volume = {146},
        pages = {323-347},
          doi = {10.1051/aas:2000272},
archivePrefix = {arXiv},
       eprint = {astro-ph/0008095},
 primaryClass = {astro-ph},
       adsurl = {https://ui.adsabs.harvard.edu/abs/2000A&AS..146..323N}
}

@ARTICLE{2023A&A...678A..71R,
       author = {{Ratzenb{\"o}ck}, Sebastian and {Gro{\ss}schedl}, Josefa E. and {Alves}, Jo{\~a}o and {Miret-Roig}, N{\'u}ria and {Bomze}, Immanuel and {Forbes}, John and {Goodman}, Alyssa and {Hacar}, {\'A}lvaro and {Lin}, Doug and {Meingast}, Stefan and {M{\"o}ller}, Torsten and {Piecka}, Martin and {Posch}, Laura and {Rottensteiner}, Alena and {Swiggum}, Cameren and {Zucker}, Catherine},
        title = "{The star formation history of the Sco-Cen association. Coherent star formation patterns in space and time}",
      journal = {\aap},
         year = 2023,
        month = oct,
       volume = {678},
          eid = {A71},
        pages = {A71},
          doi = {10.1051/0004-6361/202346901},
archivePrefix = {arXiv},
       eprint = {2302.07853},
 primaryClass = {astro-ph.SR},
       adsurl = {https://ui.adsabs.harvard.edu/abs/2023A&A...678A..71R}
}

@ARTICLE{2023A&A...677A..59R,
       author = {{Ratzenb{\"o}ck}, Sebastian and {Gro{\ss}schedl}, Josefa E. and {M{\"o}ller}, Torsten and {Alves}, Jo{\~a}o and {Bomze}, Immanuel and {Meingast}, Stefan},
        title = "{Significance mode analysis (SigMA) for hierarchical structures. An application to the Sco-Cen OB association}",
      journal = {\aap},
         year = 2023,
        month = sep,
       volume = {677},
          eid = {A59},
        pages = {A59},
          doi = {10.1051/0004-6361/202243690},
archivePrefix = {arXiv},
       eprint = {2211.14225},
 primaryClass = {astro-ph.GA},
       adsurl = {https://ui.adsabs.harvard.edu/abs/2023A&A...677A..59R}
}

@ARTICLE{2010ApJ...723.1241A,
       author = {{Andrews}, Sean M. and {Wilner}, D.~J. and {Hughes}, A.~M. and {Qi}, Chunhua and {Dullemond}, C.~P.},
        title = "{Protoplanetary Disk Structures in Ophiuchus. II. Extension to Fainter Sources}",
      journal = {\apj},
         year = 2010,
        month = nov,
       volume = {723},
       number = {2},
        pages = {1241-1254},
          doi = {10.1088/0004-637X/723/2/1241},
archivePrefix = {arXiv},
       eprint = {1007.5070},
 primaryClass = {astro-ph.SR},
       adsurl = {https://ui.adsabs.harvard.edu/abs/2010ApJ...723.1241A}
}

@ARTICLE{2020SciPy-NMeth,
  author  = {Virtanen, Pauli and Gommers, Ralf and Oliphant, Travis E. and
            Haberland, Matt and Reddy, Tyler and Cournapeau, David and
            Burovski, Evgeni and Peterson, Pearu and Weckesser, Warren and
            Bright, Jonathan and {van der Walt}, St{\'e}fan J. and
            Brett, Matthew and Wilson, Joshua and Millman, K. Jarrod and
            Mayorov, Nikolay and Nelson, Andrew R. J. and Jones, Eric and
            Kern, Robert and Larson, Eric and Carey, C J and
            Polat, {\.I}lhan and Feng, Yu and Moore, Eric W. and
            {VanderPlas}, Jake and Laxalde, Denis and Perktold, Josef and
            Cimrman, Robert and Henriksen, Ian and Quintero, E. A. and
            Harris, Charles R. and Archibald, Anne M. and
            Ribeiro, Ant{\^o}nio H. and Pedregosa, Fabian and
            {van Mulbregt}, Paul and {SciPy 1.0 Contributors}},
  title   = {{{SciPy} 1.0: Fundamental Algorithms for Scientific
            Computing in Python}},
  journal = {Nature Methods},
  year    = {2020},
  volume  = {17},
  pages   = {261--272},
  adsurl  = {https://rdcu.be/b08Wh},
  doi     = {10.1038/s41592-019-0686-2},
}

@ARTICLE{2005ApJ...631.1134A,
       author = {{Andrews}, Sean M. and {Williams}, Jonathan P.},
        title = "{Circumstellar Dust Disks in Taurus-Auriga: The Submillimeter Perspective}",
      journal = {\apj},
         year = 2005,
        month = oct,
       volume = {631},
       number = {2},
        pages = {1134-1160},
          doi = {10.1086/432712},
archivePrefix = {arXiv},
       eprint = {astro-ph/0506187},
 primaryClass = {astro-ph},
       adsurl = {https://ui.adsabs.harvard.edu/abs/2005ApJ...631.1134A}
}

@ARTICLE{2018A&A...616A...2L,
       author = {{Lindegren}, L. and {Hern{\'a}ndez}, J. and {Bombrun}, A. and {Klioner}, S. and {Bastian}, U. and {Ramos-Lerate}, M. and {de Torres}, A. and {Steidelm{\"u}ller}, H. and {Stephenson}, C. and {Hobbs}, D. and {Lammers}, U. and {Biermann}, M. and {Geyer}, R. and {Hilger}, T. and {Michalik}, D. and {Stampa}, U. and {McMillan}, P.~J. and {Casta{\~n}eda}, J. and {Clotet}, M. and {Comoretto}, G. and {Davidson}, M. and {Fabricius}, C. and {Gracia}, G. and {Hambly}, N.~C. and {Hutton}, A. and {Mora}, A. and {Portell}, J. and {van Leeuwen}, F. and {Abbas}, U. and {Abreu}, A. and {Altmann}, M. and {Andrei}, A. and {Anglada}, E. and {Balaguer-N{\'u}{\~n}ez}, L. and {Barache}, C. and {Becciani}, U. and {Bertone}, S. and {Bianchi}, L. and {Bouquillon}, S. and {Bourda}, G. and {Br{\"u}semeister}, T. and {Bucciarelli}, B. and {Busonero}, D. and {Buzzi}, R. and {Cancelliere}, R. and {Carlucci}, T. and {Charlot}, P. and {Cheek}, N. and {Crosta}, M. and {Crowley}, C. and {de Bruijne}, J. and {de Felice}, F. and {Drimmel}, R. and {Esquej}, P. and {Fienga}, A. and {Fraile}, E. and {Gai}, M. and {Garralda}, N. and {Gonz{\'a}lez-Vidal}, J.~J. and {Guerra}, R. and {Hauser}, M. and {Hofmann}, W. and {Holl}, B. and {Jordan}, S. and {Lattanzi}, M.~G. and {Lenhardt}, H. and {Liao}, S. and {Licata}, E. and {Lister}, T. and {L{\"o}ffler}, W. and {Marchant}, J. and {Martin-Fleitas}, J.-M. and {Messineo}, R. and {Mignard}, F. and {Morbidelli}, R. and {Poggio}, E. and {Riva}, A. and {Rowell}, N. and {Salguero}, E. and {Sarasso}, M. and {Sciacca}, E. and {Siddiqui}, H. and {Smart}, R.~L. and {Spagna}, A. and {Steele}, I. and {Taris}, F. and {Torra}, J. and {van Elteren}, A. and {van Reeven}, W. and {Vecchiato}, A.},
        title = "{Gaia Data Release 2. The astrometric solution}",
      journal = {\aap},
         year = 2018,
        month = aug,
       volume = {616},
          eid = {A2},
        pages = {A2},
          doi = {10.1051/0004-6361/201832727},
archivePrefix = {arXiv},
       eprint = {1804.09366},
 primaryClass = {astro-ph.IM},
       adsurl = {https://ui.adsabs.harvard.edu/abs/2018A&A...616A...2L}
}

@ARTICLE{2020AJ....160...44L,
       author = {{Luhman}, K.~L. and {Esplin}, T.~L.},
        title = "{Refining the Census of the Upper Scorpius Association with Gaia}",
      journal = {\aj},
         year = 2020,
        month = jul,
       volume = {160},
       number = {1},
          eid = {44},
        pages = {44},
          doi = {10.3847/1538-3881/ab9599},
archivePrefix = {arXiv},
       eprint = {2005.10128},
 primaryClass = {astro-ph.SR},
       adsurl = {https://ui.adsabs.harvard.edu/abs/2020AJ....160...44L}
}

@ARTICLE{2007ApJ...666..982E,
       author = {{Enoch}, Melissa L. and {Glenn}, Jason and {Evans}, II, Neal J. and {Sargent}, Anneila I. and {Young}, Kaisa E. and {Huard}, Tracy L.},
        title = "{Comparing Star Formation on Large Scales in the c2d Legacy Clouds: Bolocam 1.1 mm Dust Continuum Surveys of Serpens, Perseus, and Ophiuchus}",
      journal = {\apj},
         year = 2007,
        month = sep,
       volume = {666},
       number = {2},
        pages = {982-1001},
          doi = {10.1086/520321},
archivePrefix = {arXiv},
       eprint = {0705.3984},
 primaryClass = {astro-ph},
       adsurl = {https://ui.adsabs.harvard.edu/abs/2007ApJ...666..982E}
}

@ARTICLE{1999A&A...345..965C,
       author = {{Cambr{\'e}sy}, L.},
        title = "{Mapping of the extinction in giant molecular clouds using optical star counts}",
      journal = {\aap},
         year = 1999,
        month = may,
       volume = {345},
        pages = {965-976},
          doi = {10.48550/arXiv.astro-ph/9903149},
archivePrefix = {arXiv},
       eprint = {astro-ph/9903149},
 primaryClass = {astro-ph},
       adsurl = {https://ui.adsabs.harvard.edu/abs/1999A&A...345..965C}
}

@ARTICLE{2007ApJ...663.1149H,
       author = {{Harvey}, Paul and {Mer{\'\i}n}, Bruno and {Huard}, Tracy L. and {Rebull}, Luisa M. and {Chapman}, Nicholas and {Evans}, II, Neal J. and {Myers}, Philip C.},
        title = "{The Spitzer c2d Survey of Large, Nearby, Interstellar Clouds. IX. The Serpens YSO Population as Observed with IRAC and MIPS}",
      journal = {\apj},
         year = 2007,
        month = jul,
       volume = {663},
       number = {2},
        pages = {1149-1173},
          doi = {10.1086/518646},
archivePrefix = {arXiv},
       eprint = {0704.0009},
 primaryClass = {astro-ph},
       adsurl = {https://ui.adsabs.harvard.edu/abs/2007ApJ...663.1149H}
}

@ARTICLE{1974ApJ...191..111S,
       author = {{Strom}, S.~E. and {Grasdalen}, G.~L. and {Strom}, K.~M.},
        title = "{Infrared and optical observations of Herbig-Haro objects.}",
      journal = {\apj},
         year = 1974,
        month = jul,
       volume = {191},
        pages = {111-142},
          doi = {10.1086/152948},
       adsurl = {https://ui.adsabs.harvard.edu/abs/1974ApJ...191..111S}
}

@ARTICLE{2011A&A...528A..50D,
       author = {{Duarte-Cabral}, A. and {Dobbs}, C.~L. and {Peretto}, N. and {Fuller}, G.~A.},
        title = "{Was a cloud-cloud collision the trigger of the recent star formation in Serpens?}",
      journal = {\aap},
         year = 2011,
        month = apr,
       volume = {528},
          eid = {A50},
        pages = {A50},
          doi = {10.1051/0004-6361/201015477},
archivePrefix = {arXiv},
       eprint = {1101.2412},
 primaryClass = {astro-ph.GA},
       adsurl = {https://ui.adsabs.harvard.edu/abs/2011A&A...528A..50D}
}

@ARTICLE{2010MNRAS.409.1412G,
       author = {{Graves}, S.~F. and {Richer}, J.~S. and {Buckle}, J.~V. and {Duarte-Cabral}, A. and {Fuller}, G.~A. and {Hogerheijde}, M.~R. and {Owen}, J.~E. and {Brunt}, C. and {Butner}, H.~M. and {Cavanagh}, B. and {Chrysostomou}, A. and {Curtis}, E.~I. and {Davis}, C.~J. and {Etxaluze}, M. and {di Francesco}, J. i and {Friberg}, P. and {Friesen}, R.~K. and {Greaves}, J.~S. and {Hatchell}, J. and {Johnstone}, D. and {Matthews}, B. and {Matthews}, H. and {Matzner}, C.~D. and {Nutter}, D. and {Rawlings}, J.~M.~C. and {Roberts}, J.~F. and {Sadavoy}, S. and {Simpson}, R.~J. and {Tothill}, N.~F.~H. and {Tsamis}, Y.~G. and {Viti}, S. and {Ward-Thompson}, D. and {White}, G.~J. and {Wouterloot}, J.~G.~A. and {Yates}, J.},
        title = "{The JCMT Legacy Survey of the Gould Belt: a first look at Serpens with HARP}",
      journal = {\mnras},
         year = 2010,
        month = dec,
       volume = {409},
       number = {4},
        pages = {1412-1428},
          doi = {10.1111/j.1365-2966.2010.17140.x},
archivePrefix = {arXiv},
       eprint = {1006.0891},
 primaryClass = {astro-ph.GA},
       adsurl = {https://ui.adsabs.harvard.edu/abs/2010MNRAS.409.1412G}
}

@ARTICLE{2010A&A...523A..29D,
       author = {{Dionatos}, O. and {Nisini}, B. and {Codella}, C. and {Giannini}, T.},
        title = "{Wide field CO J = 3 {\textrightarrow} 2 mapping of the Serpens cloud core}",
      journal = {\aap},
         year = 2010,
        month = nov,
       volume = {523},
          eid = {A29},
        pages = {A29},
          doi = {10.1051/0004-6361/200913839},
archivePrefix = {arXiv},
       eprint = {1008.0365},
 primaryClass = {astro-ph.GA},
       adsurl = {https://ui.adsabs.harvard.edu/abs/2010A&A...523A..29D}
}

@ARTICLE{2018ApJ...853..169D,
       author = {{Dhabal}, Arnab and {Mundy}, Lee G. and {Rizzo}, Maxime J. and {Storm}, Shaye and {Teuben}, Peter},
        title = "{Morphology and Kinematics of Filaments in the Serpens and Perseus Molecular Clouds}",
      journal = {\apj},
         year = 2018,
        month = feb,
       volume = {853},
       number = {2},
          eid = {169},
        pages = {169},
          doi = {10.3847/1538-4357/aaa76b},
archivePrefix = {arXiv},
       eprint = {1801.03155},
 primaryClass = {astro-ph.GA},
       adsurl = {https://ui.adsabs.harvard.edu/abs/2018ApJ...853..169D}
}

@ARTICLE{2024ApJ...973..138H,
       author = {{Hsieh}, Cheng-Han and {Arce}, H{\'e}ctor G. and {Maureira}, Mar{\'\i}a Jos{\'e} and {Pineda}, Jaime E. and {Segura-Cox}, Dominique and {Mardones}, Diego and {Dunham}, Michael M. and {Arun}, Aiswarya},
        title = "{The ALMA Legacy Survey of Class 0/I Disks in Corona australis, Aquila, chaMaeleon, oPhiuchus north, Ophiuchus, Serpens (CAMPOS). I. Evolution of Protostellar Disk Radii}",
      journal = {\apj},
         year = 2024,
        month = oct,
       volume = {973},
       number = {2},
          eid = {138},
        pages = {138},
          doi = {10.3847/1538-4357/ad6152},
archivePrefix = {arXiv},
       eprint = {2404.02809},
 primaryClass = {astro-ph.SR},
       adsurl = {https://ui.adsabs.harvard.edu/abs/2024ApJ...973..138H}
}

@ARTICLE{2017AJ....154..255L,
       author = {{Law}, Charles J. and {Ricci}, Luca and {Andrews}, Sean M. and {Wilner}, David J. and {Qi}, Chunhua},
        title = "{An SMA Continuum Survey of Circumstellar Disks in the Serpens Star-forming Region}",
      journal = {\aj},
         year = 2017,
        month = dec,
       volume = {154},
       number = {6},
          eid = {255},
        pages = {255},
          doi = {10.3847/1538-3881/aa9752},
archivePrefix = {arXiv},
       eprint = {1711.01266},
 primaryClass = {astro-ph.GA},
       adsurl = {https://ui.adsabs.harvard.edu/abs/2017AJ....154..255L}
}

@ARTICLE{2023EPJP..138..225V,
       author = {{van der Marel}, Nienke},
        title = "{Transition disks: the observational revolution from SEDs to imaging}",
      journal = {European Physical Journal Plus},
         year = 2023,
        month = mar,
       volume = {138},
       number = {3},
          eid = {225},
        pages = {225},
          doi = {10.1140/epjp/s13360-022-03628-0},
archivePrefix = {arXiv},
       eprint = {2210.05539},
 primaryClass = {astro-ph.EP},
       adsurl = {https://ui.adsabs.harvard.edu/abs/2023EPJP..138..225V}
}

@ARTICLE{2014MNRAS.441.2094R,
       author = {{Rosotti}, Giovanni P. and {Dale}, James E. and {de Juan Ovelar}, Maria and {Hubber}, David A. and {Kruijssen}, J.~M. Diederik and {Ercolano}, Barbara and {Walch}, Stefanie},
        title = "{Protoplanetary disc evolution affected by star-disc interactions in young stellar clusters}",
      journal = {\mnras},
         year = 2014,
        month = jul,
       volume = {441},
       number = {3},
        pages = {2094-2110},
          doi = {10.1093/mnras/stu679},
archivePrefix = {arXiv},
       eprint = {1404.1931},
 primaryClass = {astro-ph.EP},
       adsurl = {https://ui.adsabs.harvard.edu/abs/2014MNRAS.441.2094R}
}

@ARTICLE{1993MNRAS.261..190C,
       author = {{Clarke}, C.~J. and {Pringle}, J.~E.},
        title = "{Accretion disc response to a stellar fly-by}",
      journal = {\mnras},
         year = 1993,
        month = mar,
       volume = {261},
       number = {1},
        pages = {190-202},
          doi = {10.1093/mnras/261.1.190},
       adsurl = {https://ui.adsabs.harvard.edu/abs/1993MNRAS.261..190C}
}

@ARTICLE{2023A&A...673L...2V,
       author = {{van Terwisga}, S.~E. and {Hacar}, A.},
        title = "{Survey of Orion Disks with ALMA (SODA). II. UV-driven disk mass loss in L1641 and L1647}",
      journal = {\aap},
         year = 2023,
        month = may,
       volume = {673},
          eid = {L2},
        pages = {L2},
          doi = {10.1051/0004-6361/202346135},
archivePrefix = {arXiv},
       eprint = {2304.05777},
 primaryClass = {astro-ph.EP},
       adsurl = {https://ui.adsabs.harvard.edu/abs/2023A&A...673L...2V}
}

@ARTICLE{2025A&A...701A.139R,
       author = {{Ram{\'\i}rez-Tannus}, Mar{\'\i}a Claudia and {Bik}, Arjan and {Getman}, Konstantin V. and {Waters}, Rens and {Portilla-Revelo}, Bayron and {G{\"o}ppl}, Christiane and {Winter}, Andrew J. and {Frediani}, Jenny and {Chaparro}, Germ{\'a}n and {Feigelson}, Eric D. and {Haworth}, Thomas J. and {Henning}, Thomas and {Hern{\'a}ndez}, Sebasti{\'a}n and {Lemus-Nemoc{\'o}n}, Maria Alejandra and {Kuhn}, Michael and {Preibisch}, Thomas and {Roccatagliata}, Veronica and {Sabbi}, Elena and {van Boekel}, Roy and {Zeidler}, Peter},
        title = "{XUE: JWST spectroscopy of externally irradiated disks around young intermediate-mass stars}",
      journal = {\aap},
         year = 2025,
        month = sep,
       volume = {701},
          eid = {A139},
        pages = {A139},
          doi = {10.1051/0004-6361/202555456},
archivePrefix = {arXiv},
       eprint = {2505.06093},
 primaryClass = {astro-ph.SR},
       adsurl = {https://ui.adsabs.harvard.edu/abs/2025A&A...701A.139R}
}

@ARTICLE{2024ApJ...976..132H,
       author = {{Huang}, Jane and {Ansdell}, Megan and {Birnstiel}, Tilman and {Czekala}, Ian and {Long}, Feng and {Williams}, Jonathan and {Zhang}, Shangjia and {Zhu}, Zhaohuan},
        title = "{High-resolution ALMA Observations of Richly Structured Protoplanetary Disks in {\ensuremath{\sigma}} Orionis}",
      journal = {\apj},
         year = 2024,
        month = nov,
       volume = {976},
       number = {1},
          eid = {132},
        pages = {132},
          doi = {10.3847/1538-4357/ad84df},
archivePrefix = {arXiv},
       eprint = {2410.03823},
 primaryClass = {astro-ph.EP},
       adsurl = {https://ui.adsabs.harvard.edu/abs/2024ApJ...976..132H}
}

@ARTICLE{2021ApJ...923..221O,
       author = {{Otter}, Justin and {Ginsburg}, Adam and {Ballering}, Nicholas P. and {Bally}, John and {Eisner}, J.~A. and {Goddi}, Ciriaco and {Plambeck}, Richard and {Wright}, Melvyn},
        title = "{Small Protoplanetary Disks in the Orion Nebula Cluster and OMC1 with ALMA}",
      journal = {\apj},
         year = 2021,
        month = dec,
       volume = {923},
       number = {2},
          eid = {221},
        pages = {221},
          doi = {10.3847/1538-4357/ac29c2},
archivePrefix = {arXiv},
       eprint = {2109.14592},
 primaryClass = {astro-ph.GA},
       adsurl = {https://ui.adsabs.harvard.edu/abs/2021ApJ...923..221O}
}

@ARTICLE{2022MNRAS.514.2315C,
       author = {{Coleman}, Gavin A.~L. and {Haworth}, Thomas J.},
        title = "{Dispersal of protoplanetary discs: how stellar properties and the local environment determine the pathway of evolution}",
      journal = {\mnras},
         year = 2022,
        month = aug,
       volume = {514},
       number = {2},
        pages = {2315-2332},
          doi = {10.1093/mnras/stac1513},
archivePrefix = {arXiv},
       eprint = {2204.02303},
 primaryClass = {astro-ph.EP},
       adsurl = {https://ui.adsabs.harvard.edu/abs/2022MNRAS.514.2315C}
}

@ARTICLE{2023MNRAS.522.1939Q,
       author = {{Qiao}, Lin and {Coleman}, Gavin A.~L. and {Haworth}, Thomas J.},
        title = "{Planet formation via pebble accretion in externally photoevaporating discs}",
      journal = {\mnras},
         year = 2023,
        month = jun,
       volume = {522},
       number = {2},
        pages = {1939-1950},
          doi = {10.1093/mnras/stad944},
archivePrefix = {arXiv},
       eprint = {2303.15177},
 primaryClass = {astro-ph.EP},
       adsurl = {https://ui.adsabs.harvard.edu/abs/2023MNRAS.522.1939Q}
}

@ARTICLE{2024A&A...681A..84G,
       author = {{G{\'a}rate}, Mat{\'\i}as and {Pinilla}, Paola and {Haworth}, Thomas J. and {Facchini}, Stefano},
        title = "{The external photoevaporation of structured protoplanetary disks}",
      journal = {\aap},
         year = 2024,
        month = jan,
       volume = {681},
          eid = {A84},
        pages = {A84},
          doi = {10.1051/0004-6361/202347850},
archivePrefix = {arXiv},
       eprint = {2310.20214},
 primaryClass = {astro-ph.EP},
       adsurl = {https://ui.adsabs.harvard.edu/abs/2024A&A...681A..84G}
}

@ARTICLE{2020MNRAS.492.1279S,
       author = {{Sellek}, Andrew D. and {Booth}, Richard A. and {Clarke}, Cathie J.},
        title = "{The evolution of dust in discs influenced by external photoevaporation}",
      journal = {\mnras},
         year = 2020,
        month = feb,
       volume = {492},
       number = {1},
        pages = {1279-1294},
          doi = {10.1093/mnras/stz3528},
archivePrefix = {arXiv},
       eprint = {1912.06154},
 primaryClass = {astro-ph.EP},
       adsurl = {https://ui.adsabs.harvard.edu/abs/2020MNRAS.492.1279S}
}

@ARTICLE{2014ApJ...797...76L,
       author = {{Lee}, Katherine I. and {Fern{\'a}ndez-L{\'o}pez}, Manuel and {Storm}, Shaye and {Looney}, Leslie W. and {Mundy}, Lee G. and {Segura-Cox}, Dominique and {Teuben}, Peter and {Rosolowsky}, Erik and {Arce}, H{\'e}ctor G. and {Ostriker}, Eve C. and {Shirley}, Yancy L. and {Kwon}, Woojin and {Kauffmann}, Jens and {Tobin}, John J. and {Plunkett}, Adele L. and {Pound}, Marc W. and {Salter}, Demerese M. and {Volgenau}, N.~H. and {Chen}, Che-Yu and {Tassis}, Konstantinos and {Isella}, Andrea and {Crutcher}, Richard M. and {Gammie}, Charles F. and {Testi}, Leonardo},
        title = "{CARMA Large Area Star Formation Survey: Structure and Kinematics of Dense Gas in Serpens Main}",
      journal = {\apj},
         year = 2014,
        month = dec,
       volume = {797},
       number = {2},
          eid = {76},
        pages = {76},
          doi = {10.1088/0004-637X/797/2/76},
archivePrefix = {arXiv},
       eprint = {1410.3514},
 primaryClass = {astro-ph.GA},
       adsurl = {https://ui.adsabs.harvard.edu/abs/2014ApJ...797...76L}
}

@ARTICLE{2019ApJ...871..149F,
       author = {{Francis}, Logan and {Johnstone}, Doug and {Dunham}, Michael M. and {Hunter}, Todd R. and {Mairs}, Steve},
        title = "{Identifying Variability in Deeply Embedded Protostars with ALMA and CARMA}",
      journal = {\apj},
         year = 2019,
        month = feb,
       volume = {871},
       number = {2},
          eid = {149},
        pages = {149},
          doi = {10.3847/1538-4357/aaf972},
archivePrefix = {arXiv},
       eprint = {1902.00588},
 primaryClass = {astro-ph.SR},
       adsurl = {https://ui.adsabs.harvard.edu/abs/2019ApJ...871..149F}
}

@ARTICLE{2019MNRAS.485.2895E,
       author = {{Eden}, D.~J. and {Liu}, Tie and {Kim}, Kee-Tae and {Juvela}, M. and {Liu}, S. -Y. and {Tatematsu}, K. and {Francesco}, J. Di and {Wang}, K. and {Wu}, Y. and {Thompson}, M.~A. and {Fuller}, G.~A. and {Li}, Di and {Ristorcelli}, I. and {Kang}, Sung-ju and {Hirano}, N. and {Johnstone}, D. and {Lin}, Y. and {He}, J.~H. and {Koch}, P.~M. and {Sanhueza}, Patricio and {Qin}, S. -L. and {Zhang}, Q. and {Goldsmith}, P.~F. and {Evans}, N.~J. and {Yuan}, J. and {Zhang}, C. -P. and {White}, G.~J. and {Choi}, Minho and {Lee}, Chang Won and {Toth}, L.~V. and {Mairs}, S. and {Yi}, H. -W. and {Tang}, M. and {Soam}, A. and {Peretto}, N. and {Samal}, M.~R. and {Fich}, M. and {Parsons}, H. and {Malinen}, J. and {Bendo}, G.~J. and {Rivera-Ingraham}, A. and {Liu}, H. -L. and {Wouterloot}, J. and {Li}, P.~S. and {Qian}, L. and {Rawlings}, J. and {Rawlings}, M.~G. and {Feng}, S. and {Wang}, B. and {Li}, Dalei and {Liu}, M. and {Luo}, G. and {Marston}, A.~P. and {Pattle}, K.~M. and {Pelkonen}, V. -M. and {Rigby}, A.~J. and {Zahorecz}, S. and {Zhang}, G. and {B{\H{o}}gner}, R. and {Aikawa}, Y. and {Akhter}, S. and {Alina}, D. and {Bell}, G. and {Bernard}, J. -P. and {Blain}, A. and {Bronfman}, L. and {Byun}, D. -Y. and {Chapman}, S. and {Chen}, H. -R. and {Chen}, M. and {Chen}, W. -P. and {Chen}, X. and {Chen}, Xuepeng and {Chrysostomou}, A. and {Chu}, Y. -H. and {Chung}, E.~J. and {Cornu}, D. and {Cosentino}, G. and {Cunningham}, M.~R. and {Demyk}, K. and {Drabek-Maunder}, E. and {Doi}, Y. and {Eswaraiah}, C. and {Falgarone}, E. and {Feh{\'e}r}, O. and {Fraser}, H. and {Friberg}, P. and {Garay}, G. and {Ge}, J.~X. and {Gear}, W.~K. and {Greaves}, J. and {Guan}, X. and {Harvey-Smith}, L. and {Hasegawa}, T. and {He}, Y. and {Henkel}, C. and {Hirota}, T. and {Holland}, W. and {Hughes}, A. and {Jarken}, E. and {Ji}, T. -G. and {Jimenez-Serra}, I. and {Kang}, M. and {Kawabata}, K.~S. and {Kim}, Gwanjeong and {Kim}, Jungha and {Kim}, Jongsoo and {Kim}, S. and {Koo}, B. -C. and {Kwon}, Woojin and {Kuan}, Y. -J. and {Lacaille}, K.~M. and {Lai}, S. -P. and {Lee}, C.~F. and {Lee}, J. -E. and {Lee}, Y. -U. and {Li}, H. and {Lo}, N. and {Lopez}, J.~A.~P. and {Lu}, X. and {Lyo}, A. -R. and {Mardones}, D. and {McGehee}, P. and {Meng}, F. and {Montier}, L. and {Montillaud}, J. and {Moore}, T.~J.~T. and {Morata}, O. and {Moriarty-Schieven}, G.~H. and {Ohashi}, S. and {Pak}, S. and {Park}, Geumsook and {Paladini}, R. and {Pech}, G. and {Qiu}, K. and {Ren}, Z. -Y. and {Richer}, J. and {Sakai}, T. and {Shang}, H. and {Shinnaga}, H. and {Stamatellos}, D. and {Tang}, Y. -W. and {Traficante}, A. and {Vastel}, C. and {Viti}, S. and {Walsh}, A. and {Wang}, H. and {Wang}, J. and {Ward-Thompson}, D. and {Whitworth}, A. and {Wilson}, C.~D. and {Xu}, Y. and {Yang}, J. and {Yuan}, Y. -L. and {Yuan}, L. and {Zavagno}, A. and {Zhang}, C. and {Zhang}, G. and {Zhang}, H. -W. and {Zhou}, C. and {Zhou}, J. and {Zhu}, L. and {Zuo}, P.},
        title = "{SCOPE: SCUBA-2 Continuum Observations of Pre-protostellar Evolution - survey description and compact source catalogue}",
      journal = {\mnras},
         year = 2019,
        month = may,
       volume = {485},
       number = {2},
        pages = {2895-2908},
          doi = {10.1093/mnras/stz574},
archivePrefix = {arXiv},
       eprint = {1902.10180},
 primaryClass = {astro-ph.GA},
       adsurl = {https://ui.adsabs.harvard.edu/abs/2019MNRAS.485.2895E}
}

@ARTICLE{2018A&A...615A...9P,
       author = {{Plunkett}, Adele L. and {Fern{\'a}ndez-L{\'o}pez}, Manuel and {Arce}, H{\'e}ctor G. and {Busquet}, Gemma and {Mardones}, Diego and {Dunham}, Michael M.},
        title = "{Distribution of Serpens South protostars revealed with ALMA}",
      journal = {\aap},
         year = 2018,
        month = jul,
       volume = {615},
          eid = {A9},
        pages = {A9},
          doi = {10.1051/0004-6361/201732372},
archivePrefix = {arXiv},
       eprint = {1804.02405},
 primaryClass = {astro-ph.SR},
       adsurl = {https://ui.adsabs.harvard.edu/abs/2018A&A...615A...9P}
}

@ARTICLE{2017ApJ...843...63F,
       author = {{Friesen}, Rachel K. and {Pineda}, Jaime E. and {co-PIs} and {Rosolowsky}, Erik and {Alves}, Felipe and {Chac{\'o}n-Tanarro}, Ana and {How-Huan Chen}, Hope and {Chun-Yuan Chen}, Michael and {Di Francesco}, James and {Keown}, Jared and {Kirk}, Helen and {Punanova}, Anna and {Seo}, Youngmin and {Shirley}, Yancy and {Ginsburg}, Adam and {Hall}, Christine and {Offner}, Stella S.~R. and {Singh}, Ayushi and {Arce}, H{\'e}ctor G. and {Caselli}, Paola and {Goodman}, Alyssa A. and {Martin}, Peter G. and {Matzner}, Christopher and {Myers}, Philip C. and {Redaelli}, Elena and {GAS Collaboration}},
        title = "{The Green Bank Ammonia Survey: First Results of NH$_{3}$ Mapping of the Gould Belt}",
      journal = {\apj},
         year = 2017,
        month = jul,
       volume = {843},
       number = {1},
          eid = {63},
        pages = {63},
          doi = {10.3847/1538-4357/aa6d58},
archivePrefix = {arXiv},
       eprint = {1704.06318},
 primaryClass = {astro-ph.GA},
       adsurl = {https://ui.adsabs.harvard.edu/abs/2017ApJ...843...63F}
}

@ARTICLE{2010A&A...518L.102A,
       author = {{Andr{\'e}}, Ph. and {Men'shchikov}, A. and {Bontemps}, S. and {K{\"o}nyves}, V. and {Motte}, F. and {Schneider}, N. and {Didelon}, P. and {Minier}, V. and {Saraceno}, P. and {Ward-Thompson}, D. and {di Francesco}, J. and {White}, G. and {Molinari}, S. and {Testi}, L. and {Abergel}, A. and {Griffin}, M. and {Henning}, Th. and {Royer}, P. and {Mer{\'\i}n}, B. and {Vavrek}, R. and {Attard}, M. and {Arzoumanian}, D. and {Wilson}, C.~D. and {Ade}, P. and {Aussel}, H. and {Baluteau}, J. -P. and {Benedettini}, M. and {Bernard}, J. -Ph. and {Blommaert}, J.~A.~D.~L. and {Cambr{\'e}sy}, L. and {Cox}, P. and {di Giorgio}, A. and {Hargrave}, P. and {Hennemann}, M. and {Huang}, M. and {Kirk}, J. and {Krause}, O. and {Launhardt}, R. and {Leeks}, S. and {Le Pennec}, J. and {Li}, J.~Z. and {Martin}, P.~G. and {Maury}, A. and {Olofsson}, G. and {Omont}, A. and {Peretto}, N. and {Pezzuto}, S. and {Prusti}, T. and {Roussel}, H. and {Russeil}, D. and {Sauvage}, M. and {Sibthorpe}, B. and {Sicilia-Aguilar}, A. and {Spinoglio}, L. and {Waelkens}, C. and {Woodcraft}, A. and {Zavagno}, A.},
        title = "{From filamentary clouds to prestellar cores to the stellar IMF: Initial highlights from the Herschel Gould Belt Survey}",
      journal = {\aap},
         year = 2010,
        month = jul,
       volume = {518},
          eid = {L102},
        pages = {L102},
          doi = {10.1051/0004-6361/201014666},
archivePrefix = {arXiv},
       eprint = {1005.2618},
 primaryClass = {astro-ph.GA},
       adsurl = {https://ui.adsabs.harvard.edu/abs/2010A&A...518L.102A}
}

@ARTICLE{2011ApJ...743...91O,
       author = {{Offner}, Stella S.~R. and {Lee}, Eve J. and {Goodman}, Alyssa A. and {Arce}, H{\'e}ctor},
        title = "{Radiation-hydrodynamic Simulations of Protostellar Outflows: Synthetic Observations and Data Comparisons}",
      journal = {\apj},
         year = 2011,
        month = dec,
       volume = {743},
       number = {1},
          eid = {91},
        pages = {91},
          doi = {10.1088/0004-637X/743/1/91},
archivePrefix = {arXiv},
       eprint = {1110.5640},
 primaryClass = {astro-ph.SR},
       adsurl = {https://ui.adsabs.harvard.edu/abs/2011ApJ...743...91O}
}

@ARTICLE{2023ApJ...958...98F,
       author = {{Flores}, Christian and {Ohashi}, Nagayoshi and {Tobin}, John J. and {J{\o}rgensen}, Jes K. and {Takakuwa}, Shigehisa and {Li}, Zhi-Yun and {Lin}, Zhe-Yu Daniel and {van't Hoff}, Merel L.~R. and {Plunkett}, Adele L. and {Yamato}, Yoshihide and {Sai (Insa Choi)}, Jinshi and {Koch}, Patrick M. and {Yen}, Hsi-Wei and {Aikawa}, Yuri and {Aso}, Yusuke and {de Gregorio-Monsalvo}, Itziar and {Kido}, Miyu and {Kwon}, Woojin and {Lee}, Jeong-Eun and {Lee}, Chang Won and {Looney}, Leslie W. and {Santamar{\'\i}a-Miranda}, Alejandro and {Sharma}, Rajeeb and {Thieme}, Travis J. and {Williams}, Jonathan P. and {Han}, Ilseung and {Narayanan}, Suchitra and {Lai}, Shih-Ping},
        title = "{Early Planet Formation in Embedded Disks (eDisk). XII. Accretion Streamers, Protoplanetary Disk, and Outflow in the Class I Source Oph IRS 63}",
      journal = {\apj},
         year = 2023,
        month = nov,
       volume = {958},
       number = {1},
          eid = {98},
        pages = {98},
          doi = {10.3847/1538-4357/acf7c1},
archivePrefix = {arXiv},
       eprint = {2310.14617},
 primaryClass = {astro-ph.SR},
       adsurl = {https://ui.adsabs.harvard.edu/abs/2023ApJ...958...98F}
}

@ARTICLE{2010MNRAS.408.2083L,
       author = {{Lovelace}, R.~V.~E. and {Romanova}, M.~M. and {Ustyugova}, G.~V. and {Koldoba}, A.~V.},
        title = "{One-sided outflows/jets from rotating stars with complex magnetic fields}",
      journal = {\mnras},
         year = 2010,
        month = nov,
       volume = {408},
       number = {4},
        pages = {2083-2091},
          doi = {10.1111/j.1365-2966.2010.17284.x},
archivePrefix = {arXiv},
       eprint = {1004.0385},
 primaryClass = {astro-ph.HE},
       adsurl = {https://ui.adsabs.harvard.edu/abs/2010MNRAS.408.2083L}
}

@ARTICLE{2024ApJ...972..142W,
       author = {{Wang}, Lile and {Xu}, Sheng and {Wang}, Zhenyu and {Fang}, Min and {Goodman}, Jeremy},
        title = "{Nonideal Magnetohydrodynamic Instabilities in Protoplanetary Disks: Vertical Modes and Reflection Asymmetry}",
      journal = {\apj},
         year = 2024,
        month = sep,
       volume = {972},
       number = {2},
          eid = {142},
        pages = {142},
          doi = {10.3847/1538-4357/ad5f8d},
archivePrefix = {arXiv},
       eprint = {2311.01636},
 primaryClass = {astro-ph.EP},
       adsurl = {https://ui.adsabs.harvard.edu/abs/2024ApJ...972..142W}
}

@ARTICLE{2017A&A...600A..75B,
       author = {{B{\'e}thune}, William and {Lesur}, Geoffroy and {Ferreira}, Jonathan},
        title = "{Global simulations of protoplanetary disks with net magnetic flux. I. Non-ideal MHD case}",
      journal = {\aap},
         year = 2017,
        month = apr,
       volume = {600},
          eid = {A75},
        pages = {A75},
          doi = {10.1051/0004-6361/201630056},
archivePrefix = {arXiv},
       eprint = {1612.00883},
 primaryClass = {astro-ph.EP},
       adsurl = {https://ui.adsabs.harvard.edu/abs/2017A&A...600A..75B}
}

@ARTICLE{2024ApJ...963...20T,
       author = {{Takaishi}, Daisuke and {Tsukamoto}, Yusuke and {Kido}, Miyu and {Takakuwa}, Shigehisa and {Misugi}, Yoshiaki and {Kudoh}, Yuki and {Suto}, Yasushi},
        title = "{Formation of Unipolar Outflow and Protostellar Rocket Effect in Magnetized Turbulent Molecular Cloud Cores}",
      journal = {\apj},
         year = 2024,
        month = mar,
       volume = {963},
       number = {1},
          eid = {20},
        pages = {20},
          doi = {10.3847/1538-4357/ad187a},
archivePrefix = {arXiv},
       eprint = {2401.02525},
 primaryClass = {astro-ph.SR},
       adsurl = {https://ui.adsabs.harvard.edu/abs/2024ApJ...963...20T}
}

@ARTICLE{2023ApJ...953..190K,
       author = {{Kido}, Miyu and {Takakuwa}, Shigehisa and {Saigo}, Kazuya and {Ohashi}, Nagayoshi and {Tobin}, John J. and {J{\o}rgensen}, Jes K. and {Aikawa}, Yuri and {Aso}, Yusuke and {Encalada}, Frankie J. and {Flores}, Christian and {Gavino}, Sacha and {de Gregorio-Monsalvo}, Itziar and {Han}, Ilseung and {Hirano}, Shingo and {Koch}, Patrick M. and {Kwon}, Woojin and {Lai}, Shih-Ping and {Lee}, Chang Won and {Lee}, Jeong-Eun and {Li}, Zhi-Yun and {Lin}, Zhe-Yu Daniel and {Looney}, Leslie W. and {Mori}, Shoji and {Narayanan}, Suchitra and {Plunkett}, Adele L. and {Phuong}, Nguyen Thi and {(Insa Choi)}, Jinshi Sai and {Santamar{\'\i}a-Miranda}, Alejandro and {Sharma}, Rajeeb and {Sheehan}, Patrick D. and {Thieme}, Travis J. and {Tomida}, Kengo and {van't Hoff}, Merel L.~R. and {Williams}, Jonathan P. and {Yamato}, Yoshihide and {Yen}, Hsi-Wei},
        title = "{Early Planet Formation in Embedded Disks (eDisk). VII. Keplerian Disk, Disk Substructure, and Accretion Streamers in the Class 0 Protostar IRAS 16544-1604 in CB 68}",
      journal = {\apj},
         year = 2023,
        month = aug,
       volume = {953},
       number = {2},
          eid = {190},
        pages = {190},
          doi = {10.3847/1538-4357/acdd7a},
archivePrefix = {arXiv},
       eprint = {2306.15443},
 primaryClass = {astro-ph.EP},
       adsurl = {https://ui.adsabs.harvard.edu/abs/2023ApJ...953..190K}
}

@ARTICLE{2013ApJ...774...39A,
       author = {{Arce}, H{\'e}ctor G. and {Mardones}, Diego and {Corder}, Stuartt A. and {Garay}, Guido and {Noriega-Crespo}, Alberto and {Raga}, Alejandro C.},
        title = "{ALMA Observations of the HH 46/47 Molecular Outflow}",
      journal = {\apj},
         year = 2013,
        month = sep,
       volume = {774},
       number = {1},
          eid = {39},
        pages = {39},
          doi = {10.1088/0004-637X/774/1/39},
archivePrefix = {arXiv},
       eprint = {1304.0674},
 primaryClass = {astro-ph.SR},
       adsurl = {https://ui.adsabs.harvard.edu/abs/2013ApJ...774...39A}
}

@ARTICLE{2018ApJ...863...19A,
       author = {{Aso}, Yusuke and {Hirano}, Naomi and {Aikawa}, Yuri and {Machida}, Masahiro N. and {Takakuwa}, Shigehisa and {Yen}, Hsi-Wei and {Williams}, Jonathan P.},
        title = "{The Distinct Evolutionary Nature of Two Class 0 Protostars in Serpens Main SMM4}",
      journal = {\apj},
         year = 2018,
        month = aug,
       volume = {863},
       number = {1},
          eid = {19},
        pages = {19},
          doi = {10.3847/1538-4357/aacf9b},
archivePrefix = {arXiv},
       eprint = {1806.10743},
 primaryClass = {astro-ph.SR},
       adsurl = {https://ui.adsabs.harvard.edu/abs/2018ApJ...863...19A}
}

@INPROCEEDINGS{2007prpl.conf..215B,
       author = {{Bally}, J. and {Reipurth}, B. and {Davis}, C.~J.},
        title = "{Observations of Jets and Outflows from Young Stars}",
    booktitle = {Protostars and Planets V},
         year = 2007,
       editor = {{Reipurth}, Bo and {Jewitt}, David and {Keil}, Klaus},
        month = jan,
        pages = {215},
       adsurl = {https://ui.adsabs.harvard.edu/abs/2007prpl.conf..215B}
}

@ARTICLE{2007ApJ...670.1188L,
       author = {{Lee}, Chin-Fei and {Ho}, Paul T.~P. and {Palau}, Aina and {Hirano}, Naomi and {Bourke}, Tyler L. and {Shang}, Hsien and {Zhang}, Qizhou},
        title = "{Submillimeter Arcsecond-Resolution Mapping of the Highly Collimated Protostellar Jet HH 211}",
      journal = {\apj},
         year = 2007,
        month = dec,
       volume = {670},
       number = {2},
        pages = {1188-1197},
          doi = {10.1086/522333},
archivePrefix = {arXiv},
       eprint = {0708.1365},
 primaryClass = {astro-ph},
       adsurl = {https://ui.adsabs.harvard.edu/abs/2007ApJ...670.1188L}
}

@ARTICLE{2002Msngr.109...28M,
       author = {{McCaughrean}, M. and {Zinnecker}, H. and {Andersen}, M. and {Meeus}, G. and {Lodieu}, N.},
        title = "{Standing on the shoulder of a giant: ISAAC, Antu, and star formation}",
      journal = {The Messenger},
         year = 2002,
        month = sep,
       volume = {109},
        pages = {28-36},
       adsurl = {https://ui.adsabs.harvard.edu/abs/2002Msngr.109...28M}
}

@ARTICLE{2017NatAs...1E.152L,
       author = {{Lee}, Chin-Fei and {Ho}, Paul. T.~P. and {Li}, Zhi-Yun and {Hirano}, Naomi and {Zhang}, Qizhou and {Shang}, Hsien},
        title = "{A rotating protostellar jet launched from the innermost disk of HH 212}",
      journal = {Nature Astronomy},
         year = 2017,
        month = jul,
       volume = {1},
          eid = {0152},
        pages = {0152},
          doi = {10.1038/s41550-017-0152},
archivePrefix = {arXiv},
       eprint = {1706.06343},
 primaryClass = {astro-ph.GA},
       adsurl = {https://ui.adsabs.harvard.edu/abs/2017NatAs...1E.152L}
}

@INPROCEEDINGS{2023ASPC..534..567P,
       author = {{Pascucci}, I. and {Cabrit}, S. and {Edwards}, S. and {Gorti}, U. and {Gressel}, O. and {Suzuki}, T.~K.},
        title = "{The Role of Disk Winds in the Evolution and Dispersal of Protoplanetary Disks}",
    booktitle = {Protostars and Planets VII},
         year = 2023,
       editor = {{Inutsuka}, S. and {Aikawa}, Y. and {Muto}, T. and {Tomida}, K. and {Tamura}, M.},
       series = {Astronomical Society of the Pacific Conference Series},
       volume = {534},
        month = jul,
        pages = {567},
          doi = {10.48550/arXiv.2203.10068},
archivePrefix = {arXiv},
       eprint = {2203.10068},
 primaryClass = {astro-ph.EP},
       adsurl = {https://ui.adsabs.harvard.edu/abs/2023ASPC..534..567P}
}

@ARTICLE{2017A&A...607L...6T,
       author = {{Tabone}, B. and {Cabrit}, S. and {Bianchi}, E. and {Ferreira}, J. and {Pineau des For{\^e}ts}, G. and {Codella}, C. and {Gusdorf}, A. and {Gueth}, F. and {Podio}, L. and {Chapillon}, E.},
        title = "{ALMA discovery of a rotating SO/SO$_{2}$ flow in HH212. A possible MHD disk wind?}",
      journal = {\aap},
         year = 2017,
        month = nov,
       volume = {607},
          eid = {L6},
        pages = {L6},
          doi = {10.1051/0004-6361/201731691},
archivePrefix = {arXiv},
       eprint = {1710.01401},
 primaryClass = {astro-ph.SR},
       adsurl = {https://ui.adsabs.harvard.edu/abs/2017A&A...607L...6T}
}

@ARTICLE{2023ApJ...947...25H,
       author = {{Hsieh}, Cheng-Han and {Arce}, H{\'e}ctor G. and {Li}, Zhi-Yun and {Dunham}, Michael and {Offner}, Stella and {Stephens}, Ian W. and {Stutz}, Amelia and {Megeath}, Tom and {Kong}, Shuo and {Plunkett}, Adele and {Tobin}, John J. and {Zhang}, Yichen and {Mardones}, Diego and {Pineda}, Jaime E. and {Stanke}, Thomas and {Carpenter}, John},
        title = "{The Evolution of Protostellar Outflow Cavities, Kinematics, and Angular Distribution of Momentum and Energy in Orion A: Evidence for Dynamical Cores}",
      journal = {\apj},
         year = 2023,
        month = apr,
       volume = {947},
       number = {1},
          eid = {25},
        pages = {25},
          doi = {10.3847/1538-4357/acba13},
archivePrefix = {arXiv},
       eprint = {2302.03174},
 primaryClass = {astro-ph.SR},
       adsurl = {https://ui.adsabs.harvard.edu/abs/2023ApJ...947...25H}
}

@ARTICLE{2025ApJ...987..197H,
       author = {{Hoque}, Ariful and {Baug}, Tapas and {Dewangan}, Lokesh K. and {Juvela}, Mika and {Tej}, Anandmayee and {Goldsmith}, Paul F. and {Garc{\'\i}a}, Pablo and {Stutz}, Amelia M. and {Liu}, Tie and {Lee}, Chang Won and {Xu}, Fengwei and {Sanhueza}, Patricio and {Bhadari}, N.~K. and {Tatematsu}, K. and {Liu}, Xunchuan and {Liu}, Hong-Li and {Zhang}, Yong and {Tang}, Xindi and {Garay}, Guido and {Wang}, Ke and {Zhang}, Siju and {T{\'o}th}, L. Viktor and {Nazeer}, Hafiz and {Hwang}, Jihye and {Gorai}, Prasanta and {Bronfman}, Leonardo and {Das}, Swagat Ranjan and {Sinha}, Tirthendu},
        title = "{The ALMA-ATOMS Survey: Exploring Protostellar Outflows in HC$_{3}$N}",
      journal = {\apj},
         year = 2025,
        month = jul,
       volume = {987},
       number = {2},
          eid = {197},
        pages = {197},
          doi = {10.3847/1538-4357/add928},
archivePrefix = {arXiv},
       eprint = {2505.04164},
 primaryClass = {astro-ph.GA},
       adsurl = {https://ui.adsabs.harvard.edu/abs/2025ApJ...987..197H}
}

@ARTICLE{2019ApJ...887..209A,
       author = {{Aso}, Yusuke and {Hirano}, Naomi and {Aikawa}, Yuri and {Machida}, Masahiro N. and {Ohashi}, Nagayoshi and {Saito}, Masao and {Takakuwa}, Shigehisa and {Yen}, Hsi-Wei and {Williams}, Jonathan P.},
        title = "{Protostellar Evolution in Serpens Main: Possible Origin of Disk-size Diversity}",
      journal = {\apj},
         year = 2019,
        month = dec,
       volume = {887},
       number = {2},
          eid = {209},
        pages = {209},
          doi = {10.3847/1538-4357/ab5284},
archivePrefix = {arXiv},
       eprint = {1910.13665},
 primaryClass = {astro-ph.SR},
       adsurl = {https://ui.adsabs.harvard.edu/abs/2019ApJ...887..209A}
}

@ARTICLE{2019ApJ...870...32K,
       author = {{Kuhn}, Michael A. and {Hillenbrand}, Lynne A. and {Sills}, Alison and {Feigelson}, Eric D. and {Getman}, Konstantin V.},
        title = "{Kinematics in Young Star Clusters and Associations with Gaia DR2}",
      journal = {\apj},
         year = 2019,
        month = jan,
       volume = {870},
       number = {1},
          eid = {32},
        pages = {32},
          doi = {10.3847/1538-4357/aaef8c},
archivePrefix = {arXiv},
       eprint = {1807.02115},
 primaryClass = {astro-ph.GA},
       adsurl = {https://ui.adsabs.harvard.edu/abs/2019ApJ...870...32K}
}

@ARTICLE{1998ApJ...499..758J,
       author = {{Johnstone}, Doug and {Hollenbach}, David and {Bally}, John},
        title = "{Photoevaporation of Disks and Clumps by Nearby Massive Stars: Application to Disk Destruction in the Orion Nebula}",
      journal = {\apj},
         year = 1998,
        month = may,
       volume = {499},
       number = {2},
        pages = {758-776},
          doi = {10.1086/305658},
       adsurl = {https://ui.adsabs.harvard.edu/abs/1998ApJ...499..758J}
}

@ARTICLE{2025ApJ...989....1Z,
       author = {{Zhang} and {P{\'e}rez}, Laura M. and {Pascucci}, Ilaria and {Pinilla}, Paola and {Cieza}, Lucas A. and {Carpenter}, John and {Trapman}, Leon and {Deng}, Dingshan and {Agurto-Gangas}, Carolina and {Sierra}, Anibal and {Kurtovic}, Nicol{\'a}s T. and {Ruiz-Rodriguez}, Dary A. and {Vioque}, Miguel and {Miley}, James and {Tabone}, Beno{\^\i}t and {Gonz{\'a}lez-Ruilova}, Camilo and {Anania}, Rossella and {Rosotti}, Giovanni P. and {TorresVillanueva}, Estephani and {Hogerheijde}, Michiel R. and {Schwarz}, Kamber and {Kuznetsova}, Aleksandra},
        title = "{The ALMA Survey of Gas Evolution of PROtoplanetary Disks (AGE-PRO). I. Program Overview and Summary of First Results}",
      journal = {\apj},
         year = 2025,
        month = aug,
       volume = {989},
       number = {1},
          eid = {1},
        pages = {1},
          doi = {10.3847/1538-4357/addebe},
archivePrefix = {arXiv},
       eprint = {2506.10719},
 primaryClass = {astro-ph.EP},
       adsurl = {https://ui.adsabs.harvard.edu/abs/2025ApJ...989....1Z}
}

@inproceedings{jupyter,
       booktitle = {Positioning and Power in Academic Publishing: Players, Agents and Agendas},
          editor = {Fernando Loizides and Birgit Scmidt},
           title = {Jupyter Notebooks ? a publishing format for reproducible computational workflows},
          author = {Thomas Kluyver and Benjamin Ragan-Kelley and Fernando P{\'e}rez and Brian Granger and Matthias Bussonnier and Jonathan Frederic and Kyle Kelley and Jessica Hamrick and Jason Grout and Sylvain Corlay and Paul Ivanov and Dami{\'a}n Avila and Safia Abdalla and Carol Willing and  Jupyter development team},
       publisher = {IOS Press},
            year = {2016},
           pages = {87--90},
             url = {https://eprints.soton.ac.uk/403913/}
}

@ARTICLE{matplotlib,
  author={Hunter, John D.},
  journal={Computing in Science \& Engineering}, 
  title={Matplotlib: A 2D Graphics Environment}, 
  year={2007},
  volume={9},
  number={3},
  pages={90-95},
  doi={10.1109/MCSE.2007.55}}

@ARTICLE{numpy,
  author={van der Walt, Stefan and Colbert, S. Chris and Varoquaux, Gael},
  journal={Computing in Science \& Engineering}, 
  title={The NumPy Array: A Structure for Efficient Numerical Computation}, 
  year={2011},
  volume={13},
  number={2},
  pages={22-30},
  doi={10.1109/MCSE.2011.37}}

@ARTICLE{numpy2,
       author = {{Harris}, Charles R. and {Millman}, K. Jarrod and {van der Walt}, St{\'e}fan J. and {Gommers}, Ralf and {Virtanen}, Pauli and {Cournapeau}, David and {Wieser}, Eric and {Taylor}, Julian and {Berg}, Sebastian and {Smith}, Nathaniel J. and {Kern}, Robert and {Picus}, Matti and {Hoyer}, Stephan and {van Kerkwijk}, Marten H. and {Brett}, Matthew and {Haldane}, Allan and {del R{\'\i}o}, Jaime Fern{\'a}ndez and {Wiebe}, Mark and {Peterson}, Pearu and {G{\'e}rard-Marchant}, Pierre and {Sheppard}, Kevin and {Reddy}, Tyler and {Weckesser}, Warren and {Abbasi}, Hameer and {Gohlke}, Christoph and {Oliphant}, Travis E.},
        title = "{Array programming with NumPy}",
      journal = {\nat},
         year = 2020,
        month = sep,
       volume = {585},
       number = {7825},
        pages = {357-362},
          doi = {10.1038/s41586-020-2649-2},
archivePrefix = {arXiv},
       eprint = {2006.10256},
 primaryClass = {cs.MS},
       adsurl = {https://ui.adsabs.harvard.edu/abs/2020Natur.585..357H}
}

@misc{pandas2020,
    author       = {The pandas development team},
    title        = {pandas-dev/pandas: Pandas},
    month        = feb,
    year         = 2020,
    publisher    = {Zenodo},
    version      = {latest},
    doi          = {10.5281/zenodo.3509134},
    url          = {https://doi.org/10.5281/zenodo.3509134}
}

@InProceedings{pandas2010,
  author    = { {W}es {M}c{K}inney },
  title     = { {D}ata {S}tructures for {S}tatistical {C}omputing in {P}ython },
  booktitle = { {P}roceedings of the 9th {P}ython in {S}cience {C}onference },
  pages     = { 56 - 61 },
  year      = { 2010 },
  editor    = { {S}t\'efan van der {W}alt and {J}arrod {M}illman },
  doi       = { 10.25080/Majora-92bf1922-00a }
}

@ARTICLE{astropy2018, 
author = {{The Astropy Collaboration}},
title = "{The Astropy Project: Building an Open-science Project and Status of the v2.0 Core Package}",
journal = {\aj},
archivePrefix = "arXiv",
eprint = {1801.02634},
primaryClass = "astro-ph.IM",
year = 2018,
month = sep,
volume = 156,
eid = {123},
pages = {123},
doi = {10.3847/1538-3881/aabc4f},
adsurl = {https://ui.adsabs.harvard.edu/abs/2018AJ....156..123T}}

@INCOLLECTION{2008hsf2.book..235P,
       author = {{Preibisch}, T. and {Mamajek}, E.},
        title = "{The Nearest OB Association: Scorpius-Centaurus (Sco OB2)}",
    booktitle = {Handbook of Star Forming Regions, Volume II},
         year = 2008,
       editor = {{Reipurth}, B.},
       volume = {5},
        pages = {235},
          doi = {10.48550/arXiv.0809.0407},
       adsurl = {https://ui.adsabs.harvard.edu/abs/2008hsf2.book..235P}
}

@ARTICLE{2018MNRAS.478.2700W,
       author = {{Winter}, A.~J. and {Clarke}, C.~J. and {Rosotti}, G. and {Ih}, J. and {Facchini}, S. and {Haworth}, T.~J.},
        title = "{Protoplanetary disc truncation mechanisms in stellar clusters: comparing external photoevaporation and tidal encounters}",
      journal = {\mnras},
         year = 2018,
        month = aug,
       volume = {478},
       number = {2},
        pages = {2700-2722},
          doi = {10.1093/mnras/sty984},
archivePrefix = {arXiv},
       eprint = {1804.00013},
 primaryClass = {astro-ph.SR},
       adsurl = {https://ui.adsabs.harvard.edu/abs/2018MNRAS.478.2700W}
}

@ARTICLE{2025A&A...695A..74A,
       author = {{Anania}, Rossella and {Winter}, Andrew J. and {Rosotti}, Giovanni and {Vioque}, Miguel and {Zari}, Eleonora and {Pantaleoni Gonz{\'a}lez}, Michelangelo and {Testi}, Leonardo},
        title = "{A novel method for estimating the far-ultraviolet flux, and a catalogue for disc-hosting stars in nearby star-forming regions}",
      journal = {\aap},
         year = 2025,
        month = mar,
       volume = {695},
          eid = {A74},
        pages = {A74},
          doi = {10.1051/0004-6361/202453011},
archivePrefix = {arXiv},
       eprint = {2501.18752},
 primaryClass = {astro-ph.EP},
       adsurl = {https://ui.adsabs.harvard.edu/abs/2025A&A...695A..74A}
}

@ARTICLE{1968BAN....19..421H,
       author = {{Habing}, H.~J.},
        title = "{The interstellar radiation density between 912 A and 2400 A}",
      journal = {\bain},
         year = 1968,
        month = jan,
       volume = {19},
        pages = {421},
       adsurl = {https://ui.adsabs.harvard.edu/abs/1968BAN....19..421H}
}

@ARTICLE{2025ApJ...989....8A,
       author = {{Anania}, Rossella and {Rosotti}, Giovanni P. and {G{\'a}rate}, Mat{\'\i}as and {Pinilla}, Paola and {Vioque}, Miguel and {Trapman}, Leon and {Carpenter}, John and {Zhang}, Ke and {Pascucci}, Ilaria and {Cieza}, Lucas A. and {Sierra}, Anibal and {Kurtovic}, Nicolas T. and {Miley}, James and {P{\'e}rez}, Laura M. and {Tabone}, Beno{\^\i}t and {Hogerheijde}, Michiel and {Deng}, Dingshan and {Agurto-Gangas}, Carolina and {Ruiz-Rodriguez}, Dary A. and {Gonz{\'a}lez-Ruilova}, Camilo and {TorresVillanueva}, Estephani E.},
        title = "{The ALMA Survey of Gas Evolution of PROtoplanetary Disks (AGE-PRO). VIII. The Impact of External Photoevaporation on Disk Masses and Radii in Upper Scorpius}",
      journal = {\apj},
         year = 2025,
        month = aug,
       volume = {989},
       number = {1},
          eid = {8},
        pages = {8},
          doi = {10.3847/1538-4357/adb587},
archivePrefix = {arXiv},
       eprint = {2506.10743},
 primaryClass = {astro-ph.EP},
       adsurl = {https://ui.adsabs.harvard.edu/abs/2025ApJ...989....8A}
}

@ARTICLE{2024A&A...687A..93A,
       author = {{Aru}, M. -L. and {Mauc{\'o}}, K. and {Manara}, C.~F. and {Haworth}, T.~J. and {Facchini}, S. and {McLeod}, A.~F. and {Miotello}, A. and {Petr-Gotzens}, M.~G. and {Robberto}, M. and {Rosotti}, G.~P. and {Vicente}, S. and {Winter}, A. and {Ansdell}, M.},
        title = "{Kaleidoscope of irradiated disks: MUSE observations of proplyds in the Orion Nebula Cluster. I. Sample presentation and ionization front sizes<xref rid=``FN2'' ref-type=``fn''/>}",
      journal = {\aap},
         year = 2024,
        month = jul,
       volume = {687},
          eid = {A93},
        pages = {A93},
          doi = {10.1051/0004-6361/202349004},
archivePrefix = {arXiv},
       eprint = {2403.12604},
 primaryClass = {astro-ph.SR},
       adsurl = {https://ui.adsabs.harvard.edu/abs/2024A&A...687A..93A}
}

@ARTICLE{2018ApJ...860...77E,
       author = {{Eisner}, J.~A. and {Arce}, H.~G. and {Ballering}, N.~P. and {Bally}, J. and {Andrews}, S.~M. and {Boyden}, R.~D. and {Di Francesco}, J. and {Fang}, M. and {Johnstone}, D. and {Kim}, J.~S. and {Mann}, R.~K. and {Matthews}, B. and {Pascucci}, I. and {Ricci}, L. and {Sheehan}, P.~D. and {Williams}, J.~P.},
        title = "{Protoplanetary Disk Properties in the Orion Nebula Cluster: Initial Results from Deep, High-resolution ALMA Observations}",
      journal = {\apj},
         year = 2018,
        month = jun,
       volume = {860},
       number = {1},
          eid = {77},
        pages = {77},
          doi = {10.3847/1538-4357/aac3e2},
archivePrefix = {arXiv},
       eprint = {1805.03669},
 primaryClass = {astro-ph.SR},
       adsurl = {https://ui.adsabs.harvard.edu/abs/2018ApJ...860...77E}
}

@ARTICLE{1998A&A...330..696M,
       author = {{Marconi}, A. and {Testi}, L. and {Natta}, A. and {Walmsley}, C.~M.},
        title = "{Near infrared spectra of the Orion bar}",
      journal = {\aap},
         year = 1998,
        month = feb,
       volume = {330},
        pages = {696-710},
          doi = {10.48550/arXiv.astro-ph/9710051},
archivePrefix = {arXiv},
       eprint = {astro-ph/9710051},
 primaryClass = {astro-ph},
       adsurl = {https://ui.adsabs.harvard.edu/abs/1998A&A...330..696M}
}

@ARTICLE{2021MNRAS.501.3502H,
       author = {{Haworth}, Thomas J. and {Kim}, Jinyoung S. and {Winter}, Andrew J. and {Hines}, Dean C. and {Clarke}, Cathie J. and {Sellek}, Andrew D. and {Ballabio}, Giulia and {Stapelfeldt}, Karl R.},
        title = "{Proplyds in the flame nebula NGC 2024}",
      journal = {\mnras},
         year = 2021,
        month = mar,
       volume = {501},
       number = {3},
        pages = {3502-3514},
          doi = {10.1093/mnras/staa3918},
archivePrefix = {arXiv},
       eprint = {2012.09166},
 primaryClass = {astro-ph.SR},
       adsurl = {https://ui.adsabs.harvard.edu/abs/2021MNRAS.501.3502H}
}

@ARTICLE{2020A&A...640A..27V,
       author = {{van Terwisga}, S.~E. and {van Dishoeck}, E.~F. and {Mann}, R.~K. and {Di Francesco}, J. and {van der Marel}, N. and {Meyer}, M. and {Andrews}, S.~M. and {Carpenter}, J. and {Eisner}, J.~A. and {Manara}, C.~F. and {Williams}, J.~P.},
        title = "{Protoplanetary disk masses in NGC 2024: Evidence for two populations}",
      journal = {\aap},
         year = 2020,
        month = aug,
       volume = {640},
          eid = {A27},
        pages = {A27},
          doi = {10.1051/0004-6361/201937403},
archivePrefix = {arXiv},
       eprint = {2004.13551},
 primaryClass = {astro-ph.SR},
       adsurl = {https://ui.adsabs.harvard.edu/abs/2020A&A...640A..27V}
}

@ARTICLE{2022PASP..134k4501C,
       author = {{CASA Team} and {Bean}, Ben and {Bhatnagar}, Sanjay and {Castro}, Sandra and {Donovan Meyer}, Jennifer and {Emonts}, Bjorn and {Garcia}, Enrique and {Garwood}, Robert and {Golap}, Kumar and {Gonzalez Villalba}, Justo and {Harris}, Pamela and {Hayashi}, Yohei and {Hoskins}, Josh and {Hsieh}, Mingyu and {Jagannathan}, Preshanth and {Kawasaki}, Wataru and {Keimpema}, Aard and {Kettenis}, Mark and {Lopez}, Jorge and {Marvil}, Joshua and {Masters}, Joseph and {McNichols}, Andrew and {Mehringer}, David and {Miel}, Renaud and {Moellenbrock}, George and {Montesino}, Federico and {Nakazato}, Takeshi and {Ott}, Juergen and {Petry}, Dirk and {Pokorny}, Martin and {Raba}, Ryan and {Rau}, Urvashi and {Schiebel}, Darrell and {Schweighart}, Neal and {Sekhar}, Srikrishna and {Shimada}, Kazuhiko and {Small}, Des and {Steeb}, Jan-Willem and {Sugimoto}, Kanako and {Suoranta}, Ville and {Tsutsumi}, Takahiro and {van Bemmel}, Ilse M. and {Verkouter}, Marjolein and {Wells}, Akeem and {Xiong}, Wei and {Szomoru}, Arpad and {Griffith}, Morgan and {Glendenning}, Brian and {Kern}, Jeff},
        title = "{CASA, the Common Astronomy Software Applications for Radio Astronomy}",
      journal = {\pasp},
         year = 2022,
        month = nov,
       volume = {134},
       number = {1041},
          eid = {114501},
        pages = {114501},
          doi = {10.1088/1538-3873/ac9642},
archivePrefix = {arXiv},
       eprint = {2210.02276},
 primaryClass = {astro-ph.IM},
       adsurl = {https://ui.adsabs.harvard.edu/abs/2022PASP..134k4501C}
}

@ARTICLE{2014ApJ...784...82M,
       author = {{Mann}, Rita K. and {Di Francesco}, James and {Johnstone}, Doug and {Andrews}, Sean M. and {Williams}, Jonathan P. and {Bally}, John and {Ricci}, Luca and {Hughes}, A. Meredith and {Matthews}, Brenda C.},
        title = "{ALMA Observations of the Orion Proplyds}",
      journal = {\apj},
         year = 2014,
        month = mar,
       volume = {784},
       number = {1},
          eid = {82},
        pages = {82},
          doi = {10.1088/0004-637X/784/1/82},
archivePrefix = {arXiv},
       eprint = {1403.2026},
 primaryClass = {astro-ph.SR},
       adsurl = {https://ui.adsabs.harvard.edu/abs/2014ApJ...784...82M}
}

@ARTICLE{2019ApJ...872..158A,
       author = {{Akeson}, Rachel L. and {Jensen}, Eric L.~N. and {Carpenter}, John and {Ricci}, Luca and {Laos}, Emily and {Nogueira}, Natasha F. and {Suen-Lewis}, Emma M.},
        title = "{Resolved Young Binary Systems and Their Disks}",
      journal = {\apj},
         year = 2019,
        month = feb,
       volume = {872},
       number = {2},
          eid = {158},
        pages = {158},
          doi = {10.3847/1538-4357/aaff6a},
archivePrefix = {arXiv},
       eprint = {1901.05029},
 primaryClass = {astro-ph.SR},
       adsurl = {https://ui.adsabs.harvard.edu/abs/2019ApJ...872..158A}
}

@ARTICLE{2023EPJP..138..272K,
       author = {{Kuffmeier}, Michael and {Jensen}, Sigurd S. and {Haugb{\o}lle}, Troels},
        title = "{Rejuvenating infall: a crucial yet overlooked source of mass and angular momentum}",
      journal = {European Physical Journal Plus},
         year = 2023,
        month = mar,
       volume = {138},
       number = {3},
          eid = {272},
        pages = {272},
          doi = {10.1140/epjp/s13360-023-03880-y},
archivePrefix = {arXiv},
       eprint = {2303.05261},
 primaryClass = {astro-ph.SR},
       adsurl = {https://ui.adsabs.harvard.edu/abs/2023EPJP..138..272K}
}

@ARTICLE{2016ApJ...828...46A,
       author = {{Ansdell}, M. and {Williams}, J.~P. and {van der Marel}, N. and {Carpenter}, J.~M. and {Guidi}, G. and {Hogerheijde}, M. and {Mathews}, G.~S. and {Manara}, C.~F. and {Miotello}, A. and {Natta}, A. and {Oliveira}, I. and {Tazzari}, M. and {Testi}, L. and {van Dishoeck}, E.~F. and {van Terwisga}, S.~E.},
        title = "{ALMA Survey of Lupus Protoplanetary Disks. I. Dust and Gas Masses}",
      journal = {\apj},
         year = 2016,
        month = sep,
       volume = {828},
       number = {1},
          eid = {46},
        pages = {46},
          doi = {10.3847/0004-637X/828/1/46},
archivePrefix = {arXiv},
       eprint = {1604.05719},
 primaryClass = {astro-ph.EP},
       adsurl = {https://ui.adsabs.harvard.edu/abs/2016ApJ...828...46A}
}

@ARTICLE{2022ApJ93855A,
       author = {{Anderson}, Alexa R. and {Williams}, Jonathan P. and {van der Marel}, Nienke and {Law}, Charles J. and {Ricci}, Luca and {Tobin}, John J. and {Tong}, Simin},
        title = "{Protostellar and Protoplanetary Disk Masses in the Serpens Region}",
      journal = {\apj},
         year = 2022,
        month = oct,
       volume = {938},
       number = {1},
          eid = {55},
        pages = {55},
          doi = {10.3847/1538-4357/ac8ff0},
archivePrefix = {arXiv},
       eprint = {2204.08731},
 primaryClass = {astro-ph.SR},
       adsurl = {https://ui.adsabs.harvard.edu/abs/2022ApJ...938...55A}
}

@INPROCEEDINGS{2007ASPC..376..127M,
       author = {{McMullin}, J.~P. and {Waters}, B. and {Schiebel}, D. and {Young}, W. and {Golap}, K.},
        title = "{CASA Architecture and Applications}",
    booktitle = {Astronomical Data Analysis Software and Systems XVI},
         year = 2007,
       editor = {{Shaw}, R.~A. and {Hill}, F. and {Bell}, D.~J.},
       series = {Astronomical Society of the Pacific Conference Series},
       volume = {376},
        month = oct,
        pages = {127},
       adsurl = {https://ui.adsabs.harvard.edu/abs/2007ASPC..376..127M}
}

@ARTICLE{2019ApJ...878..111H,
       author = {{Herczeg}, Gregory J. and {Kuhn}, Michael A. and {Zhou}, Xingyu and {Hatchell}, Jennifer and {Manara}, Carlo F. and {Johnstone}, Doug and {Dunham}, Michael and {Bhardwaj}, Anupam and {Jose}, Jessy and {Yuan}, Zhen},
        title = "{An Initial Overview of the Extent and Structure of Recent Star Formation within the Serpens Molecular Cloud Using Gaia Data Release 2}",
      journal = {\apj},
         year = 2019,
        month = jun,
       volume = {878},
       number = {2},
          eid = {111},
        pages = {111},
          doi = {10.3847/1538-4357/ab1d67},
archivePrefix = {arXiv},
       eprint = {1904.04085},
 primaryClass = {astro-ph.SR},
       adsurl = {https://ui.adsabs.harvard.edu/abs/2019ApJ...878..111H}
}

@ARTICLE{2020ApJ...905..162T,
       author = {{Tobin}, John J. and {Sheehan}, Patrick D. and {Reynolds}, Nickalas and {Megeath}, S. Thomas and {Osorio}, Mayra and {Anglada}, Guillem and {D{\'\i}az-Rodr{\'\i}guez}, Ana Karla and {Furlan}, Elise and {Kratter}, Kaitlin M. and {Offner}, Stella S.~R. and {Looney}, Leslie W. and {Kama}, Mihkel and {Li}, Zhi-Yun and {van't Hoff}, Merel L.~R. and {Sadavoy}, Sarah I. and {Karnath}, Nicole},
        title = "{The VLA/ALMA Nascent Disk and Multiplicity (VANDAM) Survey of Orion Protostars. IV. Unveiling the Embedded Intermediate-Mass Protostar and Disk within OMC2-FIR3/HOPS-370}",
      journal = {\apj},
         year = 2020,
        month = dec,
       volume = {905},
       number = {2},
          eid = {162},
        pages = {162},
          doi = {10.3847/1538-4357/abc5bf},
archivePrefix = {arXiv},
       eprint = {2011.01160},
 primaryClass = {astro-ph.GA},
       adsurl = {https://ui.adsabs.harvard.edu/abs/2020ApJ...905..162T}
}

@ARTICLE{2020A&A...640A..19T,
       author = {{Tychoniec}, {\L}ukasz and {Manara}, Carlo F. and {Rosotti}, Giovanni P. and {van Dishoeck}, Ewine F. and {Cridland}, Alexander J. and {Hsieh}, Tien-Hao and {Murillo}, Nadia M. and {Segura-Cox}, Dominique and {van Terwisga}, Sierk E. and {Tobin}, John J.},
        title = "{Dust masses of young disks: constraining the initial solid reservoir for planet formation}",
      journal = {\aap},
         year = 2020,
        month = aug,
       volume = {640},
          eid = {A19},
        pages = {A19},
          doi = {10.1051/0004-6361/202037851},
archivePrefix = {arXiv},
       eprint = {2006.02812},
 primaryClass = {astro-ph.EP},
       adsurl = {https://ui.adsabs.harvard.edu/abs/2020A&A...640A..19T}
}

@ARTICLE{2016A&A...592A.126V,
       author = {{van der Marel}, N. and {Verhaar}, B.~W. and {van Terwisga}, S. and {Mer{\'\i}n}, B. and {Herczeg}, G. and {Ligterink}, N.~F.~W. and {van Dishoeck}, E.~F.},
        title = "{The (w)hole survey: An unbiased sample study of transition disk candidates based on Spitzer catalogs}",
      journal = {\aap},
         year = 2016,
        month = aug,
       volume = {592},
          eid = {A126},
        pages = {A126},
          doi = {10.1051/0004-6361/201628075},
archivePrefix = {arXiv},
       eprint = {1603.07255},
 primaryClass = {astro-ph.EP},
       adsurl = {https://ui.adsabs.harvard.edu/abs/2016A&A...592A.126V}
}

@ARTICLE{2011ARA&A..49...67W,
       author = {{Williams}, Jonathan P. and {Cieza}, Lucas A.},
        title = "{Protoplanetary Disks and Their Evolution}",
      journal = {\araa},
         year = 2011,
        month = sep,
       volume = {49},
       number = {1},
        pages = {67-117},
          doi = {10.1146/annurev-astro-081710-102548},
archivePrefix = {arXiv},
       eprint = {1103.0556},
 primaryClass = {astro-ph.GA},
       adsurl = {https://ui.adsabs.harvard.edu/abs/2011ARA&A..49...67W}
}

@article{Davidson-Pilon2019, 
    doi = {10.21105/joss.01317}, 
    url = {https://doi.org/10.21105/joss.01317}, 
    year = {2019}, 
    publisher = {The Open Journal}, 
    volume = {4}, 
    number = {40}, 
    pages = {1317}, 
    author = {Cameron Davidson-Pilon}, 
    title = {lifelines: survival analysis in Python}, 
    journal = {Journal of Open Source Software} }

@ARTICLE{2018MNRAS.476.4527T,
       author = {{Tazzari}, Marco and {Beaujean}, Frederik and {Testi}, Leonardo},
        title = "{GALARIO: a GPU accelerated library for analysing radio interferometer observations}",
      journal = {\mnras},
         year = 2018,
        month = jun,
       volume = {476},
       number = {4},
        pages = {4527-4542},
          doi = {10.1093/mnras/sty409},
archivePrefix = {arXiv},
       eprint = {1709.06999},
 primaryClass = {astro-ph.IM},
       adsurl = {https://ui.adsabs.harvard.edu/abs/2018MNRAS.476.4527T}
}

@ARTICLE{2013PASP..125..306F,
       author = {{Foreman-Mackey}, Daniel and {Hogg}, David W. and {Lang}, Dustin and {Goodman}, Jonathan},
        title = "{emcee: The MCMC Hammer}",
      journal = {\pasp},
         year = 2013,
        month = mar,
       volume = {125},
       number = {925},
        pages = {306},
          doi = {10.1086/670067},
archivePrefix = {arXiv},
       eprint = {1202.3665},
 primaryClass = {astro-ph.IM},
       adsurl = {https://ui.adsabs.harvard.edu/abs/2013PASP..125..306F}
}

@ARTICLE{2021A&A...646A..46G,
       author = {{Galli}, P.~A.~B. and {Bouy}, H. and {Olivares}, J. and {Miret-Roig}, N. and {Sarro}, L.~M. and {Barrado}, D. and {Berihuete}, A. and {Bertin}, E. and {Cuillandre}, J. -C.},
        title = "{Chamaeleon DANCe. Revisiting the stellar populations of Chamaeleon I and Chamaeleon II with Gaia-DR2 data}",
      journal = {\aap},
         year = 2021,
        month = feb,
       volume = {646},
          eid = {A46},
        pages = {A46},
          doi = {10.1051/0004-6361/202039395},
archivePrefix = {arXiv},
       eprint = {2012.00329},
 primaryClass = {astro-ph.GA},
       adsurl = {https://ui.adsabs.harvard.edu/abs/2021A&A...646A..46G}
}

@ARTICLE{2016AJ....151....5M,
       author = {{Megeath}, S.~T. and {Gutermuth}, R. and {Muzerolle}, J. and {Kryukova}, E. and {Hora}, J.~L. and {Allen}, L.~E. and {Flaherty}, K. and {Hartmann}, L. and {Myers}, P.~C. and {Pipher}, J.~L. and {Stauffer}, J. and {Young}, E.~T. and {Fazio}, G.~G.},
        title = "{The Spitzer Space Telescope Survey of the Orion A and B Molecular Clouds. II. The Spatial Distribution and Demographics of Dusty Young Stellar Objects}",
      journal = {\aj},
         year = 2016,
        month = jan,
       volume = {151},
       number = {1},
          eid = {5},
        pages = {5},
          doi = {10.3847/0004-6256/151/1/5},
archivePrefix = {arXiv},
       eprint = {1511.01202},
 primaryClass = {astro-ph.GA},
       adsurl = {https://ui.adsabs.harvard.edu/abs/2016AJ....151....5M}
}

@ARTICLE{2020AJ....160..186L,
       author = {{Luhman}, K.~L.},
        title = "{A Gaia Survey for Young Stars Associated with the Lupus Clouds}",
      journal = {\aj},
         year = 2020,
        month = oct,
       volume = {160},
       number = {4},
          eid = {186},
        pages = {186},
          doi = {10.3847/1538-3881/abb12f},
archivePrefix = {arXiv},
       eprint = {2009.05123},
 primaryClass = {astro-ph.SR},
       adsurl = {https://ui.adsabs.harvard.edu/abs/2020AJ....160..186L}
}

@ARTICLE{2020AJ....159..282E,
       author = {{Esplin}, T.~L. and {Luhman}, K.~L.},
        title = "{A Survey for New Stars and Brown Dwarfs in the Ophiuchus Star-forming Complex}",
      journal = {\aj},
         year = 2020,
        month = jun,
       volume = {159},
       number = {6},
          eid = {282},
        pages = {282},
          doi = {10.3847/1538-3881/ab8dbd},
archivePrefix = {arXiv},
       eprint = {2005.10096},
 primaryClass = {astro-ph.SR},
       adsurl = {https://ui.adsabs.harvard.edu/abs/2020AJ....159..282E}
}

@ARTICLE{2009ApJS..181..321E,
       author = {{Evans}, Neal J., II and {Dunham}, Michael M. and {J{\o}rgensen}, Jes K. and {Enoch}, Melissa L. and {Mer{\'\i}n}, Bruno and {van Dishoeck}, Ewine F. and {Alcal{\'a}}, Juan M. and {Myers}, Philip C. and {Stapelfeldt}, Karl R. and {Huard}, Tracy L. and {Allen}, Lori E. and {Harvey}, Paul M. and {van Kempen}, Tim and {Blake}, Geoffrey A. and {Koerner}, David W. and {Mundy}, Lee G. and {Padgett}, Deborah L. and {Sargent}, Anneila I.},
        title = "{The Spitzer c2d Legacy Results: Star-Formation Rates and Efficiencies; Evolution and Lifetimes}",
      journal = {\apjs},
         year = 2009,
        month = apr,
       volume = {181},
       number = {2},
        pages = {321-350},
          doi = {10.1088/0067-0049/181/2/321},
archivePrefix = {arXiv},
       eprint = {0811.1059},
 primaryClass = {astro-ph},
       adsurl = {https://ui.adsabs.harvard.edu/abs/2009ApJS..181..321E}
}

@ARTICLE{2016A&A...595A...1G,
       author = {{Gaia Collaboration} and {Prusti}, T. and {de Bruijne}, J.~H.~J. and {Brown}, A.~G.~A. and {Vallenari}, A. and {Babusiaux}, C. and {Bailer-Jones}, C.~A.~L. and {Bastian}, U. and {Biermann}, M. and {Evans}, D.~W. and {Eyer}, L. and {Jansen}, F. and {Jordi}, C. and {Klioner}, S.~A. and {Lammers}, U. and {Lindegren}, L. and {Luri}, X. and {Mignard}, F. and {Milligan}, D.~J. and {Panem}, C. and {Poinsignon}, V. and {Pourbaix}, D. and {Randich}, S. and {Sarri}, G. and {Sartoretti}, P. and {Siddiqui}, H.~I. and {Soubiran}, C. and {Valette}, V. and {van Leeuwen}, F. and {Walton}, N.~A. and {Aerts}, C. and {Arenou}, F. and {Cropper}, M. and {Drimmel}, R. and {H{\o}g}, E. and {Katz}, D. and {Lattanzi}, M.~G. and {O'Mullane}, W. and {Grebel}, E.~K. and {Holland}, A.~D. and {Huc}, C. and {Passot}, X. and {Bramante}, L. and {Cacciari}, C. and {Casta{\~n}eda}, J. and {Chaoul}, L. and {Cheek}, N. and {De Angeli}, F. and {Fabricius}, C. and {Guerra}, R. and {Hern{\'a}ndez}, J. and {Jean-Antoine-Piccolo}, A. and {Masana}, E. and {Messineo}, R. and {Mowlavi}, N. and {Nienartowicz}, K. and {Ord{\'o}{\~n}ez-Blanco}, D. and {Panuzzo}, P. and {Portell}, J. and {Richards}, P.~J. and {Riello}, M. and {Seabroke}, G.~M. and {Tanga}, P. and {Th{\'e}venin}, F. and {Torra}, J. and {Els}, S.~G. and {Gracia-Abril}, G. and {Comoretto}, G. and {Garcia-Reinaldos}, M. and {Lock}, T. and {Mercier}, E. and {Altmann}, M. and {Andrae}, R. and {Astraatmadja}, T.~L. and {Bellas-Velidis}, I. and {Benson}, K. and {Berthier}, J. and {Blomme}, R. and {Busso}, G. and {Carry}, B. and {Cellino}, A. and {Clementini}, G. and {Cowell}, S. and {Creevey}, O. and {Cuypers}, J. and {Davidson}, M. and {De Ridder}, J. and {de Torres}, A. and {Delchambre}, L. and {Dell'Oro}, A. and {Ducourant}, C. and {Fr{\'e}mat}, Y. and {Garc{\'\i}a-Torres}, M. and {Gosset}, E. and {Halbwachs}, J. -L. and {Hambly}, N.~C. and {Harrison}, D.~L. and {Hauser}, M. and {Hestroffer}, D. and {Hodgkin}, S.~T. and {Huckle}, H.~E. and {Hutton}, A. and {Jasniewicz}, G. and {Jordan}, S. and {Kontizas}, M. and {Korn}, A.~J. and {Lanzafame}, A.~C. and {Manteiga}, M. and {Moitinho}, A. and {Muinonen}, K. and {Osinde}, J. and {Pancino}, E. and {Pauwels}, T. and {Petit}, J. -M. and {Recio-Blanco}, A. and {Robin}, A.~C. and {Sarro}, L.~M. and {Siopis}, C. and {Smith}, M. and {Smith}, K.~W. and {Sozzetti}, A. and {Thuillot}, W. and {van Reeven}, W. and {Viala}, Y. and {Abbas}, U. and {Abreu Aramburu}, A. and {Accart}, S. and {Aguado}, J.~J. and {Allan}, P.~M. and {Allasia}, W. and {Altavilla}, G. and {{\'A}lvarez}, M.~A. and {Alves}, J. and {Anderson}, R.~I. and {Andrei}, A.~H. and {Anglada Varela}, E. and {Antiche}, E. and {Antoja}, T. and {Ant{\'o}n}, S. and {Arcay}, B. and {Atzei}, A. and {Ayache}, L. and {Bach}, N. and {Baker}, S.~G. and {Balaguer-N{\'u}{\~n}ez}, L. and {Barache}, C. and {Barata}, C. and {Barbier}, A. and {Barblan}, F. and {Baroni}, M. and {Barrado y Navascu{\'e}s}, D. and {Barros}, M. and {Barstow}, M.~A. and {Becciani}, U. and {Bellazzini}, M. and {Bellei}, G. and {Bello Garc{\'\i}a}, A. and {Belokurov}, V. and {Bendjoya}, P. and {Berihuete}, A. and {Bianchi}, L. and {Bienaym{\'e}}, O. and {Billebaud}, F. and {Blagorodnova}, N. and {Blanco-Cuaresma}, S. and {Boch}, T. and {Bombrun}, A. and {Borrachero}, R. and {Bouquillon}, S. and {Bourda}, G. and {Bouy}, H. and {Bragaglia}, A. and {Breddels}, M.~A. and {Brouillet}, N. and {Br{\"u}semeister}, T. and {Bucciarelli}, B. and {Budnik}, F. and {Burgess}, P. and {Burgon}, R. and {Burlacu}, A. and {Busonero}, D. and {Buzzi}, R. and {Caffau}, E. and {Cambras}, J. and {Campbell}, H. and {Cancelliere}, R. and {Cantat-Gaudin}, T. and {Carlucci}, T. and {Carrasco}, J.~M. and {Castellani}, M. and {Charlot}, P. and {Charnas}, J. and {Charvet}, P. and {Chassat}, F. and {Chiavassa}, A. and {Clotet}, M. and {Cocozza}, G. and {Collins}, R.~S. and {Collins}, P. and {Costigan}, G. and {Crifo}, F. and {Cross}, N.~J.~G. and {Crosta}, M. and {Crowley}, C. and {Dafonte}, C. and {Damerdji}, Y. and {Dapergolas}, A. and {David}, P. and {David}, M. and {De Cat}, P. and {de Felice}, F. and {de Laverny}, P. and {De Luise}, F. and {De March}, R. and {de Martino}, D. and {de Souza}, R. and {Debosscher}, J. and {del Pozo}, E. and {Delbo}, M. and {Delgado}, A. and {Delgado}, H.~E. and {di Marco}, F. and {Di Matteo}, P. and {Diakite}, S. and {Distefano}, E. and {Dolding}, C. and {Dos Anjos}, S. and {Drazinos}, P. and {Dur{\'a}n}, J. and {Dzigan}, Y. and {Ecale}, E. and {Edvardsson}, B. and {Enke}, H. and {Erdmann}, M. and {Escolar}, D. and {Espina}, M. and {Evans}, N.~W. and {Eynard Bontemps}, G. and {Fabre}, C. and {Fabrizio}, M. and {Faigler}, S. and {Falc{\~a}o}, A.~J. and {Farr{\`a}s Casas}, M. and {Faye}, F. and {Federici}, L. and {Fedorets}, G. and {Fern{\'a}ndez-Hern{\'a}ndez}, J. and {Fernique}, P. and {Fienga}, A. and {Figueras}, F. and {Filippi}, F. and {Findeisen}, K. and {Fonti}, A. and {Fouesneau}, M. and {Fraile}, E. and {Fraser}, M. and {Fuchs}, J. and {Furnell}, R. and {Gai}, M. and {Galleti}, S. and {Galluccio}, L. and {Garabato}, D. and {Garc{\'\i}a-Sedano}, F. and {Gar{\'e}}, P. and {Garofalo}, A. and {Garralda}, N. and {Gavras}, P. and {Gerssen}, J. and {Geyer}, R. and {Gilmore}, G. and {Girona}, S. and {Giuffrida}, G. and {Gomes}, M. and {Gonz{\'a}lez-Marcos}, A. and {Gonz{\'a}lez-N{\'u}{\~n}ez}, J. and {Gonz{\'a}lez-Vidal}, J.~J. and {Granvik}, M. and {Guerrier}, A. and {Guillout}, P. and {Guiraud}, J. and {G{\'u}rpide}, A. and {Guti{\'e}rrez-S{\'a}nchez}, R. and {Guy}, L.~P. and {Haigron}, R. and {Hatzidimitriou}, D. and {Haywood}, M. and {Heiter}, U. and {Helmi}, A. and {Hobbs}, D. and {Hofmann}, W. and {Holl}, B. and {Holland}, G. and {Hunt}, J.~A.~S. and {Hypki}, A. and {Icardi}, V. and {Irwin}, M. and {Jevardat de Fombelle}, G. and {Jofr{\'e}}, P. and {Jonker}, P.~G. and {Jorissen}, A. and {Julbe}, F. and {Karampelas}, A. and {Kochoska}, A. and {Kohley}, R. and {Kolenberg}, K. and {Kontizas}, E. and {Koposov}, S.~E. and {Kordopatis}, G. and {Koubsky}, P. and {Kowalczyk}, A. and {Krone-Martins}, A. and {Kudryashova}, M. and {Kull}, I. and {Bachchan}, R.~K. and {Lacoste-Seris}, F. and {Lanza}, A.~F. and {Lavigne}, J. -B. and {Le Poncin-Lafitte}, C. and {Lebreton}, Y. and {Lebzelter}, T. and {Leccia}, S. and {Leclerc}, N. and {Lecoeur-Taibi}, I. and {Lemaitre}, V. and {Lenhardt}, H. and {Leroux}, F. and {Liao}, S. and {Licata}, E. and {Lindstr{\o}m}, H.~E.~P. and {Lister}, T.~A. and {Livanou}, E. and {Lobel}, A. and {L{\"o}ffler}, W. and {L{\'o}pez}, M. and {Lopez-Lozano}, A. and {Lorenz}, D. and {Loureiro}, T. and {MacDonald}, I. and {Magalh{\~a}es Fernandes}, T. and {Managau}, S. and {Mann}, R.~G. and {Mantelet}, G. and {Marchal}, O. and {Marchant}, J.~M. and {Marconi}, M. and {Marie}, J. and {Marinoni}, S. and {Marrese}, P.~M. and {Marschalk{\'o}}, G. and {Marshall}, D.~J. and {Mart{\'\i}n-Fleitas}, J.~M. and {Martino}, M. and {Mary}, N. and {Matijevi{\v{c}}}, G. and {Mazeh}, T. and {McMillan}, P.~J. and {Messina}, S. and {Mestre}, A. and {Michalik}, D. and {Millar}, N.~R. and {Miranda}, B.~M.~H. and {Molina}, D. and {Molinaro}, R. and {Molinaro}, M. and {Moln{\'a}r}, L. and {Moniez}, M. and {Montegriffo}, P. and {Monteiro}, D. and {Mor}, R. and {Mora}, A. and {Morbidelli}, R. and {Morel}, T. and {Morgenthaler}, S. and {Morley}, T. and {Morris}, D. and {Mulone}, A.~F. and {Muraveva}, T. and {Musella}, I. and {Narbonne}, J. and {Nelemans}, G. and {Nicastro}, L. and {Noval}, L. and {Ord{\'e}novic}, C. and {Ordieres-Mer{\'e}}, J. and {Osborne}, P. and {Pagani}, C. and {Pagano}, I. and {Pailler}, F. and {Palacin}, H. and {Palaversa}, L. and {Parsons}, P. and {Paulsen}, T. and {Pecoraro}, M. and {Pedrosa}, R. and {Pentik{\"a}inen}, H. and {Pereira}, J. and {Pichon}, B. and {Piersimoni}, A.~M. and {Pineau}, F. -X. and {Plachy}, E. and {Plum}, G. and {Poujoulet}, E. and {Pr{\v{s}}a}, A. and {Pulone}, L. and {Ragaini}, S. and {Rago}, S. and {Rambaux}, N. and {Ramos-Lerate}, M. and {Ranalli}, P. and {Rauw}, G. and {Read}, A. and {Regibo}, S. and {Renk}, F. and {Reyl{\'e}}, C. and {Ribeiro}, R.~A. and {Rimoldini}, L. and {Ripepi}, V. and {Riva}, A. and {Rixon}, G. and {Roelens}, M. and {Romero-G{\'o}mez}, M. and {Rowell}, N. and {Royer}, F. and {Rudolph}, A. and {Ruiz-Dern}, L. and {Sadowski}, G. and {Sagrist{\`a} Sell{\'e}s}, T. and {Sahlmann}, J. and {Salgado}, J. and {Salguero}, E. and {Sarasso}, M. and {Savietto}, H. and {Schnorhk}, A. and {Schultheis}, M. and {Sciacca}, E. and {Segol}, M. and {Segovia}, J.~C. and {Segransan}, D. and {Serpell}, E. and {Shih}, I. -C. and {Smareglia}, R. and {Smart}, R.~L. and {Smith}, C. and {Solano}, E. and {Solitro}, F. and {Sordo}, R. and {Soria Nieto}, S. and {Souchay}, J. and {Spagna}, A. and {Spoto}, F. and {Stampa}, U. and {Steele}, I.~A. and {Steidelm{\"u}ller}, H. and {Stephenson}, C.~A. and {Stoev}, H. and {Suess}, F.~F. and {S{\"u}veges}, M. and {Surdej}, J. and {Szabados}, L. and {Szegedi-Elek}, E. and {Tapiador}, D. and {Taris}, F. and {Tauran}, G. and {Taylor}, M.~B. and {Teixeira}, R. and {Terrett}, D. and {Tingley}, B. and {Trager}, S.~C. and {Turon}, C. and {Ulla}, A. and {Utrilla}, E. and {Valentini}, G. and {van Elteren}, A. and {Van Hemelryck}, E. and {van Leeuwen}, M. and {Varadi}, M. and {Vecchiato}, A. and {Veljanoski}, J. and {Via}, T. and {Vicente}, D. and {Vogt}, S. and {Voss}, H. and {Votruba}, V. and {Voutsinas}, S. and {Walmsley}, G. and {Weiler}, M. and {Weingrill}, K. and {Werner}, D. and {Wevers}, T. and {Whitehead}, G. and {Wyrzykowski}, {\L}. and {Yoldas}, A. and {{\v{Z}}erjal}, M. and {Zucker}, S. and {Zurbach}, C. and {Zwitter}, T. and {Alecu}, A. and {Allen}, M. and {Allende Prieto}, C. and {Amorim}, A. and {Anglada-Escud{\'e}}, G. and {Arsenijevic}, V. and {Azaz}, S. and {Balm}, P. and {Beck}, M. and {Bernstein}, H. -H. and {Bigot}, L. and {Bijaoui}, A. and {Blasco}, C. and {Bonfigli}, M. and {Bono}, G. and {Boudreault}, S. and {Bressan}, A. and {Brown}, S. and {Brunet}, P. -M. and {Bunclark}, P. and {Buonanno}, R. and {Butkevich}, A.~G. and {Carret}, C. and {Carrion}, C. and {Chemin}, L. and {Ch{\'e}reau}, F. and {Corcione}, L. and {Darmigny}, E. and {de Boer}, K.~S. and {de Teodoro}, P. and {de Zeeuw}, P.~T. and {Delle Luche}, C. and {Domingues}, C.~D. and {Dubath}, P. and {Fodor}, F. and {Fr{\'e}zouls}, B. and {Fries}, A. and {Fustes}, D. and {Fyfe}, D. and {Gallardo}, E. and {Gallegos}, J. and {Gardiol}, D. and {Gebran}, M. and {Gomboc}, A. and {G{\'o}mez}, A. and {Grux}, E. and {Gueguen}, A. and {Heyrovsky}, A. and {Hoar}, J. and {Iannicola}, G. and {Isasi Parache}, Y. and {Janotto}, A. -M. and {Joliet}, E. and {Jonckheere}, A. and {Keil}, R. and {Kim}, D. -W. and {Klagyivik}, P. and {Klar}, J. and {Knude}, J. and {Kochukhov}, O. and {Kolka}, I. and {Kos}, J. and {Kutka}, A. and {Lainey}, V. and {LeBouquin}, D. and {Liu}, C. and {Loreggia}, D. and {Makarov}, V.~V. and {Marseille}, M.~G. and {Martayan}, C. and {Martinez-Rubi}, O. and {Massart}, B. and {Meynadier}, F. and {Mignot}, S. and {Munari}, U. and {Nguyen}, A. -T. and {Nordlander}, T. and {Ocvirk}, P. and {O'Flaherty}, K.~S. and {Olias Sanz}, A. and {Ortiz}, P. and {Osorio}, J. and {Oszkiewicz}, D. and {Ouzounis}, A. and {Palmer}, M. and {Park}, P. and {Pasquato}, E. and {Peltzer}, C. and {Peralta}, J. and {P{\'e}turaud}, F. and {Pieniluoma}, T. and {Pigozzi}, E. and {Poels}, J. and {Prat}, G. and {Prod'homme}, T. and {Raison}, F. and {Rebordao}, J.~M. and {Risquez}, D. and {Rocca-Volmerange}, B. and {Rosen}, S. and {Ruiz-Fuertes}, M.~I. and {Russo}, F. and {Sembay}, S. and {Serraller Vizcaino}, I. and {Short}, A. and {Siebert}, A. and {Silva}, H. and {Sinachopoulos}, D. and {Slezak}, E. and {Soffel}, M. and {Sosnowska}, D. and {Strai{\v{z}}ys}, V. and {ter Linden}, M. and {Terrell}, D. and {Theil}, S. and {Tiede}, C. and {Troisi}, L. and {Tsalmantza}, P. and {Tur}, D. and {Vaccari}, M. and {Vachier}, F. and {Valles}, P. and {Van Hamme}, W. and {Veltz}, L. and {Virtanen}, J. and {Wallut}, J. -M. and {Wichmann}, R. and {Wilkinson}, M.~I. and {Ziaeepour}, H. and {Zschocke}, S.},
        title = "{The Gaia mission}",
      journal = {\aap},
         year = 2016,
        month = nov,
       volume = {595},
          eid = {A1},
        pages = {A1},
          doi = {10.1051/0004-6361/201629272},
archivePrefix = {arXiv},
       eprint = {1609.04153},
 primaryClass = {astro-ph.IM},
       adsurl = {https://ui.adsabs.harvard.edu/abs/2016A&A...595A...1G}
}

@ARTICLE{2018A&A...616A...1G,
       author = {{Gaia Collaboration} and {Brown}, A.~G.~A. and {Vallenari}, A. and {Prusti}, T. and {de Bruijne}, J.~H.~J. and {Babusiaux}, C. and {Bailer-Jones}, C.~A.~L. and {Biermann}, M. and {Evans}, D.~W. and {Eyer}, L. and {Jansen}, F. and {Jordi}, C. and {Klioner}, S.~A. and {Lammers}, U. and {Lindegren}, L. and {Luri}, X. and {Mignard}, F. and {Panem}, C. and {Pourbaix}, D. and {Randich}, S. and {Sartoretti}, P. and {Siddiqui}, H.~I. and {Soubiran}, C. and {van Leeuwen}, F. and {Walton}, N.~A. and {Arenou}, F. and {Bastian}, U. and {Cropper}, M. and {Drimmel}, R. and {Katz}, D. and {Lattanzi}, M.~G. and {Bakker}, J. and {Cacciari}, C. and {Casta{\~n}eda}, J. and {Chaoul}, L. and {Cheek}, N. and {De Angeli}, F. and {Fabricius}, C. and {Guerra}, R. and {Holl}, B. and {Masana}, E. and {Messineo}, R. and {Mowlavi}, N. and {Nienartowicz}, K. and {Panuzzo}, P. and {Portell}, J. and {Riello}, M. and {Seabroke}, G.~M. and {Tanga}, P. and {Th{\'e}venin}, F. and {Gracia-Abril}, G. and {Comoretto}, G. and {Garcia-Reinaldos}, M. and {Teyssier}, D. and {Altmann}, M. and {Andrae}, R. and {Audard}, M. and {Bellas-Velidis}, I. and {Benson}, K. and {Berthier}, J. and {Blomme}, R. and {Burgess}, P. and {Busso}, G. and {Carry}, B. and {Cellino}, A. and {Clementini}, G. and {Clotet}, M. and {Creevey}, O. and {Davidson}, M. and {De Ridder}, J. and {Delchambre}, L. and {Dell'Oro}, A. and {Ducourant}, C. and {Fern{\'a}ndez-Hern{\'a}ndez}, J. and {Fouesneau}, M. and {Fr{\'e}mat}, Y. and {Galluccio}, L. and {Garc{\'\i}a-Torres}, M. and {Gonz{\'a}lez-N{\'u}{\~n}ez}, J. and {Gonz{\'a}lez-Vidal}, J.~J. and {Gosset}, E. and {Guy}, L.~P. and {Halbwachs}, J. -L. and {Hambly}, N.~C. and {Harrison}, D.~L. and {Hern{\'a}ndez}, J. and {Hestroffer}, D. and {Hodgkin}, S.~T. and {Hutton}, A. and {Jasniewicz}, G. and {Jean-Antoine-Piccolo}, A. and {Jordan}, S. and {Korn}, A.~J. and {Krone-Martins}, A. and {Lanzafame}, A.~C. and {Lebzelter}, T. and {L{\"o}ffler}, W. and {Manteiga}, M. and {Marrese}, P.~M. and {Mart{\'\i}n-Fleitas}, J.~M. and {Moitinho}, A. and {Mora}, A. and {Muinonen}, K. and {Osinde}, J. and {Pancino}, E. and {Pauwels}, T. and {Petit}, J. -M. and {Recio-Blanco}, A. and {Richards}, P.~J. and {Rimoldini}, L. and {Robin}, A.~C. and {Sarro}, L.~M. and {Siopis}, C. and {Smith}, M. and {Sozzetti}, A. and {S{\"u}veges}, M. and {Torra}, J. and {van Reeven}, W. and {Abbas}, U. and {Abreu Aramburu}, A. and {Accart}, S. and {Aerts}, C. and {Altavilla}, G. and {{\'A}lvarez}, M.~A. and {Alvarez}, R. and {Alves}, J. and {Anderson}, R.~I. and {Andrei}, A.~H. and {Anglada Varela}, E. and {Antiche}, E. and {Antoja}, T. and {Arcay}, B. and {Astraatmadja}, T.~L. and {Bach}, N. and {Baker}, S.~G. and {Balaguer-N{\'u}{\~n}ez}, L. and {Balm}, P. and {Barache}, C. and {Barata}, C. and {Barbato}, D. and {Barblan}, F. and {Barklem}, P.~S. and {Barrado}, D. and {Barros}, M. and {Barstow}, M.~A. and {Bartholom{\'e} Mu{\~n}oz}, S. and {Bassilana}, J. -L. and {Becciani}, U. and {Bellazzini}, M. and {Berihuete}, A. and {Bertone}, S. and {Bianchi}, L. and {Bienaym{\'e}}, O. and {Blanco-Cuaresma}, S. and {Boch}, T. and {Boeche}, C. and {Bombrun}, A. and {Borrachero}, R. and {Bossini}, D. and {Bouquillon}, S. and {Bourda}, G. and {Bragaglia}, A. and {Bramante}, L. and {Breddels}, M.~A. and {Bressan}, A. and {Brouillet}, N. and {Br{\"u}semeister}, T. and {Brugaletta}, E. and {Bucciarelli}, B. and {Burlacu}, A. and {Busonero}, D. and {Butkevich}, A.~G. and {Buzzi}, R. and {Caffau}, E. and {Cancelliere}, R. and {Cannizzaro}, G. and {Cantat-Gaudin}, T. and {Carballo}, R. and {Carlucci}, T. and {Carrasco}, J.~M. and {Casamiquela}, L. and {Castellani}, M. and {Castro-Ginard}, A. and {Charlot}, P. and {Chemin}, L. and {Chiavassa}, A. and {Cocozza}, G. and {Costigan}, G. and {Cowell}, S. and {Crifo}, F. and {Crosta}, M. and {Crowley}, C. and {Cuypers}, J. and {Dafonte}, C. and {Damerdji}, Y. and {Dapergolas}, A. and {David}, P. and {David}, M. and {de Laverny}, P. and {De Luise}, F. and {De March}, R. and {de Martino}, D. and {de Souza}, R. and {de Torres}, A. and {Debosscher}, J. and {del Pozo}, E. and {Delbo}, M. and {Delgado}, A. and {Delgado}, H.~E. and {Di Matteo}, P. and {Diakite}, S. and {Diener}, C. and {Distefano}, E. and {Dolding}, C. and {Drazinos}, P. and {Dur{\'a}n}, J. and {Edvardsson}, B. and {Enke}, H. and {Eriksson}, K. and {Esquej}, P. and {Eynard Bontemps}, G. and {Fabre}, C. and {Fabrizio}, M. and {Faigler}, S. and {Falc{\~a}o}, A.~J. and {Farr{\`a}s Casas}, M. and {Federici}, L. and {Fedorets}, G. and {Fernique}, P. and {Figueras}, F. and {Filippi}, F. and {Findeisen}, K. and {Fonti}, A. and {Fraile}, E. and {Fraser}, M. and {Fr{\'e}zouls}, B. and {Gai}, M. and {Galleti}, S. and {Garabato}, D. and {Garc{\'\i}a-Sedano}, F. and {Garofalo}, A. and {Garralda}, N. and {Gavel}, A. and {Gavras}, P. and {Gerssen}, J. and {Geyer}, R. and {Giacobbe}, P. and {Gilmore}, G. and {Girona}, S. and {Giuffrida}, G. and {Glass}, F. and {Gomes}, M. and {Granvik}, M. and {Gueguen}, A. and {Guerrier}, A. and {Guiraud}, J. and {Guti{\'e}rrez-S{\'a}nchez}, R. and {Haigron}, R. and {Hatzidimitriou}, D. and {Hauser}, M. and {Haywood}, M. and {Heiter}, U. and {Helmi}, A. and {Heu}, J. and {Hilger}, T. and {Hobbs}, D. and {Hofmann}, W. and {Holland}, G. and {Huckle}, H.~E. and {Hypki}, A. and {Icardi}, V. and {Jan{\ss}en}, K. and {Jevardat de Fombelle}, G. and {Jonker}, P.~G. and {Juh{\'a}sz}, {\'A}. L. and {Julbe}, F. and {Karampelas}, A. and {Kewley}, A. and {Klar}, J. and {Kochoska}, A. and {Kohley}, R. and {Kolenberg}, K. and {Kontizas}, M. and {Kontizas}, E. and {Koposov}, S.~E. and {Kordopatis}, G. and {Kostrzewa-Rutkowska}, Z. and {Koubsky}, P. and {Lambert}, S. and {Lanza}, A.~F. and {Lasne}, Y. and {Lavigne}, J. -B. and {Le Fustec}, Y. and {Le Poncin-Lafitte}, C. and {Lebreton}, Y. and {Leccia}, S. and {Leclerc}, N. and {Lecoeur-Taibi}, I. and {Lenhardt}, H. and {Leroux}, F. and {Liao}, S. and {Licata}, E. and {Lindstr{\o}m}, H.~E.~P. and {Lister}, T.~A. and {Livanou}, E. and {Lobel}, A. and {L{\'o}pez}, M. and {Managau}, S. and {Mann}, R.~G. and {Mantelet}, G. and {Marchal}, O. and {Marchant}, J.~M. and {Marconi}, M. and {Marinoni}, S. and {Marschalk{\'o}}, G. and {Marshall}, D.~J. and {Martino}, M. and {Marton}, G. and {Mary}, N. and {Massari}, D. and {Matijevi{\v{c}}}, G. and {Mazeh}, T. and {McMillan}, P.~J. and {Messina}, S. and {Michalik}, D. and {Millar}, N.~R. and {Molina}, D. and {Molinaro}, R. and {Moln{\'a}r}, L. and {Montegriffo}, P. and {Mor}, R. and {Morbidelli}, R. and {Morel}, T. and {Morris}, D. and {Mulone}, A.~F. and {Muraveva}, T. and {Musella}, I. and {Nelemans}, G. and {Nicastro}, L. and {Noval}, L. and {O'Mullane}, W. and {Ord{\'e}novic}, C. and {Ord{\'o}{\~n}ez-Blanco}, D. and {Osborne}, P. and {Pagani}, C. and {Pagano}, I. and {Pailler}, F. and {Palacin}, H. and {Palaversa}, L. and {Panahi}, A. and {Pawlak}, M. and {Piersimoni}, A.~M. and {Pineau}, F. -X. and {Plachy}, E. and {Plum}, G. and {Poggio}, E. and {Poujoulet}, E. and {Pr{\v{s}}a}, A. and {Pulone}, L. and {Racero}, E. and {Ragaini}, S. and {Rambaux}, N. and {Ramos-Lerate}, M. and {Regibo}, S. and {Reyl{\'e}}, C. and {Riclet}, F. and {Ripepi}, V. and {Riva}, A. and {Rivard}, A. and {Rixon}, G. and {Roegiers}, T. and {Roelens}, M. and {Romero-G{\'o}mez}, M. and {Rowell}, N. and {Royer}, F. and {Ruiz-Dern}, L. and {Sadowski}, G. and {Sagrist{\`a} Sell{\'e}s}, T. and {Sahlmann}, J. and {Salgado}, J. and {Salguero}, E. and {Sanna}, N. and {Santana-Ros}, T. and {Sarasso}, M. and {Savietto}, H. and {Schultheis}, M. and {Sciacca}, E. and {Segol}, M. and {Segovia}, J.~C. and {S{\'e}gransan}, D. and {Shih}, I. -C. and {Siltala}, L. and {Silva}, A.~F. and {Smart}, R.~L. and {Smith}, K.~W. and {Solano}, E. and {Solitro}, F. and {Sordo}, R. and {Soria Nieto}, S. and {Souchay}, J. and {Spagna}, A. and {Spoto}, F. and {Stampa}, U. and {Steele}, I.~A. and {Steidelm{\"u}ller}, H. and {Stephenson}, C.~A. and {Stoev}, H. and {Suess}, F.~F. and {Surdej}, J. and {Szabados}, L. and {Szegedi-Elek}, E. and {Tapiador}, D. and {Taris}, F. and {Tauran}, G. and {Taylor}, M.~B. and {Teixeira}, R. and {Terrett}, D. and {Teyssandier}, P. and {Thuillot}, W. and {Titarenko}, A. and {Torra Clotet}, F. and {Turon}, C. and {Ulla}, A. and {Utrilla}, E. and {Uzzi}, S. and {Vaillant}, M. and {Valentini}, G. and {Valette}, V. and {van Elteren}, A. and {Van Hemelryck}, E. and {van Leeuwen}, M. and {Vaschetto}, M. and {Vecchiato}, A. and {Veljanoski}, J. and {Viala}, Y. and {Vicente}, D. and {Vogt}, S. and {von Essen}, C. and {Voss}, H. and {Votruba}, V. and {Voutsinas}, S. and {Walmsley}, G. and {Weiler}, M. and {Wertz}, O. and {Wevers}, T. and {Wyrzykowski}, {\L}. and {Yoldas}, A. and {{\v{Z}}erjal}, M. and {Ziaeepour}, H. and {Zorec}, J. and {Zschocke}, S. and {Zucker}, S. and {Zurbach}, C. and {Zwitter}, T.},
        title = "{Gaia Data Release 2. Summary of the contents and survey properties}",
      journal = {\aap},
         year = 2018,
        month = aug,
       volume = {616},
          eid = {A1},
        pages = {A1},
          doi = {10.1051/0004-6361/201833051},
archivePrefix = {arXiv},
       eprint = {1804.09365},
 primaryClass = {astro-ph.GA},
       adsurl = {https://ui.adsabs.harvard.edu/abs/2018A&A...616A...1G}
}

@ARTICLE{2023A&A...674A...1G,
       author = {{Gaia Collaboration} and {Vallenari}, A. and {Brown}, A.~G.~A. and {Prusti}, T. and {de Bruijne}, J.~H.~J. and {Arenou}, F. and {Babusiaux}, C. and {Biermann}, M. and {Creevey}, O.~L. and {Ducourant}, C. and {Evans}, D.~W. and {Eyer}, L. and {Guerra}, R. and {Hutton}, A. and {Jordi}, C. and {Klioner}, S.~A. and {Lammers}, U.~L. and {Lindegren}, L. and {Luri}, X. and {Mignard}, F. and {Panem}, C. and {Pourbaix}, D. and {Randich}, S. and {Sartoretti}, P. and {Soubiran}, C. and {Tanga}, P. and {Walton}, N.~A. and {Bailer-Jones}, C.~A.~L. and {Bastian}, U. and {Drimmel}, R. and {Jansen}, F. and {Katz}, D. and {Lattanzi}, M.~G. and {van Leeuwen}, F. and {Bakker}, J. and {Cacciari}, C. and {Casta{\~n}eda}, J. and {De Angeli}, F. and {Fabricius}, C. and {Fouesneau}, M. and {Fr{\'e}mat}, Y. and {Galluccio}, L. and {Guerrier}, A. and {Heiter}, U. and {Masana}, E. and {Messineo}, R. and {Mowlavi}, N. and {Nicolas}, C. and {Nienartowicz}, K. and {Pailler}, F. and {Panuzzo}, P. and {Riclet}, F. and {Roux}, W. and {Seabroke}, G.~M. and {Sordo}, R. and {Th{\'e}venin}, F. and {Gracia-Abril}, G. and {Portell}, J. and {Teyssier}, D. and {Altmann}, M. and {Andrae}, R. and {Audard}, M. and {Bellas-Velidis}, I. and {Benson}, K. and {Berthier}, J. and {Blomme}, R. and {Burgess}, P.~W. and {Busonero}, D. and {Busso}, G. and {C{\'a}novas}, H. and {Carry}, B. and {Cellino}, A. and {Cheek}, N. and {Clementini}, G. and {Damerdji}, Y. and {Davidson}, M. and {de Teodoro}, P. and {Nu{\~n}ez Campos}, M. and {Delchambre}, L. and {Dell'Oro}, A. and {Esquej}, P. and {Fern{\'a}ndez-Hern{\'a}ndez}, J. and {Fraile}, E. and {Garabato}, D. and {Garc{\'\i}a-Lario}, P. and {Gosset}, E. and {Haigron}, R. and {Halbwachs}, J. -L. and {Hambly}, N.~C. and {Harrison}, D.~L. and {Hern{\'a}ndez}, J. and {Hestroffer}, D. and {Hodgkin}, S.~T. and {Holl}, B. and {Jan{\ss}en}, K. and {Jevardat de Fombelle}, G. and {Jordan}, S. and {Krone-Martins}, A. and {Lanzafame}, A.~C. and {L{\"o}ffler}, W. and {Marchal}, O. and {Marrese}, P.~M. and {Moitinho}, A. and {Muinonen}, K. and {Osborne}, P. and {Pancino}, E. and {Pauwels}, T. and {Recio-Blanco}, A. and {Reyl{\'e}}, C. and {Riello}, M. and {Rimoldini}, L. and {Roegiers}, T. and {Rybizki}, J. and {Sarro}, L.~M. and {Siopis}, C. and {Smith}, M. and {Sozzetti}, A. and {Utrilla}, E. and {van Leeuwen}, M. and {Abbas}, U. and {{\'A}brah{\'a}m}, P. and {Abreu Aramburu}, A. and {Aerts}, C. and {Aguado}, J.~J. and {Ajaj}, M. and {Aldea-Montero}, F. and {Altavilla}, G. and {{\'A}lvarez}, M.~A. and {Alves}, J. and {Anders}, F. and {Anderson}, R.~I. and {Anglada Varela}, E. and {Antoja}, T. and {Baines}, D. and {Baker}, S.~G. and {Balaguer-N{\'u}{\~n}ez}, L. and {Balbinot}, E. and {Balog}, Z. and {Barache}, C. and {Barbato}, D. and {Barros}, M. and {Barstow}, M.~A. and {Bartolom{\'e}}, S. and {Bassilana}, J. -L. and {Bauchet}, N. and {Becciani}, U. and {Bellazzini}, M. and {Berihuete}, A. and {Bernet}, M. and {Bertone}, S. and {Bianchi}, L. and {Binnenfeld}, A. and {Blanco-Cuaresma}, S. and {Blazere}, A. and {Boch}, T. and {Bombrun}, A. and {Bossini}, D. and {Bouquillon}, S. and {Bragaglia}, A. and {Bramante}, L. and {Breedt}, E. and {Bressan}, A. and {Brouillet}, N. and {Brugaletta}, E. and {Bucciarelli}, B. and {Burlacu}, A. and {Butkevich}, A.~G. and {Buzzi}, R. and {Caffau}, E. and {Cancelliere}, R. and {Cantat-Gaudin}, T. and {Carballo}, R. and {Carlucci}, T. and {Carnerero}, M.~I. and {Carrasco}, J.~M. and {Casamiquela}, L. and {Castellani}, M. and {Castro-Ginard}, A. and {Chaoul}, L. and {Charlot}, P. and {Chemin}, L. and {Chiaramida}, V. and {Chiavassa}, A. and {Chornay}, N. and {Comoretto}, G. and {Contursi}, G. and {Cooper}, W.~J. and {Cornez}, T. and {Cowell}, S. and {Crifo}, F. and {Cropper}, M. and {Crosta}, M. and {Crowley}, C. and {Dafonte}, C. and {Dapergolas}, A. and {David}, M. and {David}, P. and {de Laverny}, P. and {De Luise}, F. and {De March}, R. and {De Ridder}, J. and {de Souza}, R. and {de Torres}, A. and {del Peloso}, E.~F. and {del Pozo}, E. and {Delbo}, M. and {Delgado}, A. and {Delisle}, J. -B. and {Demouchy}, C. and {Dharmawardena}, T.~E. and {Di Matteo}, P. and {Diakite}, S. and {Diener}, C. and {Distefano}, E. and {Dolding}, C. and {Edvardsson}, B. and {Enke}, H. and {Fabre}, C. and {Fabrizio}, M. and {Faigler}, S. and {Fedorets}, G. and {Fernique}, P. and {Fienga}, A. and {Figueras}, F. and {Fournier}, Y. and {Fouron}, C. and {Fragkoudi}, F. and {Gai}, M. and {Garcia-Gutierrez}, A. and {Garcia-Reinaldos}, M. and {Garc{\'\i}a-Torres}, M. and {Garofalo}, A. and {Gavel}, A. and {Gavras}, P. and {Gerlach}, E. and {Geyer}, R. and {Giacobbe}, P. and {Gilmore}, G. and {Girona}, S. and {Giuffrida}, G. and {Gomel}, R. and {Gomez}, A. and {Gonz{\'a}lez-N{\'u}{\~n}ez}, J. and {Gonz{\'a}lez-Santamar{\'\i}a}, I. and {Gonz{\'a}lez-Vidal}, J.~J. and {Granvik}, M. and {Guillout}, P. and {Guiraud}, J. and {Guti{\'e}rrez-S{\'a}nchez}, R. and {Guy}, L.~P. and {Hatzidimitriou}, D. and {Hauser}, M. and {Haywood}, M. and {Helmer}, A. and {Helmi}, A. and {Sarmiento}, M.~H. and {Hidalgo}, S.~L. and {Hilger}, T. and {H{\l}adczuk}, N. and {Hobbs}, D. and {Holland}, G. and {Huckle}, H.~E. and {Jardine}, K. and {Jasniewicz}, G. and {Jean-Antoine Piccolo}, A. and {Jim{\'e}nez-Arranz}, {\'O}. and {Jorissen}, A. and {Juaristi Campillo}, J. and {Julbe}, F. and {Karbevska}, L. and {Kervella}, P. and {Khanna}, S. and {Kontizas}, M. and {Kordopatis}, G. and {Korn}, A.~J. and {K{\'o}sp{\'a}l}, {\'A}. and {Kostrzewa-Rutkowska}, Z. and {Kruszy{\'n}ska}, K. and {Kun}, M. and {Laizeau}, P. and {Lambert}, S. and {Lanza}, A.~F. and {Lasne}, Y. and {Le Campion}, J. -F. and {Lebreton}, Y. and {Lebzelter}, T. and {Leccia}, S. and {Leclerc}, N. and {Lecoeur-Taibi}, I. and {Liao}, S. and {Licata}, E.~L. and {Lindstr{\o}m}, H.~E.~P. and {Lister}, T.~A. and {Livanou}, E. and {Lobel}, A. and {Lorca}, A. and {Loup}, C. and {Madrero Pardo}, P. and {Magdaleno Romeo}, A. and {Managau}, S. and {Mann}, R.~G. and {Manteiga}, M. and {Marchant}, J.~M. and {Marconi}, M. and {Marcos}, J. and {Marcos Santos}, M.~M.~S. and {Mar{\'\i}n Pina}, D. and {Marinoni}, S. and {Marocco}, F. and {Marshall}, D.~J. and {Martin Polo}, L. and {Mart{\'\i}n-Fleitas}, J.~M. and {Marton}, G. and {Mary}, N. and {Masip}, A. and {Massari}, D. and {Mastrobuono-Battisti}, A. and {Mazeh}, T. and {McMillan}, P.~J. and {Messina}, S. and {Michalik}, D. and {Millar}, N.~R. and {Mints}, A. and {Molina}, D. and {Molinaro}, R. and {Moln{\'a}r}, L. and {Monari}, G. and {Mongui{\'o}}, M. and {Montegriffo}, P. and {Montero}, A. and {Mor}, R. and {Mora}, A. and {Morbidelli}, R. and {Morel}, T. and {Morris}, D. and {Muraveva}, T. and {Murphy}, C.~P. and {Musella}, I. and {Nagy}, Z. and {Noval}, L. and {Oca{\~n}a}, F. and {Ogden}, A. and {Ordenovic}, C. and {Osinde}, J.~O. and {Pagani}, C. and {Pagano}, I. and {Palaversa}, L. and {Palicio}, P.~A. and {Pallas-Quintela}, L. and {Panahi}, A. and {Payne-Wardenaar}, S. and {Pe{\~n}alosa Esteller}, X. and {Penttil{\"a}}, A. and {Pichon}, B. and {Piersimoni}, A.~M. and {Pineau}, F. -X. and {Plachy}, E. and {Plum}, G. and {Poggio}, E. and {Pr{\v{s}}a}, A. and {Pulone}, L. and {Racero}, E. and {Ragaini}, S. and {Rainer}, M. and {Raiteri}, C.~M. and {Rambaux}, N. and {Ramos}, P. and {Ramos-Lerate}, M. and {Re Fiorentin}, P. and {Regibo}, S. and {Richards}, P.~J. and {Rios Diaz}, C. and {Ripepi}, V. and {Riva}, A. and {Rix}, H. -W. and {Rixon}, G. and {Robichon}, N. and {Robin}, A.~C. and {Robin}, C. and {Roelens}, M. and {Rogues}, H.~R.~O. and {Rohrbasser}, L. and {Romero-G{\'o}mez}, M. and {Rowell}, N. and {Royer}, F. and {Ruz Mieres}, D. and {Rybicki}, K.~A. and {Sadowski}, G. and {S{\'a}ez N{\'u}{\~n}ez}, A. and {Sagrist{\`a} Sell{\'e}s}, A. and {Sahlmann}, J. and {Salguero}, E. and {Samaras}, N. and {Sanchez Gimenez}, V. and {Sanna}, N. and {Santove{\~n}a}, R. and {Sarasso}, M. and {Schultheis}, M. and {Sciacca}, E. and {Segol}, M. and {Segovia}, J.~C. and {S{\'e}gransan}, D. and {Semeux}, D. and {Shahaf}, S. and {Siddiqui}, H.~I. and {Siebert}, A. and {Siltala}, L. and {Silvelo}, A. and {Slezak}, E. and {Slezak}, I. and {Smart}, R.~L. and {Snaith}, O.~N. and {Solano}, E. and {Solitro}, F. and {Souami}, D. and {Souchay}, J. and {Spagna}, A. and {Spina}, L. and {Spoto}, F. and {Steele}, I.~A. and {Steidelm{\"u}ller}, H. and {Stephenson}, C.~A. and {S{\"u}veges}, M. and {Surdej}, J. and {Szabados}, L. and {Szegedi-Elek}, E. and {Taris}, F. and {Taylor}, M.~B. and {Teixeira}, R. and {Tolomei}, L. and {Tonello}, N. and {Torra}, F. and {Torra}, J. and {Torralba Elipe}, G. and {Trabucchi}, M. and {Tsounis}, A.~T. and {Turon}, C. and {Ulla}, A. and {Unger}, N. and {Vaillant}, M.~V. and {van Dillen}, E. and {van Reeven}, W. and {Vanel}, O. and {Vecchiato}, A. and {Viala}, Y. and {Vicente}, D. and {Voutsinas}, S. and {Weiler}, M. and {Wevers}, T. and {Wyrzykowski}, {\L}. and {Yoldas}, A. and {Yvard}, P. and {Zhao}, H. and {Zorec}, J. and {Zucker}, S. and {Zwitter}, T.},
        title = "{Gaia Data Release 3. Summary of the content and survey properties}",
      journal = {\aap},
         year = 2023,
        month = jun,
       volume = {674},
          eid = {A1},
        pages = {A1},
          doi = {10.1051/0004-6361/202243940},
archivePrefix = {arXiv},
       eprint = {2208.00211},
 primaryClass = {astro-ph.GA},
       adsurl = {https://ui.adsabs.harvard.edu/abs/2023A&A...674A...1G}
}

@ARTICLE{2022AJ....163...64E,
       author = {{Esplin}, T.~L. and {Luhman}, K.~L.},
        title = "{A Census of Stars and Disks in Corona Australis}",
      journal = {\aj},
         year = 2022,
        month = feb,
       volume = {163},
       number = {2},
          eid = {64},
        pages = {64},
          doi = {10.3847/1538-3881/ac3e64},
archivePrefix = {arXiv},
       eprint = {2111.14903},
 primaryClass = {astro-ph.SR},
       adsurl = {https://ui.adsabs.harvard.edu/abs/2022AJ....163...64E}
}

@ARTICLE{2019A&A...626A..11C,
       author = {{Cazzoletti}, P. and {Manara}, C.~F. and {Liu}, H. Baobab and {van Dishoeck}, E.~F. and {Facchini}, S. and {Alcal{\`a}}, J.~M. and {Ansdell}, M. and {Testi}, L. and {Williams}, J.~P. and {Carrasco-Gonz{\'a}lez}, C. and {Dong}, R. and {Forbrich}, J. and {Fukagawa}, M. and {Galv{\'a}n-Madrid}, R. and {Hirano}, N. and {Hogerheijde}, M. and {Hasegawa}, Y. and {Muto}, T. and {Pinilla}, P. and {Takami}, M. and {Tamura}, M. and {Tazzari}, M. and {Wisniewski}, J.~P.},
        title = "{ALMA survey of Class II protoplanetary disks in Corona Australis: a young region with low disk masses}",
      journal = {\aap},
         year = 2019,
        month = jun,
       volume = {626},
          eid = {A11},
        pages = {A11},
          doi = {10.1051/0004-6361/201935273},
archivePrefix = {arXiv},
       eprint = {1904.02409},
 primaryClass = {astro-ph.EP},
       adsurl = {https://ui.adsabs.harvard.edu/abs/2019A&A...626A..11C}
}
\onecolumn

\begin{appendix}
\section{Other detections in the survey}\label{appendix:tent_det}
We show the previously known millimeter sources that are detected in this survey but are not on the primary target list of this survey in Fig. \ref{fig:nd_gallery}. Their coordinates and the references that reported them are listed in Table \ref{tb:nd_table}. We also show the newly identified millimeter sources that are not included in \textsc{SIMBAD} in Fig. \ref{fig:ni_gallery}.

We cross-match sources listed in Table \ref{tb:nd_table} with those in the Spitzer c2d \citep{2009ApJS..181..321E} and Gould Belt \citep{2015ApJS..220...11D} surveys, and measure their infrared spectral slopes between $3.6$ and $24~\mathrm{\mu m}$, as described in Section \ref{sec:sample} and in \citet{2022ApJ93855A}. After excluding sources located outside the sky region shown in Fig. \ref{fig:scatter} and sources with Gaia DR3 memberships but with distances outside the range of 275-725 pc, this yields 26 sources with known infrared spectral slopes, consisting of 10 Class I objects, 7 Flat Spectrum objects and 9 Class II objects. Only 7 of the 26 sources have known Gaia parallaxes. We therefore adopt the median distance of the  sub-cluster to which each source is closest in the 2D projected sky plane. This assignment yields 15 sources in Serpens Main, 10 sources in Serpens South and 1 source in Serpens Northeast. We measure their disk dust masses using the method described in Section \ref{subsec:disk_mass}. We then combine these 26 sources with the primary targets of this survey to re-assess the cumulative disk dust mass distributions. The result is shown in Fig. \ref{fig:cumulative_distr_wnt}. We perform log-rank tests to assess whether the inclusion of these additional detected targets results in cumulative disk dust mass distributions that are statistically significantly different from those derived from the primary targets of this survey alone. We find no statistically significant differences, with p-values of $\gtrsim 0.7$ for all three infrared classifications.

\begin{figure*}[ht!]
    \centering
    \includegraphics[width=\textwidth]{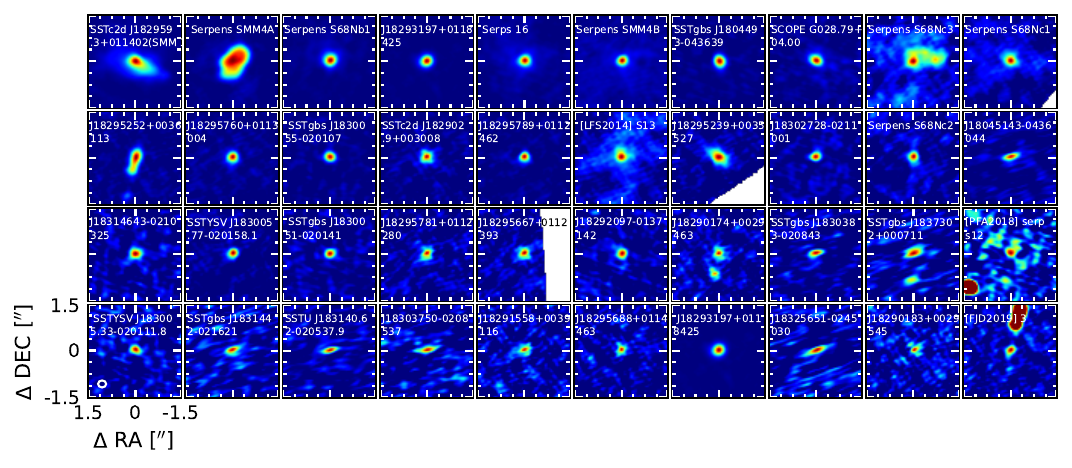}
    \caption{1.3-mm continuum YSO candidates that have been observed before but are not part of our samples selected in Section \ref{sec:sample} are plotted in boxes of ${3^{\prime\prime}}\times {3^{\prime\prime}}$. Incomplete images are shown for sources located at the edge of the field of view. Targets are sorted by the continuum flux measured by aperture analysis. A representative beam of $0.^{\prime\prime}25 \times 0.^{\prime\prime}22$ ($89.^{\circ}07$) is shown at the left corner of the figure.}
    \label{fig:nd_gallery}
\end{figure*}

\begin{table}[htb!]
\renewcommand{\arraystretch}{1.1}
\centering
\caption{Summary of previously detected millimeter sources shown in Fig. \ref{fig:nd_gallery}. }\label{tb:nd_table}
\begin{tabular}{cccccc}
\hline
\hline
 & Source name & RA & Dec & Survey & Ref. \\
 \hline
1&SSTc2d J182959.3+011402(SMM 3)&277.4972&1.2334&Spitzer& \\
2& Serpens SMM4A&277.4863&1.2209&ALMA& 1\\
3&Serpens S68Nb1&277.4563&1.2864&ALMA& 2 \\
4&J18293197+0118425&277.3833&1.3118&2MASS/Spitzer/WISE/IRAS& \\
5& Serps 16&277.5100&-2.0470&ALMA& 3\\  
6&Serpens SMM4B &277.4855&1.2198&ALMA& 1\\
7&SSTgbs J1804493-043639&271.2055&-4.6111&Spitzer& \\
8&SCOPE G028.79+04.00&277.4068&-1.8499&JCMT& 4\\
9&Serpens S68Nc3&277.4536&1.2845&ALMA& 2 \\
10&Serpens S68Nc1&277.4530&1.2821&ALMA& 2,5\\
11&J18295252+0036113&277.4689&0.6032&2MASS/Spitzer& \\
12&J18295760+0113004&277.4899&1.2167&2MASS/Spitzer& \\
13& SSTgbs J1830055-020107&277.5230&-2.0189&Spitzer& \\
14&SSTc2d J182902.9+003008 &277.2623&0.5022&Spitzer& \\
15&J18295789+0112462 &277.4912&1.2128&2MASS/Spitzer& \\
16& [LFS2014] S13&277.4926&1.2208&CARMA& 6\\
17&J18295239+0035527&277.4684&0.5980&2MASS/Spitzer& \\
18&J18302728-0211001&277.6137&-2.1834&2MASS/Spitzer& \\
19&Serpens S68Nc2&277.4541&1.2853&ALMA& 2\\
20&J18045143-0436044&271.2144&-4.6013&2MASS/Spitzer& \\
21&J18314643-0210325&277.9435&-2.1758&2MASS/Spitzer& \\
22&SSTYSV J183005.77-020158.1  &277.5241&-2.0328&Spitzer& \\
23& SSTgbs J1830051-020141&277.5216&-2.0283&Spitzer& \\
24&J18295781+0112280 &277.4909&1.2077&2MASS/Spitzer& \\
25&J18295667+0112393&277.4862&1.2108&2MASS/Spitzer& \\
26&J18292097-0137142&277.3374&-1.6207&2MASS/Spitzer& \\
27&J18290174+0029463&277.2573&0.4962&2MASS/Spitzer/WISE& \\
28&SSTgbs J1830383-020843&277.6597&-2.1455&Spitzer& \\
29&SSTgbs J1837302+000711&279.3761&0.1197&Spitzer& \\
30&[PFA2018] serps12 &277.5052&-2.0617&ALMA& 3\\
31& SSTYSV J183005.33-020111.8&277.5223&-2.0200&Spitzer& \\
32&SSTgbs J1831442-021621&277.9344&-2.2726&Spitzer& \\
33&SSTU J183140.62-020537.9&277.9192&-2.0939&Spitzer& \\
34&J18303750-0208537&277.6563&-2.1483&2MASS/Spitzer& \\
35&J18291558+0039116&277.3150&0.6532&2MASS/Spitzer& \\
36&J18295688+0114463&277.4870&1.2461&2MASS/Spitzer& \\
37& J18293197+0118425&277.3832&1.3118&2MASS/Spitzer/IRAS& \\
38&J18325651-0245030&278.2355&-2.7509&2MASS/Spitzer& \\
39&J18290183+0029545&277.2577&0.4984&2MASS/Spitzer& \\
40&[FJD2019] 3 &277.4690&0.6029&ALMA/CARMA& 5\\
 \hline
\end{tabular}
\tablefoot{The instruments used for observations are listed for sources that have only been observed in millimeter, and corresponding references are added in the last column. References: 1. \citet{2018ApJ...863...19A}; 2. \citet{2019ApJ...887..209A}; 3. \citet{2018A&A...615A...9P}; 4. \citet{2019MNRAS.485.2895E}; 5. \citet{2019ApJ...871..149F}; 6. \citet{2014ApJ...797...76L}.}
\end{table}

\begin{figure*}[ht!]
    \centering
    \includegraphics[width=\textwidth]{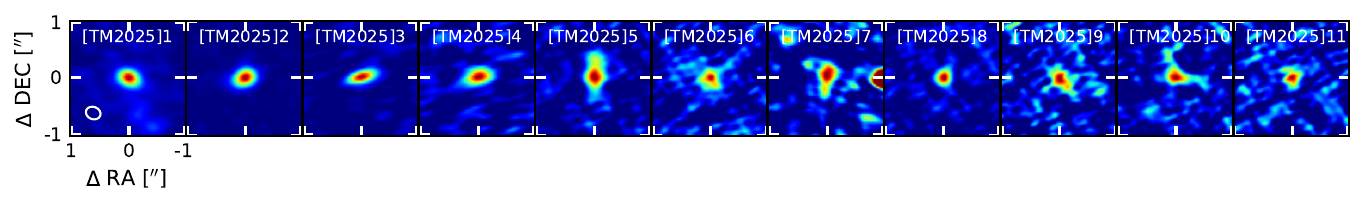}
    \caption{1.3-mm continuum millimeter sources that are newly detected in this survey and that have not been reported before. The gallery is sorted by their fluxes and each source is presented in a ${2}^{\prime\prime}\times {2}^{\prime\prime}$ box. A representative beam of $0.^{\prime\prime}26 \times 0.^{\prime\prime}21$ ($-67.^{\circ}86$) is shown at the left corner of the figure. We named them by [TM2025] plus their orders by their continuum fluxes.}
    \label{fig:ni_gallery}
\end{figure*}

\begin{figure}
    \centering
    \includegraphics[width=0.43\linewidth]{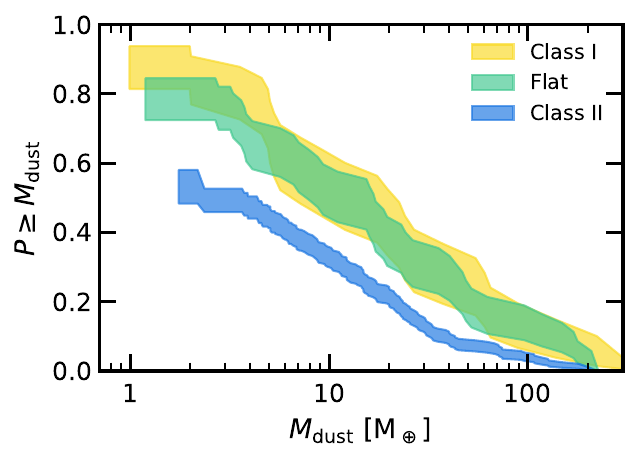}
    \caption{Cumulative disk dust mass distributions for Class I, Flat spectrum and Class II young stellar objects in the Serpens star-forming region. Similar as Fig. \ref{fig:cum_mass_distri_multi}, but including 24 sources from Table \ref{tb:nd_table} that have known fluxes at $3.6$ and $24~\mathrm{\mu m}$ from Spitzer c2d \citep{2009ApJS..181..321E} and Gould belt \citep{2015ApJS..220...11D} surveys.}
    \label{fig:cumulative_distr_wnt}
\end{figure}

\section{Modelling of transition disk candidates}\label{appendix:TD_candidate}
We show the results of visibility modeling for disks listed in Table \ref{tb:TD_candidate} in Fig. \ref{fig:TD_cand_vis}, except for J18295533+0049391 and J18273858-0402289, which are shown in Fig. \ref{fig:2td_vis}. The best-fit parameters of these models are listed in Table \ref{tb:mcmc_params}, and the corresponding synthetic observations of models are shown in Fig. \ref{fig:TD_cand_model}. The posterior corner plots for J18295533+0049391 and J18273858-0402289 are shown in Fig. \ref{fig:TD1829} and Fig. \ref{fig:TD1827}, respectively.

\begin{figure*}[ht!]
    \centering
    \includegraphics[width=0.75\textwidth]{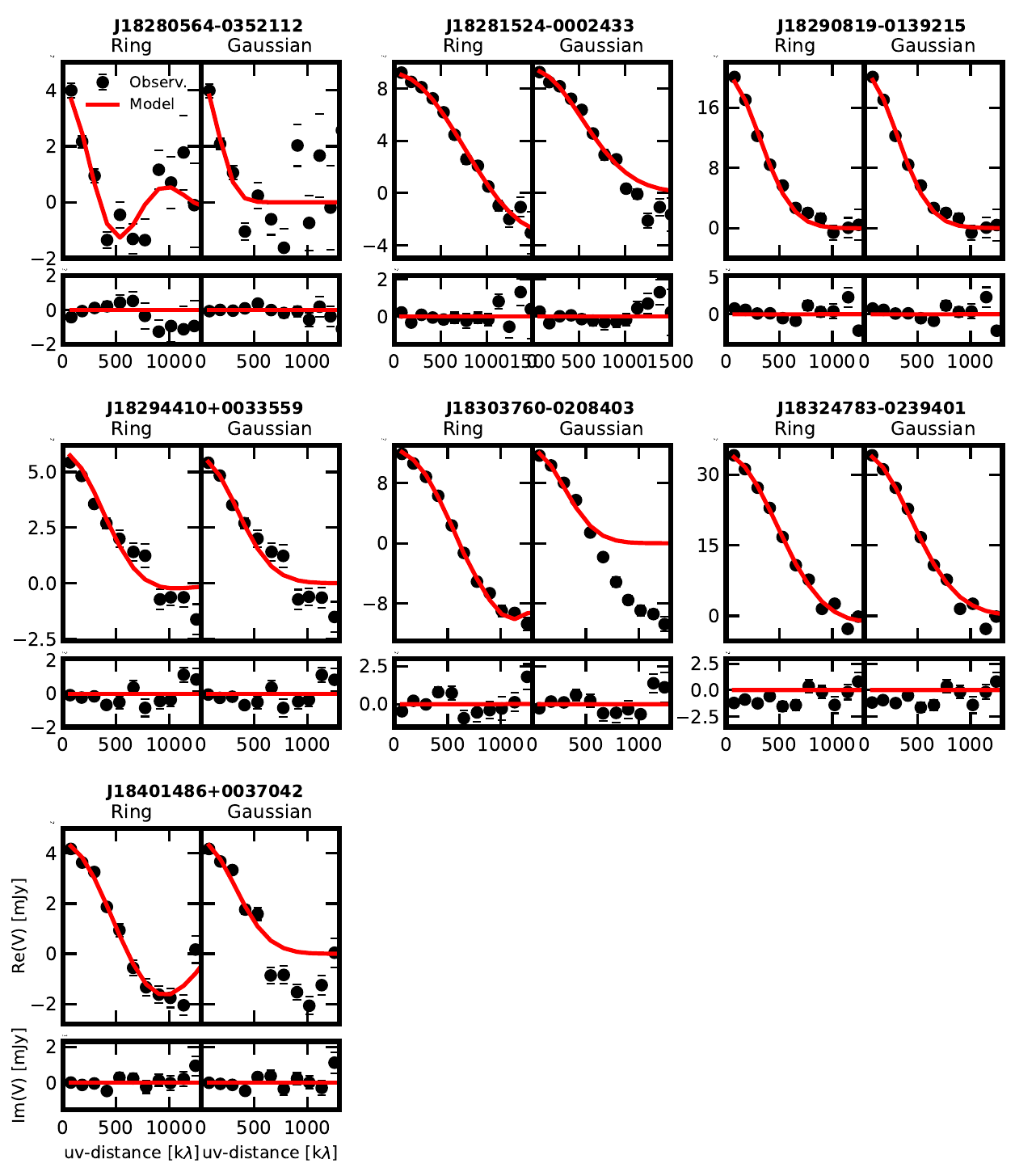}
    \caption{Comparison of visibility models (red lines) for the best-fit ring (left) and Gaussian (right) models to the binned observed visibility (black dots).}
    \label{fig:TD_cand_vis}
\end{figure*}
\begin{figure*}[ht!]
    \centering
    \includegraphics[width=0.78\textwidth]{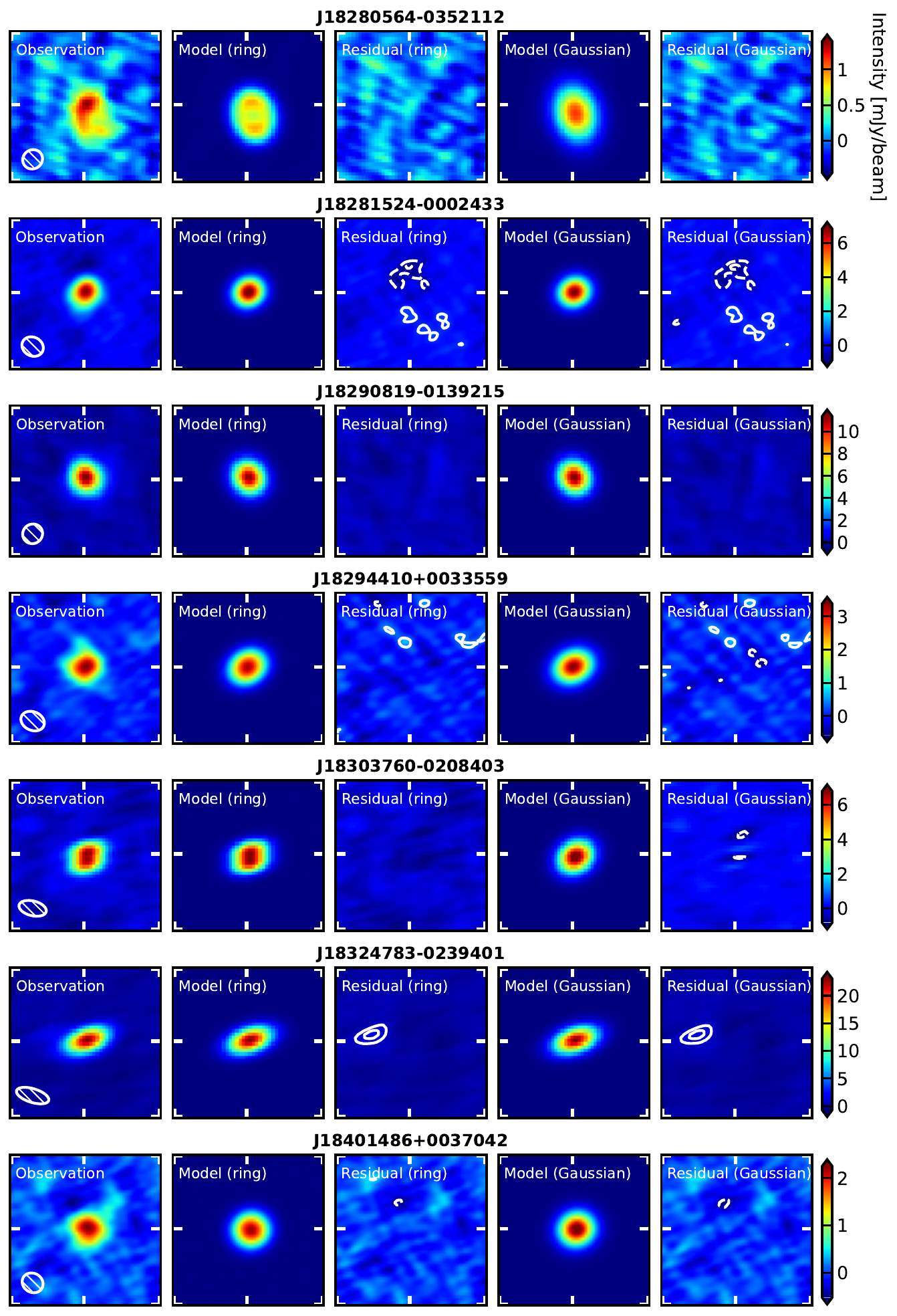}
    \caption{1.3-mm continuum imaging for transition disk candidates: observations (column 1), synthetic observations of the best-fit ring model (column 2), and residuals between the ring models and the observations (column 3). Columns 4 and 5 show synthetic observations of the best-fit Gaussian models (column 4) and residuals between the Gaussian model and observations (column 5). The over-plotted contours in Columns 3 and 5 indicate the 5, 8 and 11 times of the rms in the image (dashed lines for negative value and solid lines for positive value).}
    \label{fig:TD_cand_model}
\end{figure*}

\begin{landscape}
\begingroup
\setlength{\tabcolsep}{2pt}
\renewcommand{\arraystretch}{1.1}
\begin{longtable}{cccccccccc}
\caption{The best-fit parameters for ring and Gaussian models of transition disk candidates.}\label{tb:mcmc_params}
\\
\hline
\hline
2MASS ID  & Models   & $I_r$ & $R_r$              & $R_w$ & $i$ & PA & dRA & dDec &$r_\mathrm{cav}$\\[1em]
&&[Jy~sr$^{-1}$]&[${}^{\prime\prime}$]&[${}^{\prime\prime}$]&[${}^{\circ}$]&[${}^{\circ}$]&[${}^{\prime\prime}$]&[${}^{\prime\prime}$]& [AU]\\
\hline
\endfirsthead
\hline\hline
2MASS ID          & Models   & $I_r$ & $R_r$              & $R_w$ & $i$ & PA & dRA & dDec &$r_\mathrm{cav}$\\[1em]
&&[Jy~sr$^{-1}$]&[${}^{\prime\prime}$]&[${}^{\prime\prime}$]&[${}^{\circ}$]&[${}^{\circ}$]&[${}^{\prime\prime}$]&[${}^{\prime\prime}$]& [AU]\\
\hline
\endhead
\\
\multirow{2}{*}{18295533+0049391}& Ring &$9.82\substack{+0.006\\-0.007}$&
$0.20\substack{+0.001\\-0.001}$&
$0.08\substack{+0.002\\-0.002}$&
$26.71\substack{+0.932\\-1.022}$&
$29.31\substack{+2.385\\-2.121}$&
$0.02\substack{+0.001\\-0.001}$&
$-0.02\substack{+0.001\\-0.001}$& $94.19\substack{+1.14\\-1.14}$\\ \cline{2-2} & Gaussian &  -- & -- & -- & -- & -- & -- &  --&--\\
\\
\multirow{2}{*}{18273858-0402289} & Ring &$9.79\substack{+0.057\\-0.046}$&$0.19\substack{+0.003\\-0.003}$&$0.04\substack{+0.005\\-0.005}$& $30.67\substack{+2.374\\-3.078}$&$61.44\substack{+5.981\\-6.163}$&$0.04\substack{+0.003\\-0.003}$&$0.01\substack{+0.003\\-0.003}$& $86.85\substack{+2.54\\-2.54}$\\ \cline{2-2} & Gaussian &  -- & -- & -- & -- &  -- &  -- &  -- &  --\\
\\
\multirow{2}{*}{18280564-0352112}& Ring &$9.20\substack{+0.360\\-0.174}$&$0.23\substack{+0.015\\-0.016}$&$0.04\substack{+0.025\\-0.024}$&$42.79\substack{+5.625\\-9.920}$&$11.24\substack{+10.907\\-7.068}$&$0.08\substack{+0.013\\-0.013}$&$0.03\substack{+0.015\\-0.012}$& $100.03\substack{+12.82\\-13.04}$\\\cline{2-2}&Gaussian&$8.98\substack{+0.052\\-0.050}$&--&$0.21\substack{+0.024\\-0.025}$&$47.03\substack{+7.438\\-14.024}$&$13.52\substack{+15.508\\-8.805}$&$0.13\substack{+0.016\\-0.015}$&$0.06\substack{+0.020\\-0.019}$&  --\\
\\
\multirow{2}{*}{18281524-0002433}& Ring &$10.43\substack{+0.142\\-0.127}$&
$0.07\substack{+0.002\\-0.003}$&
$0.01\substack{+0.005\\-0.004}$&
$22.93\substack{+5.878\\-11.353}$&
$77.42\substack{+12.977\\-14.363}$&
$0.01\substack{+0.001\\-0.001}$&
$-0.00\substack{+0.001\\-0.001}$&$21.76\substack{+1.48\\-1.64}$
\\\cline{2-2} & Gaussian& $10.24\substack{+0.015\\-0.015}$&--&
$0.06\substack{+0.002\\-0.002}$&
$15.32\substack{+9.082\\-10.700}$&
$66.53\substack{+30.834\\-26.358}$&
$0.01\substack{+0.002\\-0.001}$&
$-0.00\substack{+0.001\\-0.001}$&  --\\
\\

\multirow{2}{*}{18290819-0139215}& Ring &$10.16\substack{+0.015\\-0.018}$&$0.01\substack{+0.008\\-0.004}$&$0.10\substack{+0.003\\-0.004}$&$33.46\substack{+2.912\\-3.072}$&$29.00\substack{+4.547\\-5.047}$&$0.02\substack{+0.002\\-0.002}$&$0.02\substack{+0.002\\-0.002}$&$5.01\substack{+4.10\\-2.18}$\\\cline{2-2} & Gaussian& $10.17\substack{+0.011\\-0.011}$&--&$0.11\substack{+0.002\\-0.002}$&$33.51\substack{+2.956\\-3.203}$&$29.14\substack{+5.340\\-5.408}$&$0.02\substack{+0.001\\-0.002}$&$0.02\substack{+0.002\\-0.002}$&  --\\
\\
\multirow{2}{*}{18294410+0033559}& Ring &$10.06\substack{+0.088\\-0.077}$&
$0.06\substack{+0.027\\-0.036}$&
$0.07\substack{+0.017\\-0.019}$&
$73.26\substack{+3.125\\-3.140}$&
$34.60\substack{+2.538\\-2.590}$&
$0.01\substack{+0.003\\-0.002}$&
$0.01\substack{+0.003\\-0.003}$&$27.98\substack{+12.62\\-16.81}$\\\cline{2-2} & Gaussian& $10.14\substack{+0.077\\-0.067}$&--&
$0.10\substack{+0.004\\-0.004}$&
$73.90\substack{+2.794\\-2.828}$&
$34.59\substack{+2.598\\-2.796}$&
$0.01\substack{+0.002\\-0.002}$&
$0.01\substack{+0.003\\-0.003}$&  --\\
\\
\multirow{2}{*}{18303760-0208403}& Ring &$11.12\substack{+0.044\\-0.036}$&$0.11\substack{+0.002\\-0.002}$&$0.00\substack{+0.001\\-0.001}$&$53.16\substack{+1.723\\-1.773}$&$0.47\substack{+0.767\\-0.345}$&$-0.00\substack{+0.002\\-0.002}$&$-0.00\substack{+0.001\\-0.001}$&$55.13\substack{+9.54\\-9.54}$\\\cline{2-2} & Gaussian& $10.54\substack{+0.230\\-0.115}$&--&
$0.11\substack{+0.002\\-0.002}$&
$78.51\substack{+4.911\\-3.538}$&
$0.94\substack{+1.122\\-0.667}$&
$-0.00\substack{+0.002\\-0.002}$&
$-0.00\substack{+0.002\\-0.002}$&  --\\
\\
\multirow{2}{*}{18324783-0239401}& Ring &$10.59\substack{+0.017\\-0.017}$&$0.04\substack{+0.005\\-0.006}$&$0.06\substack{+0.003\\-0.003}$&$32.09\substack{+2.759\\-3.464}$&$4.36\substack{+4.596\\-2.970}$&$-0.00\substack{+0.001\\-0.001}$&$0.01\substack{+0.001\\-0.001}$&$25.58\substack{+7.51\\-7.80}$\\\cline{2-2} & Gaussian& $10.70\substack{+0.014\\-0.015}$&--&
$0.07\substack{+0.001\\-0.001}$&
$34.23\substack{+2.779\\-4.005}$&
$4.83\substack{+4.606\\-3.134}$&
$-0.00\substack{+0.001\\-0.001}$&
$0.01\substack{+0.001\\-0.001}$&  --\\
\\
\multirow{2}{*}{18401486+0037042}& Ring &$10.01\substack{+0.164\\-0.174}$&
$0.13\substack{+0.004\\-0.004}$&
$0.01\substack{+0.006\\-0.004}$&
$34.39\substack{+5.307\\-6.493}$&
$28.39\substack{+8.646\\-8.323}$&
$0.02\substack{+0.004\\-0.004}$&
$0.00\substack{+0.003\\-0.003}$&$81.20\substack{+12.90\\-12.90}$\\\cline{2-2} & Gaussian& $9.50\substack{+0.025\\-0.027}$&--&
$0.10\substack{+0.005\\-0.005}$&
$28.62\substack{+7.737\\-13.780}$&
$40.29\substack{+19.767\\-17.761}$&
$0.02\substack{+0.005\\-0.004}$&
$0.00\substack{+0.004\\-0.004}$&  --\\
\hline
\end{longtable}
\endgroup
\end{landscape}

\begin{figure}
    \centering
    \includegraphics[width=0.6\textwidth]{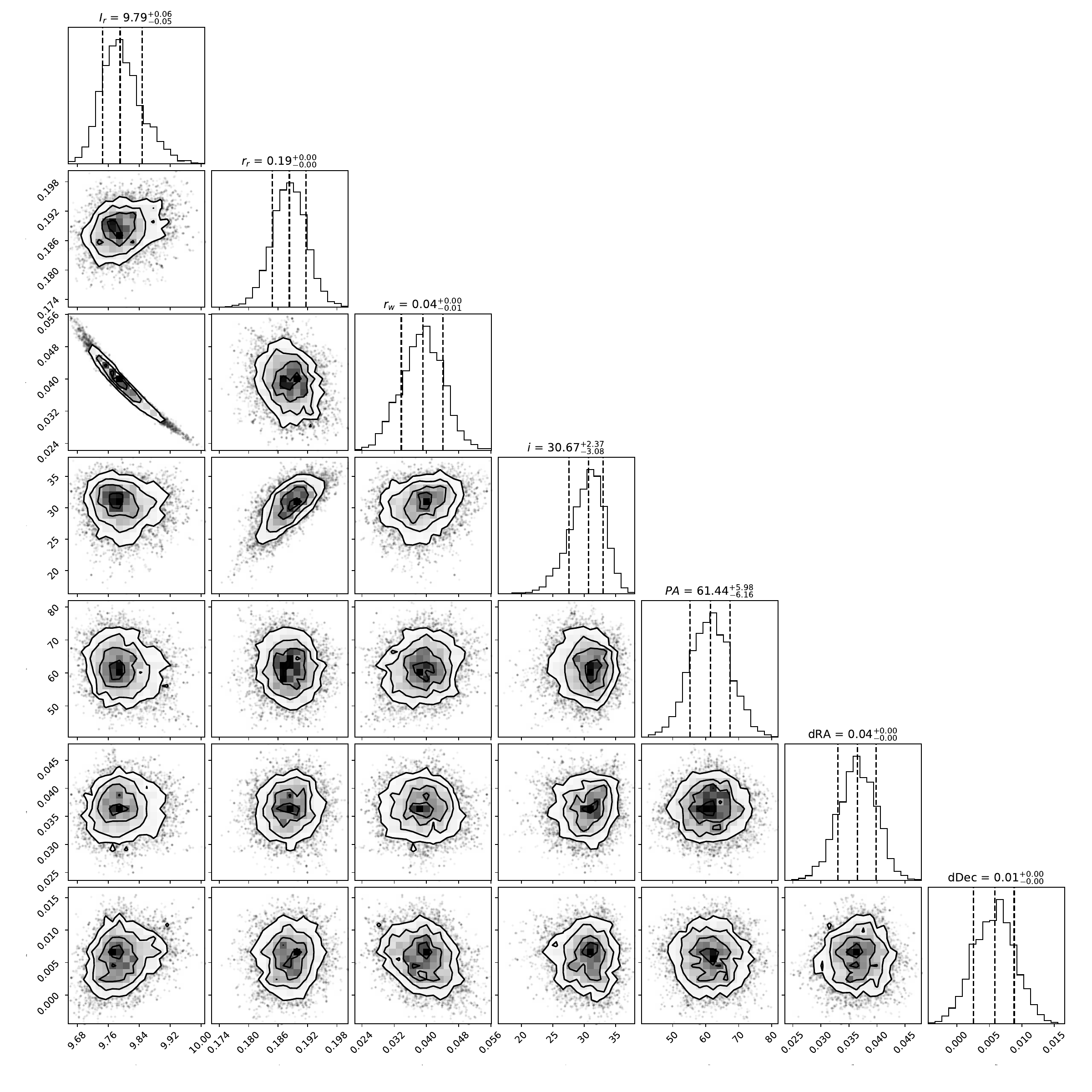}
    \caption{Posterior of the transition disk candidate 18273858-0402289. The banana-shape posterior shows the degeneracy between the ring intensity $I$ and the ring width $r_w$. Best-fit parameters are listed in Table \ref{tb:mcmc_params}.}
    \label{fig:TD1827}
\end{figure}

\begin{figure}
    \centering
    \includegraphics[width=0.6\textwidth]{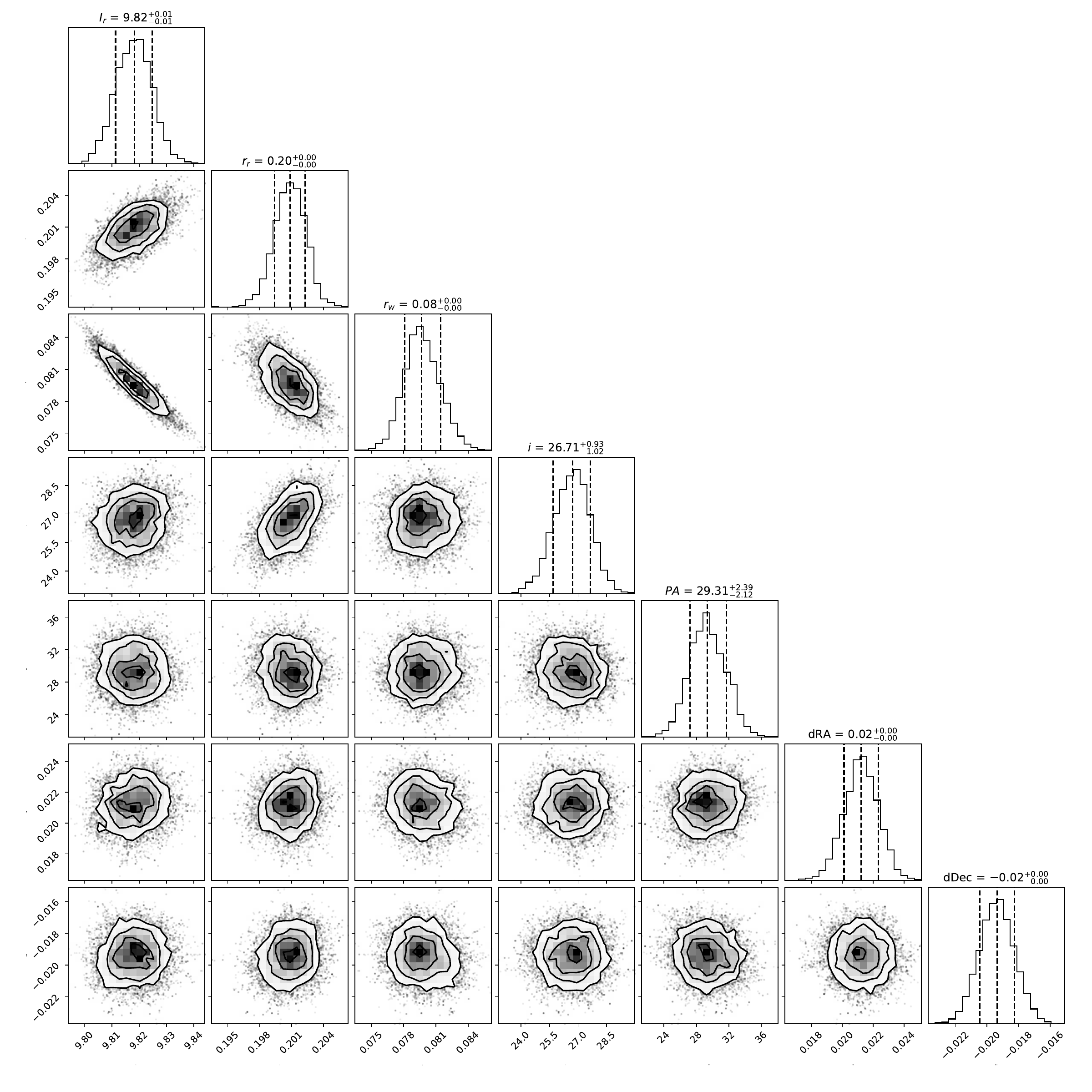}
    \caption{Posterior of the transition disk candidate 18295533+0049391. The elongated posterior shows the degeneracy between the ring intensity $I$ and the ring width $r_w$. Best-fit parameters are listed in Table \ref{tb:mcmc_params}.}
    \label{fig:TD1829}
\end{figure}

\section{Distributions of distances and disk dust masses for targets in different star-forming regions}
\subsection{Distributions of disk distances}
We plot the distributions of distances for disks in each star-forming region in Figure \ref{fig:distance}. Distances shown here are from disks with known Gaia DR3 parallaxes, and are the median values. Serpens appears more extended along the light of sight than other star-forming regions. We list in Table \ref{tb:known_distance} the number of sources with and without known Gaia DR3 parallaxes, the number of sources with parallaxes but with $\overline{\omega}/\sigma(\overline{\omega})<5$, and the number of sources outside the adopted distance range for each star-forming region.
\begin{figure*}
    \centering
    \includegraphics[width=0.99\textwidth]{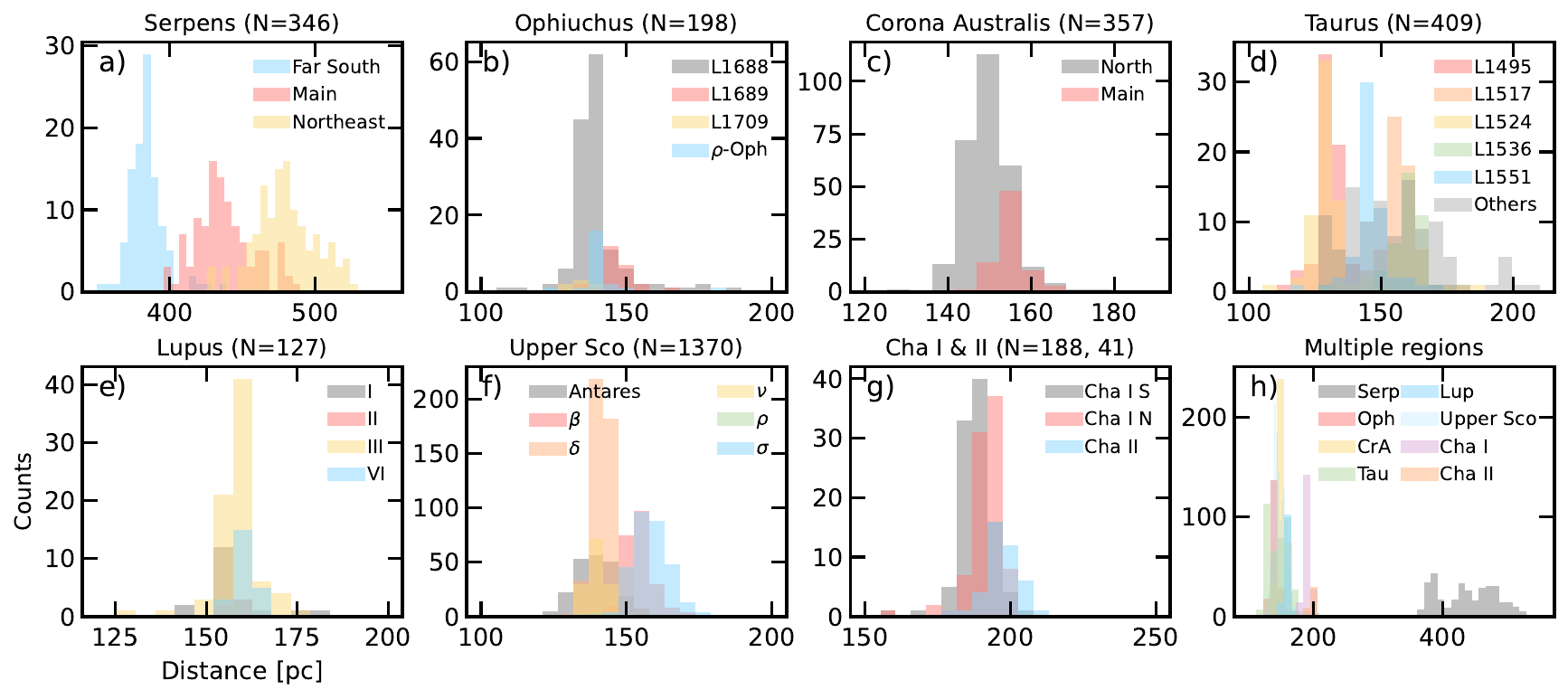}
    \caption{Distributions of known median distances derived from Gaia DR3 parallaxes for YSOs in star-forming regions: Serpens (a), Ophiuchus (b), Corona Australis (c), Taurus (d), Lupus (e), Upper Sco (f), Cha I \& II (g) and summary of them (h). The number of YSOs with known distances in each region is also noted in the title. Sub-clusters in Taurus with less than $30$ members, including B209N, HD28354, L1489/L1498, L1521/B213, L1527, L1544, L1558 and T Tau, are shown in gray as a whole.}
    \label{fig:distance}
\end{figure*}

\renewcommand{\arraystretch}{1.03}
\begin{table}[]
\begin{center}
\caption{Numbers and fractions of targets with unknown Gaia parallaxes, either being masked or no Gaia identifications in star-forming regions.}\label{tb:known_distance}
\begin{tabular}{cccccc}
\hline
\hline
Regions & Sub-cluster & \# known & \# unknown & \#$\mathrm{SNR<5}$ & \# removed \\
\hline
Serpens&Serpens-NE & 127 & 0 & 0 & 0\\
&Serpens-Main & 112 & 0 & 0 & 0\\
&Serpens-Far South & 107 & 0 & 0& 0\\

Ophiuchus &Oph-$\rho$ Oph & 23 & 4 & 1 & 0\\
&Oph-L1709 & 7 & 0 & 0 & 0\\
&Oph-L1689 & 25 & 3 & 0 & 0\\
&Oph-L1688 & 143 & 69 & 2 & 2\\

Lupus &Lupus-I & 20 & 3 & 0 & 0\\
&Lupus-II & 5 & 1 & 0 & 0\\
&Lupus-III & 78 & 7 & 1 & 1\\
&Lupus-IV & 24 & 5 & 0 & 0\\

Chamaeleon & Cham I-N & 87 & 0 & 0 & 0\\
& Cham I-S & 101 & 0 & 0 & 0\\
& Cham II & 41 & 0 & 0 & 0\\       

Corona Australis & CrA core & 16 & 18 & 0 & 0 \\
& CrA Main & 77 & 3 & 1 & 2 \\
& CrA North & 277 & 0 & 0 & 0 \\
& CrA (ALMA) & 82 & 21 & 1 & 2 \\
Taurus & B209N & 6  & 0  & 0 & 0\\
 & HD28354 & 16  & 0  & 0 & 0\\
 & L1489/L1498 & 6  & 0  & 0 & 0\\
 & L1495/B209 & 69  & 2  & 0 & 0\\
 & L1517 & 62  & 0  & 0 & 0\\
 & L1521/B213 & 28  & 0  & 0 & 0\\
 & {\small L1524/L1529/B215} & 66  & 0  & 0 & 0\\
 & L1527 & 27  & 1  & 0 & 0\\
 & L1536 & 42  & 0  & 1 & 0\\
 & L1544 & 15  & 0  & 0 & 0\\
 & L1551 & 58  & 1  & 0 & 0\\
 & L1558 & 9  & 0  & 0 & 0\\
 & T Tau & 5  & 0  & 0 & 0\\

Upper Sco & $\nu$ Sco & 135 & 0 & 0 & 0 \\
   & $\delta$ Sco & 489 & 0 & 0 & 0 \\
   & $\beta$ Sco & 223 & 0 & 0 & 0 \\
   & $\sigma$ Sco & 300 & 0 & 0 & 0 \\
   & $\rho$ Sco & 9 & 0 & 0 & 0 \\
   & Antares & 214 & 0 & 0 & 0 \\
        
\hline
\end{tabular}
\end{center}
\end{table}

\subsection{Disk dust mass distributions}\label{appendix:disk mass distri}

We compare the disk dust mass distributions for Class II disks in regions with a similar age of $<3~\mathrm{Myr}$. This includes Corona Australis \citep{2019A&A...626A..11C}, Ophiuchus \citep{2019ApJ...875L...9W}, Serpens (this work), Lupus \citep{2018ApJ...859...21A} and Chamaeleon I \citep{2016ApJ...831..125P}. The disk dust masses are computed under the assumptions of optically thin, following the method described in Section \ref{subsec:disk_mass}. Targets without Gaia parallaxes are assigned with the mean distance of the entire star-forming region.
\begin{figure}
    \centering
    \includegraphics[width=0.4\textwidth]{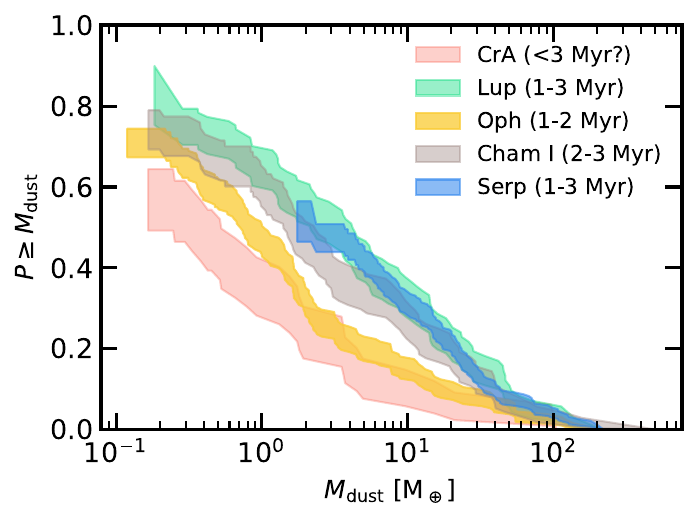}
    \caption{Cumulative disk dust mass distributions (see Section \ref{subsec:disk_mass}) for Class II disks in Corona Australis \citep[denoted as CrA, $<3~$Myr, e.g.][though more recent studies suggest a more evolved age as discussed in Section \ref{subsec:comparison_with_stellar_density}]{2005A&A...429..543N, 2008ApJ...687.1145S, 2009PASP..121..350M, 2011ApJ...736..137S, 2019A&A...626A..11C}, Lupus \citep[denoted as Lup, $1-3~$Myr, e.g.][]{2008hsf2.book..295C, 2018ApJ...859...21A}, Ophiuchus \citep[denoted as Oph, $1-2$~Myr, e.g.][]{1999ApJ...525..440L, 2005AJ....130.1733W, 2019ApJ...875L...9W, 2019MNRAS.482..698C}, Chamaeleon I \citep[denoted as Cham I, $2-3~$Myr, e.g.][]{2004ApJ...602..816L, 2008ApJ...675.1375L, 2016ApJ...831..125P} and Serpens \citep[denoted as Serp, $1-3~$Myr, e.g.][]{2013ApJ...762..128O, 2019ApJ...878..111H, 2022ApJ93855A}.}
    \label{fig:cum_mass_distri_multi}
\end{figure}
\end{appendix}

\end{document}